\documentclass{article}

\usepackage{arxiv}

\usepackage[utf8]{inputenc} 
\usepackage[T1]{fontenc}    
\usepackage{hyperref}       
\hypersetup{colorlinks,allcolors=black}
\usepackage{url}            
\usepackage{booktabs}       
\usepackage{amsfonts}       
\usepackage{nicefrac}       
\usepackage{microtype}      
\usepackage{lipsum}		
\usepackage{graphicx}
\usepackage{natbib}
\usepackage{doi}
\usepackage{comment}
\usepackage{subcaption}
\usepackage{xcolor}
\usepackage{soul}
\usepackage{newtxtext}
\usepackage{newtxmath}

\title{Dynamical stability and flow regimes in a stably stratified valley-shaped cavity heated from below}

\author{{Patrick J.~Stofanak, Cheng-Nian Xiao, \& Inanc Senocak\thanks{senocak@pitt.edu} } \\
	Department of Mechanical Engineering and Materials Science\\
	University of Pittsburgh\\
	Pittsburgh, PA 15261 \\
}

\date{}

\renewcommand{\shorttitle}{Stability and flow regimes in heated triangular cavities}

\hypersetup{
pdftitle={Dynamical stability and flow regimes in a stably stratified valley-shaped cavity heated from below},
pdfauthor={Patrick J.~Stofanak, Cheng-Nian Xiao, Inanc Senocak},
pdfkeywords={atmospheric flows, flow instability, slope flows, stratified flows, valley flows},
}

\begin{document}
\maketitle

\begin{abstract}
We investigate the three-dimensional stability of a stably stratified fluid in a valley-shaped  cavity heated from below using linear stability analysis and direct numerical simulations. First, we describe the quiescent, pure-conduction flow state and derive a dimensionless, necessary condition that establishes a lower bound for the critical value at which this state becomes unstable, applicable to any slope angle $\alpha$ of the valley-shaped cavity. Next, we examine the sequence of flow regimes for slope angle of $\alpha = 30^{\circ}$ and Prandtl number of $Pr = 7$, including two-dimensional steady states, the onset of a Hopf bifurcation, and the development of both steady and oscillatory three-dimensional structures preceding the transition to fully unsteady, chaotic flow. Our results demonstrate that, although the governing nonlinear equations formally depend on two dimensionless parameters at a fixed slope angle and Prandtl number, the flow dynamics across a broad range of flow regimes effectively collapse onto a single dimensionless parameter--the composite stratification parameter, $\Pi_c$, as supported by the equations of linear stability analysis. However, as the flow progress toward a chaotic state, an increasing dependence on the additional parameter, $\Pi_h$, emerges. We construct a regime map of the observed flow states as a function of $\Pi_c$ and $\Pi_h$, and confirm the transition to chaotic flow through the computation of Lyapunov exponents. Moreover, we find that across all flow regimes, asymmetric circulation remains the dominant flow pattern in the cavity. Even in the fully unsteady, chaotic regime, this asymmetric pattern persists in the time-averaged field. Finally, we quantify heat transfer within the cavity using the Nusselt number scaling of $Nu \sim \Pi_{c}^{0.43}$, or equivalently $Nu \sim Ra^{0.275}$, across all the flow regimes identified in this study. This result further establishes $\Pi_c$ as a key dimensionless parameter governing the flow dynamics preceeding the chaotic regime.
\end{abstract}


\section{Introduction}
\label{sec:intro}

During the nighttime, surface radiative cooling leads to downslope (katabatic) flows that accumulate at the valley floor, forming a stably stratified layer, or cold pool, that persists throughout the night. After sunrise, surface heating initiates upslope (anabatic) flows, which gradually erode the cold pool and lead to the development of the convective boundary layer as the day progresses. This transition, known as the morning transition, plays a crucial role in phenomena such as pollutant transport, fog, and frost formation. However, numerical weather prediction models struggle to accurately capture this period due to the interplay of processes from both the convective and stable boundary layers \citep{angevine2020transition}. This challenge is further exacerbated in valleys, where complex terrain introduces additional uncertainties due to inadequate parameterizations \citep{serafin2018exchange}. Therefore, gaining a deeper understanding of the fundamental processes governing the morning transition in valleys is a critical first step toward improving their representation in numerical weather prediction models.

Numerous studies have examined large-scale idealized valley flow systems across a wide range of valley geometries using mesoscale numerical weather prediction models. Some have investigated valley-plain topographies, where the valley opens into a plain, leading to thermally driven along-valley circulations \citep{rampanelli2004mechanisms, schmidli2013daytime}. 
\citet{shapiro2025revisiting} developed an analytical model for steady flow over a periodically repeating system of identical sloping valleys subjected to uniform heating or cooling. Their approach builds on the Prandtl model for slope flows, extending it to capture the flow dynamics within this periodic valley configuration. The model predicts symmetric flow patterns characterized by pairs of counter-rotating vortices on either side of each valley.

Others have focused exclusively on two-dimensional valleys, analyzing only cross-valley flow.
\citet{serafin2010daytime} used large-eddy simulations (LES) to study daytime circulation in idealized valleys with both narrow and wide bases. Their findings highlight the interaction between thermal convection up the valley walls and compensating ``top-down" heating of the valley core due to subsidence. They also observed that while flow symmetry is largely maintained along the valley walls, asymmetric motions develop in the valley core. Similarly, \citet{leukauf2015impact} examined an idealized valley geometry with sinusoidal slopes and a narrow valley floor, focusing on how surface heating influences the breakup of initial stratification. They classified different flow regimes based on the strength of surface heat flux and the resulting stratification breakup. Additionally, \citet{leukauf2016quantifying} studied pollutant transport in a heated idealized valley, observing intrusions from the slopes into the stable core above the convective boundary layer. They also employed the dimensionless breakup parameter $B$ to quantify the combined effects of surface heating and initial stratification.

Several prior studies have further simplified the idealized valley geometry by considering a capped, triangular V-shaped valley. This approach facilitates a more detailed examination of small-scale dynamics, enabling researchers to extract precise flow field information through direct numerical simulation (DNS) or small-scale laboratory experiments. An example of this approach is the study by \citet{princevac2008morning}, which investigates the breakup of stratification in a V-shaped tank filled with initially stratified water. The experiment subjects the bottom walls to heating, allowing for a detailed analysis of how the stratification evolves under these conditions.
They introduce the dimensionless breakup parameter $B$ to classify the breakup mechanisms as a function of both $B$ and the slope angle of the valley walls. A similar dimensionless breakup parameter was proposed by \citet{leukauf2016quantifying}, defined as the ratio of the energy required to break up an inversion to the cumulative surface sensible heat flux up to the time of breakup. This parameter facilitated the identification of similarity across simulations with varying stability conditions and forcing amplitudes.

Most numerical studies on natural convection in triangular cavities have concentrated on attic-shaped geometries \citep{saha2011review, ridouane2006formation}. In contrast, relatively few have investigated natural convection in valley-shaped geometries pertinent to the morning transition in valley flows, as exemplified by \citet{princevac2008morning}. \citet{bhowmick2018natural} examined natural convection in a V-shaped cavity containing initially stratified water, following a setup similar to that of \citeauthor{princevac2008morning}. However, their numerical simulations were restricted to two-dimensional (2D) computations and applied imposed temperature boundary conditions, rather than the flux conditions utilized in the experiments of \citeauthor{princevac2008morning}. Consequently, the Rayleigh number served as the primary control parameter, complicating direct comparisons with the breakup parameter from the experiments.
Other studies have explored natural convection in V-shaped cavities without considering stratification. \citet{bhowmick2019transition, bhowmick2022chaotic} adopted 2D computations with imposed temperature boundary conditions and identified a series of bifurcations leading to chaotic flow, including pitchfork and Hopf bifurcations. A corresponding experimental study by \citet{wang2021experimental} similarly observed the occurrence of a Hopf bifurcation. \citet{cui2015three} considered natural convection in an attic-shaped triangular cavity heated from below and cooled from above, revealing complex 3D flow structures such as longitudinal rolls. Their investigation adopted three-dimensional (3D) computations.


Although previous studies have advanced our understanding of natural convection in triangular cavities--mostly through 2D computations--a comprehensive grasp of three-dimensional flows in stably stratified idealized valleys remains incomplete. Moreover, conventional dimensionless parameters used to characterize these flows either do not sufficiently capture the full spectrum of flow regimes or have restricted applicability.
To address these gaps, this research investigates the three-dimensional dynamics of stably stratified flows in idealized valleys heated from below. Building on the experimental study by \citet{princevac2008morning}, we adopt the same V-shaped valley geometry with a stably stratified ambient and impose a constant heat flux condition on the valley walls to better replicate the conditions observed during the morning transition.
Furthermore, our formulation aligns with Prandtl’s idealized representation of mountain and valley flows \cite{prandtl1952essentials}.

Beyond studies of natural convection in triangular cavities, our work also builds on previous research on the dynamics of stratified, convective flows in rectangular closed domains. For instance, \citet{yalim2019parametrically} investigated the dynamics of a stably stratified fluid in a square cavity subjected to vertical oscillations, identifying the primary instabilities and the sequence of bifurcations that occur as the forcing parameter increases. Similarly, \citet{grayer2020dynamics} examined the stability and dynamics of flow in a differentially heated square cavity with varying inclination angles. They classified the 2D stability using the dimensionless buoyancy parameter $R_N$, which quantifies the relative influence of buoyancy and viscous forces. Their study determined the critical $R_N$ value at which steady flow persists, as well as the primary instabilities and limit cycles that emerge over a range of $R_N$ and inclination angles.
Extending this work to three dimensions, \citet{shen2025three} investigated 3D instabilities in the same differentially heated cavity, identifying the critical $R_N$ for 3D instability and characterizing the structure of the dominant instability modes. Additionally, \citet{jiang2024critical} studied the full transition from laminar to chaotic flow above a circular heated surface with water as the working fluid, outlining the sequence of bifurcations and intermediate states, including periodic and puffing states, leading to chaotic flow.

Recent studies on anabatic and katabatic Prandtl slope flows in infinitely wide domains have revealed new flow instabilities, including stationary longitudinal vortices and traveling transverse waves \citep{xiao2019stability, xiao2020stability}. These investigations also introduced a new set of dimensionless parameters governing slope-flow stability, which are expected to be relevant for similar stratified slope-flow problems. Building on this foundation, our study extends Prandtl slope flows to valley flows, incorporating both the assumption of constant background stratification and the newly introduced dimensionless parameters. Additionally, \citet{xiao2022impact} demonstrated that the combined effect of constant ambient stratification and surface cooling distinctly alters turbulent open channel flow dynamics compared to stratification driven solely by surface cooling. This further reinforces our decision to assume a constant background stratification in the valley geometry.

In our previous investigation of flows in V-shaped valleys \citep{stofanak2024unusual}, we used linear stability analysis (LSA) and three-dimensional direct numerical simulations (DNS) to characterize the two-dimensional (2D) flow states in the valley. This analysis revealed a distinctive bifurcation diagram, featuring a perfect pitchfork bifurcation that gives rise to two asymmetric states, as well as an unusual bifurcation leading to unstable upslope and downslope symmetric states. Additionally, we identified a unique self-organizing three-dimensional (3D) instability that ultimately returns to a steady 2D state in the long-time limit \citep{stofanak2025self}.
In this study, we extend our analysis to investigate three-dimensional states using both LSA and DNS, providing a comprehensive characterization of the transition from the quiescent pure conduction state to the fully 3D chaotic state. Additionally, we delineate the flow regimes associated with each state based on our established set of dimensionless parameters, gaining key insights into the flow dynamics.


\section {Problem description and methods} \label{sec:prob_des}

\subsection{Problem description}

\begin{figure}
    \centering
    \includegraphics[width=0.65\textwidth]{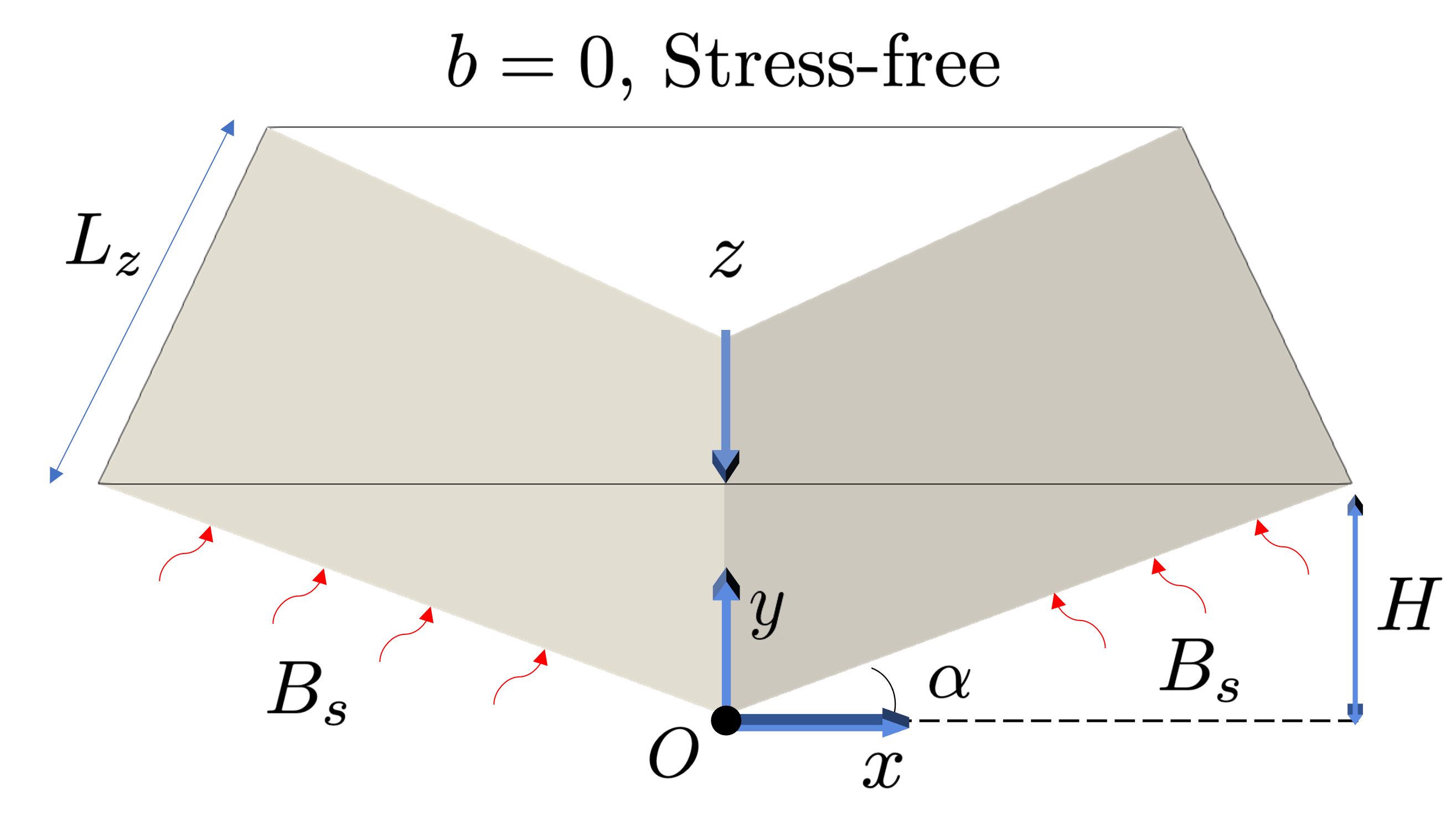}\hfill
    \caption{Schematic of the 3D valley geometry with key parameters and coordinate axes. The 2D valley lies in the $x$-$y$ plane, and the $z$ axis points out of the page.}
    \label{fig:ValleySchematic}
\end{figure}

We consider a V-shaped cavity filled with a stably stratified fluid and heated from below, serving as an idealized model of valley flows. A schematic of the 3D geometry is shown in figure \ref{fig:ValleySchematic} with key parameters, including the height of the valley $H$, the slope angle of the valley walls relative to horizontal $\alpha$, and the length in the $z$ direction $L_z$. The V-shaped valley lies in the $x$-$y$ plane with horizontal velocity $u$ and vertical velocity $v$, and the geometry is homogeneous along the $z$ direction with spanwise velocity $w$. In our previous studies, we used linear stability analysis (LSA) and Navier-Stokes simulations to investigate the 2D states \citep{stofanak2024unusual}, and the primary 3D instability at low parameter values \citep{stofanak2025self}. Here, we use the same combination of LSA and Navier-Stokes simulations to investigate all 3D states of the valley and the dependence on the dimensionless parameter space. For all simulations, a minimum $L_z$ of twice the height is used, with $L_z$ extending up to 16 times the height for some simulations.

We define buoyancy as the scaled perturbation potential temperature $\Theta$ as $b = g \left( \Theta - \Theta_e \right) / \Theta_r$, where $\Theta_e$ is the background or environmental potential temperature that varies in the direction of gravity only and $\Theta_r$ is a reference potential temperature. Following the Prandtl model for slope flows \citep{prandtl1952essentials}, a constant background stratification is given by the buoyancy frequency, or Brunt-V\"ais\"al\"a frequency, defined as $N = \sqrt{ \left(g / \Theta_r\right) d \Theta_e / d y}$. Therefore, the buoyancy represents a perturbation from an assumed constant, stable background state. This is relevant to the scenario of an isolated valley early in the morning transition when a strong inversion still exists \citep{whiteman2000mountain}.
The buoyancy boundary conditions are shown in figure \ref{fig:ValleySchematic}, with a constant positive buoyancy flux representing surface heating applied on both bottom walls defined as $B_s = \beta \partial b/ \partial n$, where $\beta$ is the thermal diffusivity and $n$ is the direction normal to the sloping boundaries. A constant $b = 0$ is imposed on the top boundary. For velocity, no-slip conditions are imposed on the bottom walls, and free slip (stress-free) condition is imposed on the top boundary.

The governing equations are given by the Navier-Stokes equations with the Oberbeck-Boussinesq approximation, written as
\begin{equation} \label{eq:cont}
    \nabla \cdot \mathbf{u} = 0,
\end{equation}
\begin{equation} \label{eq:mom}
    \frac{\partial \mathbf{u}}{\partial t} + \mathbf{u} \cdot \nabla \mathbf{u} = - \nabla p + \nu \nabla^2 \mathbf{u} + b \mathbf{g},
\end{equation}
\begin{equation} \label{eq:buoy}
    \frac{\partial b}{\partial t} + \mathbf{u} \cdot \nabla b = \beta \nabla^2 b - N^2 \mathbf{g} \cdot \mathbf{u},
\end{equation}
where $p$ is the specific pressure, $\nu$ is the kinematic viscosity, $\beta$ is the thermal diffusivity, and $\mathbf{g}$ is the effective gravity vector which acts only in the $y$ direction. Additionally, we note that the last term in Equation \ref{eq:buoy} is due to the assumption of the constant background stratification, consistent with the Prandtl model \citep{prandtl1952essentials}.

\subsection{Dimensionless parameters} \label{sec:dim_par}

In our prior studies \citep{stofanak2024unusual,stofanak2025self}, we used the Buckingham $\pi$ theorem to determine that the dimensionless parameter space governing the present problem depends on four parameters. These were defined as 
\begin{equation}
    \Pi_s = \frac{B_s}{\beta N^2}, \quad \Pi_h = \frac{N H^2}{\beta}, \quad Pr = \frac{\nu}{\beta}, \quad \alpha,
\end{equation}
where $Pr$ is the Prandtl number and $\alpha$ is the slope angle. The stratification perturbation parameter $\Pi_s$, first introduced to characterize the stability of Prandtl's model of katabatic and anabatic slope flow \citep{xiao2019stability,xiao2020stability}, represents the ratio between imposed surface buoyancy flux $B_s$ and the stable background stratification. In slope flows, increasing $\Pi_s$ leads to more dynamically unstable and eventually turbulent flow. This parameter can be easily overlooked because the Brunt-V\"ais\"al\"a frequency $N$ can be easily overlooked in the governing equations, but our formulation in terms of buoyancy makes the importance of this term more clear in Equation \ref{eq:buoy}. The parameter $\Pi_h$ was introduced in our prior study, and is related to the buoyancy number $R_N$ used in prior studies of stratified, convective cavity flows \citep{grayer2020dynamics,shen2025three} by $\Pi_h = Pr R_N$. In our definition, $\Pi_h$ represents the ratio between the diffusive timescale $H^2/\beta$ and the stratification timescale $1/N$. Our prior studies have not investigated the dependence of the flow state on $\Pi_h$, and we intend to investigate this in the present work.

We also note that our study is the first of stratified flow in a V-shaped valley that has uncovered the dependence on four dimensionless parameters. The prior study of \citet{princevac2008morning} introduced the breakup parameter $B = N^3 H^2 / B_s$, along with Prandtl number and slope angle. However, we notice that this breakup parameter can be written in terms of our new parameter space by $B = \Pi_h / \Pi_s$. This reveals that four dimensionless parameters are needed to fully describe the dynamics in a stably stratified, heated V-shaped valley.

\subsection{Linear stability analysis}

To perform linear stability analysis, we linearize the Navier-Stokes equations around an arbitrary 3D base flow given by $U(x,y,z), V(x,y,z), W(x,y,z), P(x,y,z), B(x,y,z)$. We then assume a small disturbance $\mathbf{q}$ is added to this base flow of the form
\begin{equation} \label{eq:disturbance}
    \mathbf{q}(x, y, z, t) = \mathbf{\hat{q}} \exp \left(\omega t \right),
\end{equation}
where $\mathbf{\hat{q}} = \left[\hat{u}(x,y,z), \hat{v}(x,y,z), \hat{w}(x,y,z), \hat{p}(x,y,z), \hat{b}(x,y,z) \right]$ represents the 3D disturbance quantities. Substituting this expression for the disturbances into the linearized Navier Stokes equations gives a set of equations for the disturbance quantities and $\omega$ dependent on the base flow.
These equations can be written as a generalized eigenvalue problem of the form
\begin{equation}
    \mathbf{A} \mathbf{\hat{q}}(x,y,z) = \omega \mathbf{B} \mathbf{\hat{q}}(x,y,z),
\end{equation}
where $\omega$ is the eigenvalue and the disturbance quantities $\mathbf{\hat{q}}$ are the eigenvectors. Solving this eigenvalue problem represent a global stability analysis in which the base flow and disturbance quantities depends on all three spatial dimensions. In this paper, we perform bi-global stability analysis in which the base flow and disturbance quantities depend on only two spatial dimensions and are assumed periodic in the third. In the case of bi-global stability analysis, the disturbance is of the form $\mathbf{q}(x, y, z, t) = \mathbf{\hat{q}}(x,y) \exp \left(ik_z z + \omega t \right)$, where $k_z$ is the wavenumber in the homogeneous $z$ direction, and the solution to the eigenvalue problem further depends on the choice of wavenumber.

In our investigation of all possible flow states in a V-shaped valley heated from the sides, we start from the most basic state which we call the pure conduction state, or the zero flow state. Given the boundary conditions for buoyancy, this state can be determined from the Navier Stokes equations assuming all velocity components are zero. This gives an exact solution for the buoyancy and pressure as
\begin{equation} \label{eq:exact_buoy_norm}
B(y) = \frac{B_s}{\beta \cos \alpha} \left( H - y \right), \qquad P(y) = \frac{- B_s}{2 \beta \cos \alpha} \left( y - H \right)^2, 
\end{equation}
where $B_s$ is the imposed surface buoyancy flux.
This is used as the base state as the starting point of our LSA. Substitution of this base state as well as the assumption of periodicity in the third direction into the LSA equations leads to a simplified set of equations given by
\begin{align}
\frac{\partial \hat{u}}{\partial x} & + \frac{\partial \hat{v}}{\partial y} + i k_z \hat{w} = 0, \label{eq:lsa_zeroState_1}\\
\omega \hat{u} & = - \frac{\partial \hat{p}}{\partial x} + \nu \left(   \frac{\partial^2\hat{u}}{\partial x^2}+\frac{\partial^2\hat{u}}{\partial y^2} - k_{z}^{2} \hat{u} \right),\\ 
\omega \hat{v} & = - \frac{\partial \hat{p}}{\partial y} + \nu \left( \frac{\partial^2\hat{v}}{\partial x^2}+\frac{\partial^2\hat{v}}{\partial y^2} - k_{z}^{2} \hat{v} \right) ,\\
\omega \hat{w} & = - i k_z \hat{p} + \nu \left( \frac{\partial^2\hat{w}}{\partial x^2}+\frac{\partial^2\hat{w}}{\partial y^2} - k_{z}^{2} \hat{w} \right) ,\\
\omega \hat{b} & = \beta \left( \frac{\partial^2\hat{b}}{\partial x^2}+\frac{\partial^2\hat{b}}{\partial y^2} - k_{z}^{2} \hat{b} \right) + \left( \frac{B_s}{\beta \cos \alpha} - N^2 \right) \hat{v}. \label{eq:lsa_zeroState_2}
\end{align} 
From these linearized equations, it is clear that for the same slope angle, kinematic viscosity, and thermal diffusivity, the system of equations depend only on the value of $\left((B_s/\beta \cos \alpha) - N^2 \right)$. The first term in this difference, $B_s/\beta \cos \alpha$ represents the unstable vertical buoyancy gradient due to the heating of the bottom walls, whereas $N^2$ represents the imposed stable background buoyancy gradient. Therefore, this term represents the total vertical buoyancy gradient between the combined effects of the surface heating and the background stratification.

Because of this, we choose the following scales to normalize all dimensional quantities:
\begin{equation} \label{eq:scales}
    l_0 = H, \quad u_0 = H \sqrt{G_y - N^2}, \quad b_0 = \frac{\beta^2}{H^3}, \quad p_0 = \frac{\beta^2}{H^2} , \quad t_{\mathrm{0,c}} = \frac{1}{\sqrt{G_y - N^2}}, 
\end{equation}
where $G_y = B_s/ \beta \cos \alpha$. The convective time scale $t_{\mathrm{0,c}}$ is defined as the ratio of the characteristic length scale to the characteristic velocity scale, i.e., $t_{\mathrm{0,c}}=l_0 / u_0$. Alternatively, a diffusion timescale can be defined as $t_{0,d} = H^2/\beta$, and a buoyancy timescale can be defined as $t_{\mathrm{0,b}} = 1/N$, but here we use the convective timescale, $t_{\mathrm{0,c}}$ to nondimensionalize the equations and results below. However, because the diffusion timescale, $t_{\mathrm{0,d}}$, is much longer than both other timescales, with $O(t_{\mathrm{0, c}})\sim 1$, $O(t_{\mathrm{0, b}})\sim 1$, and $O(t_{\mathrm{0, d}})\sim 1000$, the diffusion timescale is used as a guide to determine the periods to run simulations in order to come to a steady or statistically steady state.

Using the defined scales, we nondimensionalize the linearized equations for the pure conduction base state, equations \ref{eq:lsa_zeroState_1}-\ref{eq:lsa_zeroState_2}, to the following
\begin{align}
\frac{\partial \hat{u}}{\partial x} & + \frac{\partial \hat{v}}{\partial y} + i k_z \hat{w} = 0, \\
\omega \hat{u} & = - \frac{1}{\Pi_{c}^{2}} \frac{\partial \hat{p}}{\partial x} + \frac{Pr}{\Pi_c} \left(   \frac{\partial^2\hat{u}}{\partial x^2}+\frac{\partial^2\hat{u}}{\partial y^2} - k_{z}^{2} \hat{u} \right),\\ 
\omega \hat{v} & = - \frac{1}{\Pi_{c}^{2}} \frac{\partial \hat{p}}{\partial y} + \frac{Pr}{\Pi_c} \left( \frac{\partial^2\hat{v}}{\partial x^2}+\frac{\partial^2\hat{v}}{\partial y^2} - k_{z}^{2} \hat{v} \right) ,\\
\omega \hat{w} & = - \frac{1}{\Pi_{c}^{2}} i k_z \hat{p} + \frac{Pr}{\Pi_c} \left( \frac{\partial^2\hat{w}}{\partial x^2}+\frac{\partial^2\hat{w}}{\partial y^2} - k_{z}^{2} \hat{w} \right) ,\\
\omega \hat{b} & = \frac{1}{\Pi_c} \left( \frac{\partial^2\hat{b}}{\partial x^2}+\frac{\partial^2\hat{b}}{\partial y^2} - k_{z}^{2} \hat{b} \right) + \Pi_{c}^{2} \hat{v}.
\end{align} 
where we define the new dimensionless parameter $\Pi_c$ as 
\begin{equation} \label{eq:PIc}
    \Pi_c = \frac{H^2 \sqrt{G_y - N^2}}{\beta}.
\end{equation}
Based on the linearized equations and assuming a constant slope angle, the solution to the eigenvalue problem depends on only two dimensionless parameters, the Prandtl number $Pr$, and the newly introduced $\Pi_c$, which we refer to as the composite stratification parameter. Now nondimensionalizing the full Navier-Stokes equations using the same scales, we obtain
\begin{equation} \label{eq:cont_nondim}
    \nabla \cdot \mathbf{u} = 0,
\end{equation}
\begin{equation} \label{eq:mom_nondim}
    \frac{\partial \mathbf{u}}{\partial t} + \mathbf{u} \cdot \nabla \mathbf{u} = - \frac{1}{\Pi_{c}^{2}} \nabla p + \frac{Pr}{\Pi_c} \nabla^2 \mathbf{u} + \frac{1}{\Pi_{c}^{2}} b \mathbf{g},
\end{equation}
\begin{equation} \label{eq:buoy_nondim}
    \frac{\partial b}{\partial t} + \mathbf{u} \cdot \nabla b = \frac{1}{\Pi_c} \nabla^2 b + \Pi_{h}^{2} \mathbf{g} \cdot \mathbf{u}.
\end{equation}
For a constant Prandtl number, $Pr$, and slope angle, $\alpha$, the linearized equations under a pure conduction base flow depend solely on the composite stratification parameter, $\Pi_c$. However, in the nonlinear equations, an additional dependence on the buoyancy parameter, $\Pi_h$, arises. As a result, the complete parameter space for this problem is defined by four variables: $\Pi_c$, $\Pi_h$, $Pr$, and $\alpha$. 

Although the slope angle does not explicitly appear in the linearized or nonlinear equations, its influence arises through the boundary conditions. This dependence is due to its effect on both the angle itself and the effective surface area for heating. Specifically, at the bottom walls, the boundary condition is given as a heat flux normal to the wall, meaning $\partial b/\partial n$ is fixed, where the normal direction is defined in terms of slope angle as the unit vector $\mathbf{n} = \left( \pm \sin \alpha, \cos \alpha \right)$, where the sign is positive for the left slope and negative for the right slope. The buoyancy parameter, which we have used in prior studies \citep{stofanak2024unusual}, was introduced in \S \ref{sec:dim_par}, whereas the composite stratification parameter, $\Pi_c$, is newly introduced in this study. It can be interpreted as the product of an internal Reynolds number and Prandtl number. Specifically, if a Reynolds number for the flow is defined through the internal scales as $Re = u_0 l_0 / \nu$, then the composite stratification parameter can be viewed as $\Pi_c = Re Pr$, which is the P\'eclet number based on the internal flow scales. The advantage of using  $\Pi_c$ is that it can be defined \textit{a priori} based on external control parameters.

\subsection{Numerical methods}

We use the open-source, spectral/hp element code \textit{Nektar++} \citep{cantwell2015nektar++,moxey2020nektar++}, with a finite element discretization of 31 elements along each bottom wall of the valley and polynomial order of four. A Fourier expansion method is used in the third direction with 16 Fourier modes per nondimensional length. A finer mesh was used for the highest $\Pi_c$ cases reported in this paper, with the finest mesh using 41 elements along each bottom wall, polynomial order of four, and 32 Fourier modes per nondimensional length in the homogeneous direction. The eigenvalue problem of the LSA is solved using the modified Arnoldi method included in Nektar++, with a typical Krylov subspace of between 32 and 512, and are converged to a residual of less than $10^{-6}$. Our LSA and DNS results are validated by comparison to prior published results for Prandtl slope flow. Specifically, we perform LSA and DNS of Prandtl slope flow and obtain the same results are described in \citet{xiao2020stability} which used a separate in-house code. For our valley geometry, we show the mesh independence for a case of intermediate parameter values in table \ref{tab:mesh_sen}, which shows the total kinetic energy of the steady state and $u$ velocity at a point as a function of increasing mesh resolution. Additionally, more details of validation of our numerical solver is provided in our previously published results \citep{stofanak2024unusual,stofanak2025self}.

\begin{table}
  \begin{center}
\def~{\hphantom{0}}
  \begin{tabular}{l|cc}
       Number of elements  & Total kinetic energy & $u$ velocity at (0, 0.9, 0) \\[3pt]
       15   &  $1.7806 \times 10^{-3}$ & $-5.5467 \times 10^{-2}$  \\
       66   &  $1.7836 \times 10^{-3}$ & $-5.5661 \times 10^{-2}$ \\
       231  &  $1.7931 \times 10^{-3}$ & $-5.6466 \times 10^{-2}$ \\
       496  &  $1.7932 \times 10^{-3}$ & $-5.6481 \times 10^{-2}$ \\
       861  &  $1.7933 \times 10^{-3}$ & $-5.6485 \times 10^{-2}$ \\
       1,326 & $1.7933 \times 10^{-3}$ & $-5.6485 \times 10^{-2}$ \\
  \end{tabular}
  \caption{Total kinetic energy and $u$ velocity at point (0, 0.9, 0) of 2D asymmetric steady state flow at $\Pi_c = 297$ and $\Pi_h = 1500$ for increasing mesh resolution. Velocity is normalized by $u_0=0.2122$, and kinetic energy is normalized by $u_{0}^{2}$.}
  \label{tab:mesh_sen}
  \end{center}
\end{table}

\section{Results} \label{sec:results}

We now present results of the primary instabilities and flow states observed in the heated valley geometry. First we focus only on increasing the $\Pi_c$ parameter, and the dependence on $\Pi_h$ and Prandtl number will be investigated in later sections.

\subsection{Classification of flow states for increasing $\Pi_c$}

\subsubsection{Pure-conduction state}

We begin with linear stability analysis (LSA) to identify the primary instabilities before proceeding to determine the corresponding steady states. As stated in \S \ref{sec:prob_des}, we apply our linear stability analysis (LSA) to a quiescent base state that represents pure conduction, characterized by a linear buoyancy profile, a quadratic pressure distribution, and a zero velocity field. In our previous work \citep{stofanak2024unusual}, we used the dimensionless parameter $\Pi_s = B_s / \beta N^2$ to determine the critical value of the instability at a fixed $\Pi_h = 1500$. However, given our new analysis in \S \ref{sec:prob_des} showing that the linearized equations with the pure conduction base flow depends only on the parameter $\Pi_c$, we can now give a critical value at which the base state becomes unstable for any combination of values $\Pi_s$ and $\Pi_h$, given a constant Prandtl number. For any values of parameters $\Pi_h$ and $\Pi_s$, we can determine $\Pi_c$ as 
\begin{equation}
    \Pi_c = \Pi_h \sqrt{\frac{\Pi_s}{\cos \alpha} - 1}. \label{eq:PIc_2}
\end{equation}
In \cite{stofanak2024unusual}, we have shown that the pure conduction state becomes unstable at a critical value of approximately $\Pi_s = 0.8751$ at $\Pi_h = 1500$. Therefore, the corresponding critical value of $\Pi_c$ is 153.5. Below this value, the pure conduction state is linearly stable for all $\Pi_h$ values.

A distinctive feature of the $\Pi_c$ parameter, as expressed in Eq. \ref{eq:PIc_2}, is that it can assume imaginary values when falling below the critical instability threshold. This specifically arises when the stable background stratification, $N^2$, surpasses the unstable vertical buoyancy gradient, $G_y$, induced by surface heating. In these cases, motion cannot develop because surface heating fails to overcome the stabilizing effect of the background stratification, leaving the system in a quiescent state as the only possible solution. Given this fact, and the definition of $\Pi_c$ in terms of $\Pi_h$ and $\Pi_s$, we can see that a necessary condition for critical value for convective motion in the valley can be given by
\begin{equation}
    \Pi_s > \cos \alpha.
\end{equation}
When this is not the case, the term in the square root of Equation \ref{eq:PIc_2} is negative and $\Pi_c$ would be imaginary. Therefore, for any slope angle $\alpha$, a minimum $\Pi_s$ value must be met to initiate instability, regardless of the $\Pi_h$ value. This can be demonstrated at $\alpha = 30^{\circ}$, where the critical value is $\Pi_s = 0.8751 > \cos \alpha$. We have confirmed this through LSA and DNS at $\alpha = 10^{\circ}$ as well, where the critical value must exceed 0.985. This also reveals that the critical $\Pi_s$ for instability increases for lower slope angles.
We note that this expression assumes a non-zero background stratification, but can be written in a more general form as $B_s/\beta > N^2 \cos \alpha$.

\subsubsection{Self-organizing two-dimensional flow state}

\begin{figure*}
    \centering
    \begin{subfigure}[b]{0.5\textwidth}
        \centering
        \includegraphics[width=\textwidth]{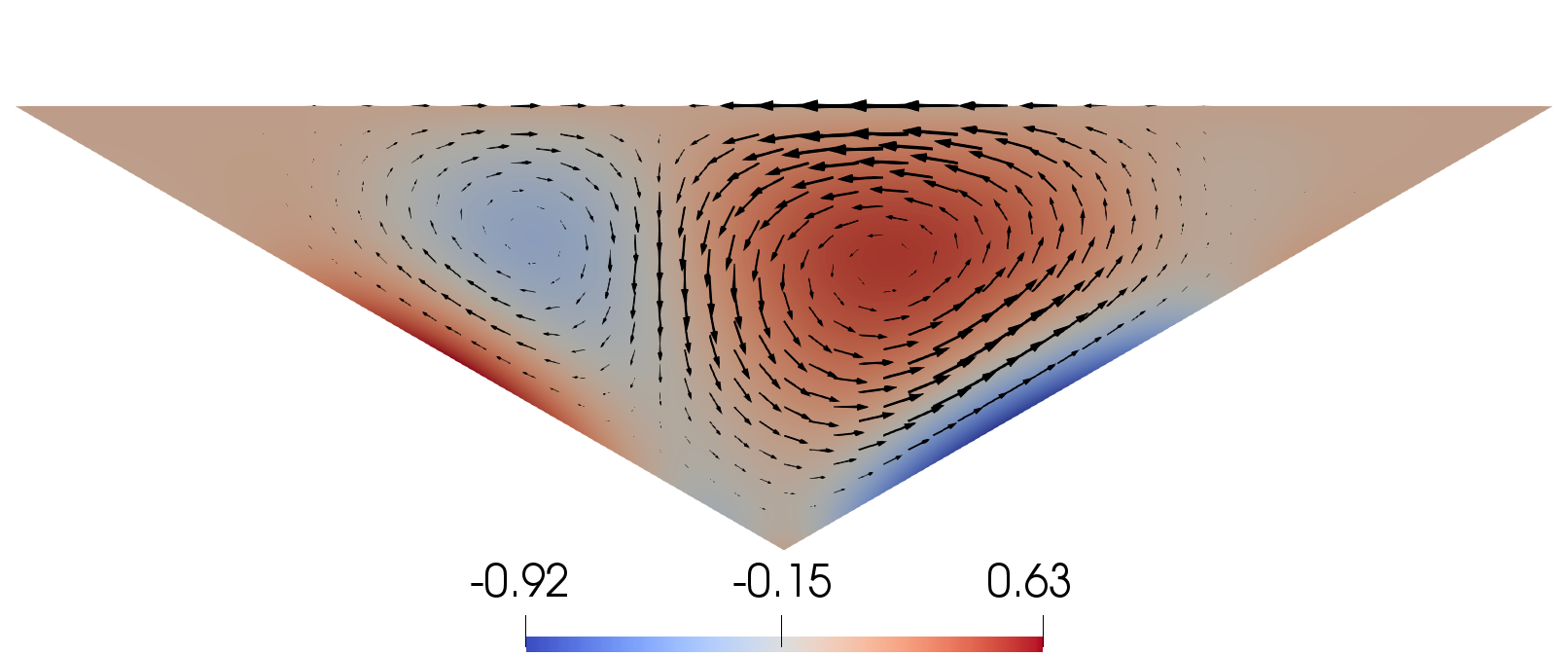}
        \caption{}
        \label{fig:2dSS_a}
    \end{subfigure}%
    ~ 
    \begin{subfigure}[b]{0.5\textwidth}
        \centering
        \includegraphics[width=\textwidth]{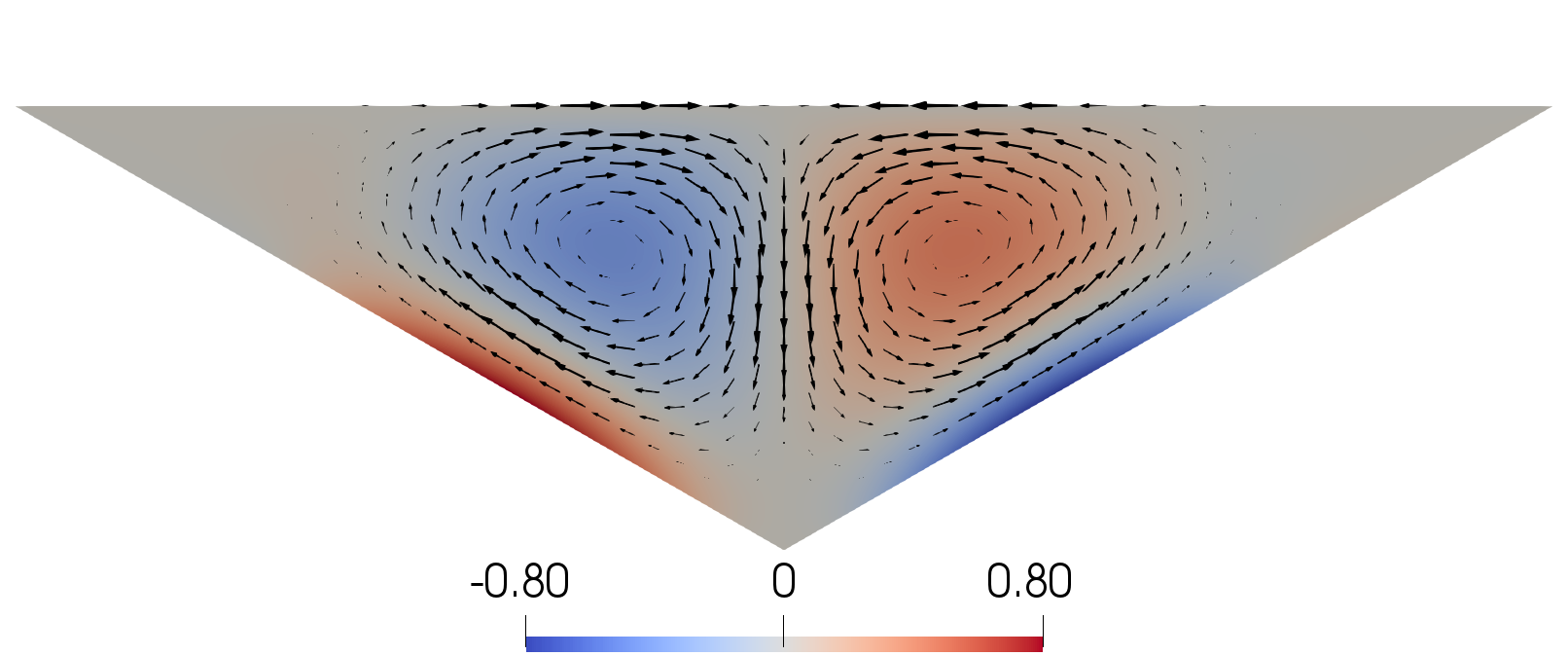}
        \caption{}
    \end{subfigure}
    \caption{Visualization of (a) asymmetric and (b) symmetric 2D steady states for $\Pi_c = 297$ and $\Pi_h = 1500$, colored by normalized vorticity and showing velocity vectors.}
    \label{fig:2dSS}
\end{figure*}

Above $\Pi_c = 153.5$, two distinct 2D instabilities emerge: one representing a 2D asymmetric state, and the other representing a 2D symmetric state. The steady-state solutions for the asymmetric and symmetric states are shown in figure \ref{fig:2dSS} for $\Pi_c = 297$ and $\Pi_h = 1500$. Secondary linear stability analysis of the 2D states shows that the symmetric state is unstable to the asymmetric state. Additional details of the 2D states and the bifurcation diagram can be found in \citet{stofanak2024unusual}. 

\begin{figure}
\centering
\includegraphics[width=0.55\textwidth]{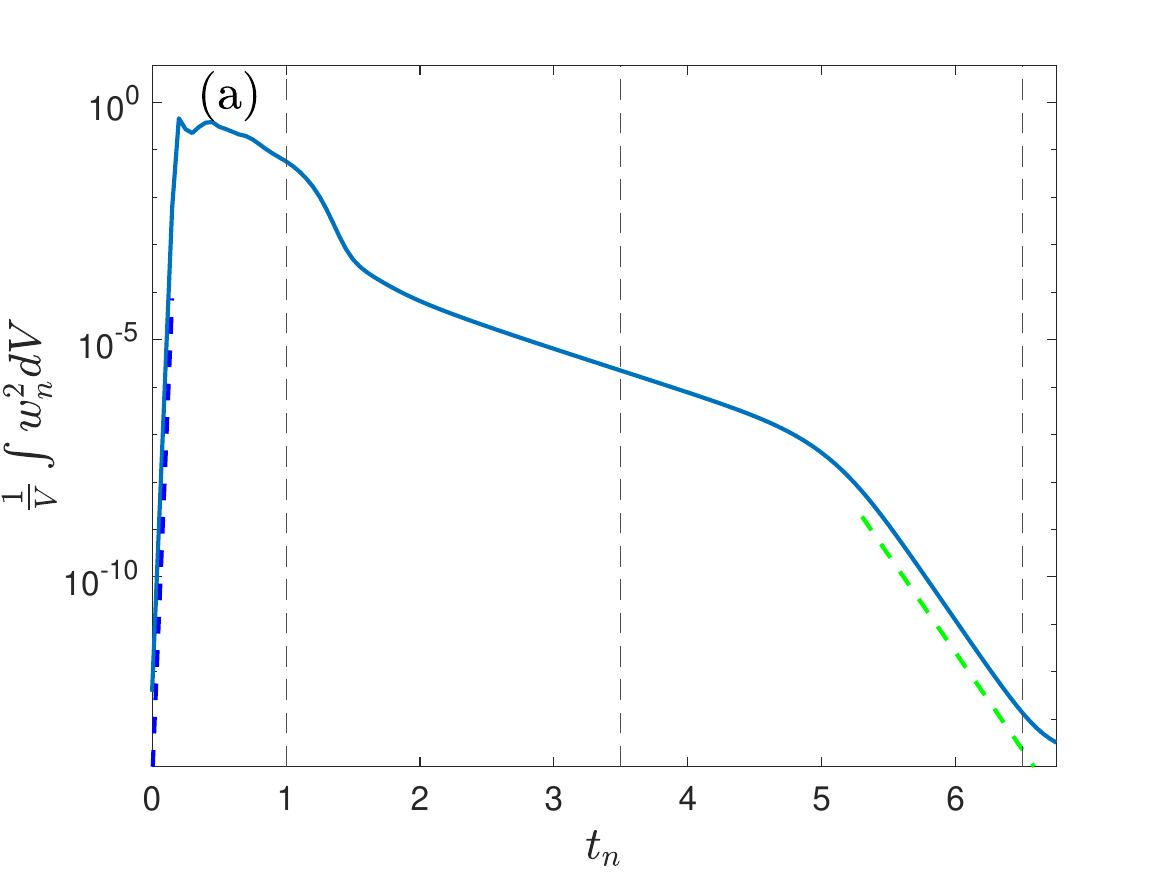} 
\includegraphics[width=0.8\textwidth]{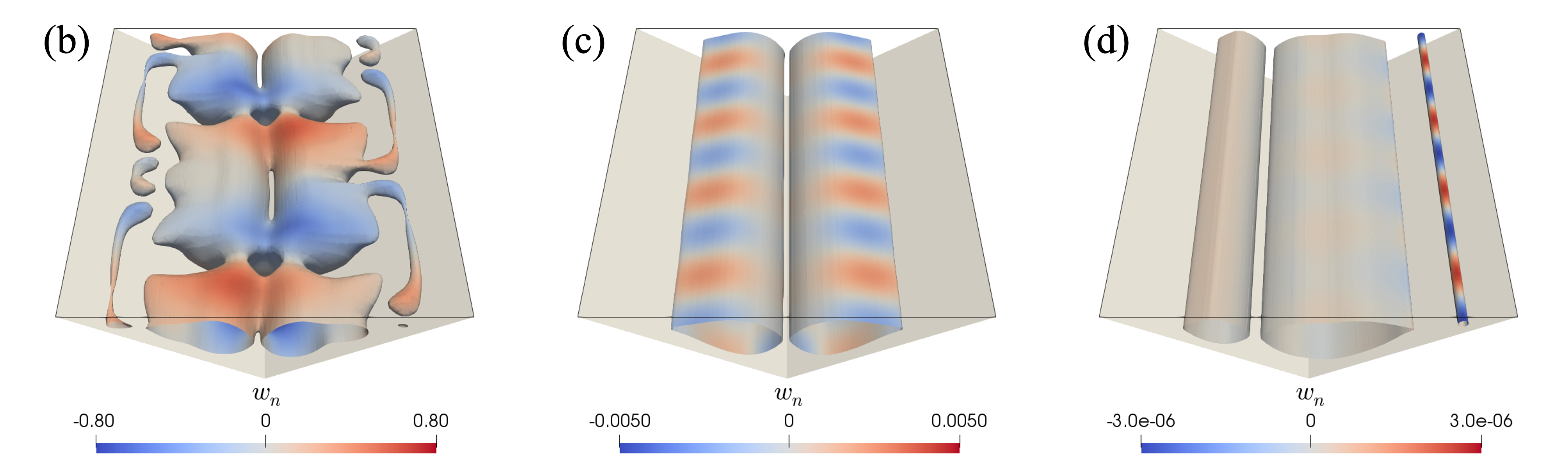}
\caption{(a) Normalized, volume-averaged $w^2$ velocity over time for $\Pi_c = 934, \Pi_h = 3000$, and $L_z = 4$, along with growth/decay rates predicted by LSA at two different points in the evolution shown with colored dashed lines. Vertical dashed lines represent the times at which the flow field is depicted with the Q-criterion in (b) $t_n = 1$, (c) $t_n = 3.5$, and (d) $t_n = 6.5$, taken as 4\% of the maximum, and are colored with span-wise velocity $w$ . }
\label{fig:wSquared}
\end{figure}

We also obtain an unstable 3D instability to the pure conduction base flow, but in nonlinear Navier-Stokes simulations, the 3D flow eventually self-organizes to the 2D asymmetric state, and we are not able to obtain a steady 3D state corresponding to this instability. We have analyzed this self-organization behavior in depth in our prior work for the case of $\Pi_c = 297$ \citep{stofanak2025self}. Here, we confirm that this same self-organization behavior occurs over the entire range in which the 3D instability exists. In other words, over the parameter range $153.5 \lesssim \Pi_c \lesssim 940$, LSA indicates a dominant 3D instability, but in nonlinear simulations, after the initial exponential growth of the 3D flow structures, they self-organize back to the 2D asymmetric state. To support this, figure \ref{fig:wSquared} shows the total volume-averaged span-wise velocity $w$ squared over time for the case of $\Pi_c = 934$, $L_z = 4$ with a small initial 3D disturbance. This follows the same trend described in our prior work for $\Pi_c = 297$, including the initial exponential growth of the 3D instability, an initial period of fast decay of $w^2$ followed by a longer period of slower decay, and finally a period of fast decay as the state converges to the 2D asymmetric flow state. In our prior work, we explained this behavior through the dominance of viscous dissipation over buoyant production after the nonlinear saturation of the instability, and the secondary stability of the 2D symmetric and asymmetric steady states. This example further confirms that even at significantly larger $\Pi_c$ values, the same self-organization behavior persists. 
Therefore, the asymmetric circulation state is the only stable equilibrium state in the parameter range $153.5 \lesssim \Pi_c \lesssim 940$. A more detailed analysis of the 3D instability and the subsequent self-organization can be found in \citet{stofanak2025self}.

\begin{figure*}
    \centering
    \begin{subfigure}[b]{0.5\textwidth}
        \centering
        \includegraphics[width=\textwidth]{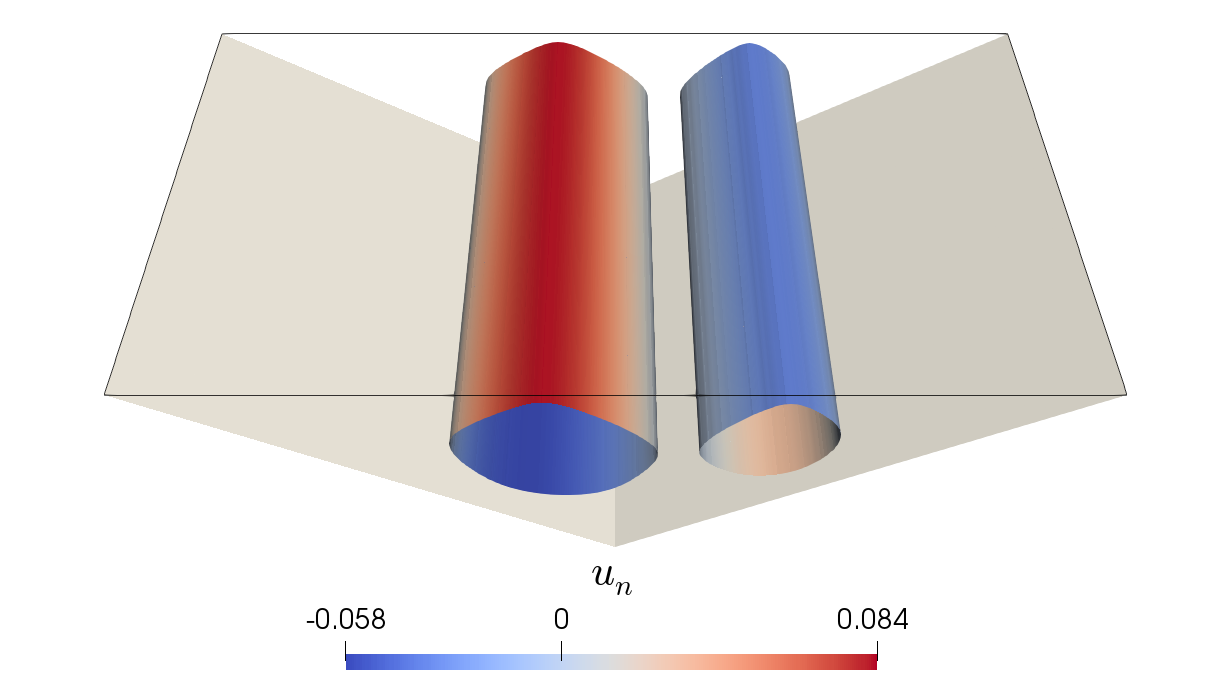}
        \caption{}
        \label{fig:2dSS_3d_a}
    \end{subfigure}%
    ~ 
    \begin{subfigure}[b]{0.5\textwidth}
        \centering
        \includegraphics[width=\textwidth]{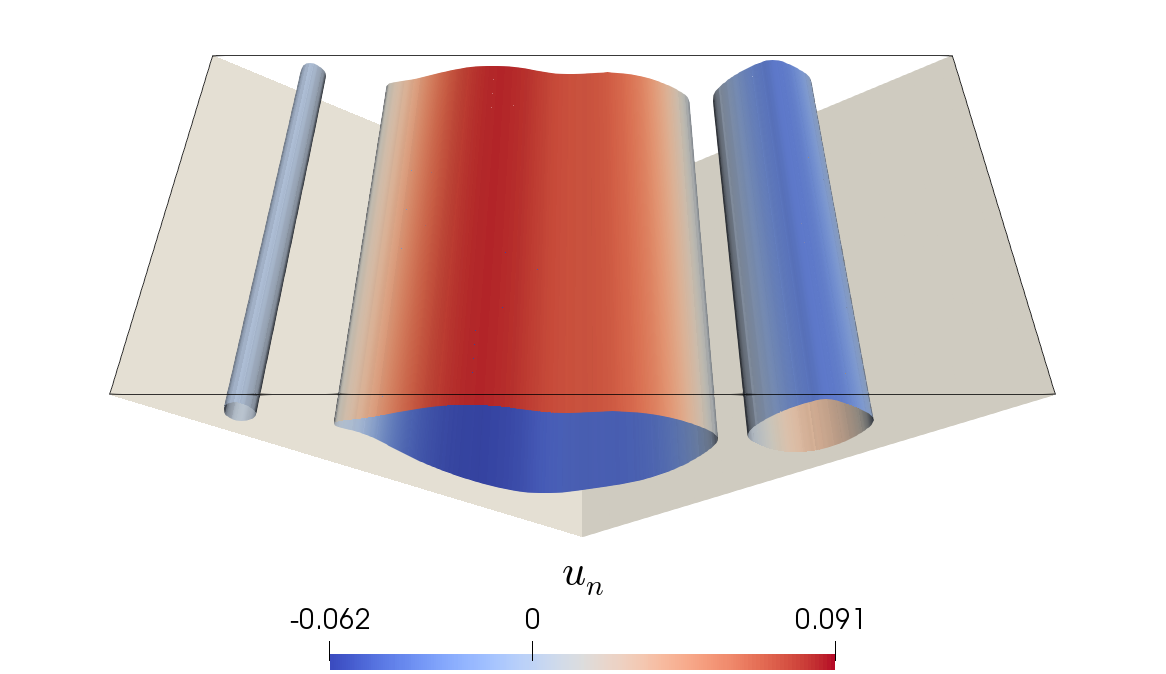}
        \caption{}
        \label{fig:2dSS_3d_b}
    \end{subfigure}
    \caption{Self-organizing 2D asymmetric state. (a) Visualization of $Q$ criterion for $\Pi_c = 297$, $\Pi_h = 1500$ and (b) $Q$ criterion for $\Pi_c = 931$ and $\Pi_h = 1500$.}
    \label{fig:2dSS_3d}
\end{figure*}

Figure \ref{fig:2dSS_3d} presents 3D visualizations of the 2D asymmetric state for two values of $\Pi_c$. This clarifies the structure of the 2D asymmetric state and facilitates comparison with the 3D states that emerge at higher parameter values later in this study. Specifically, at larger $\Pi_c$ values, the 2D asymmetric state consists of three circulation rolls: a primary roll in the center of the valley, a counter-rotating roll in the opposite corner, and a weaker counter-rotating roll within the main circulation’s corner. Identifying these three circulations provides a foundation for interpreting the 3D states observed at higher $\Pi_c$ values.

\subsubsection{Hopf bifurcation: oscillating asymmetric circulation state} \label{sec:Hopf}

\begin{figure*}
    \centering
    \begin{subfigure}[b]{0.5\textwidth}
        \centering
        \raisebox{5mm}{\includegraphics[width=\textwidth]{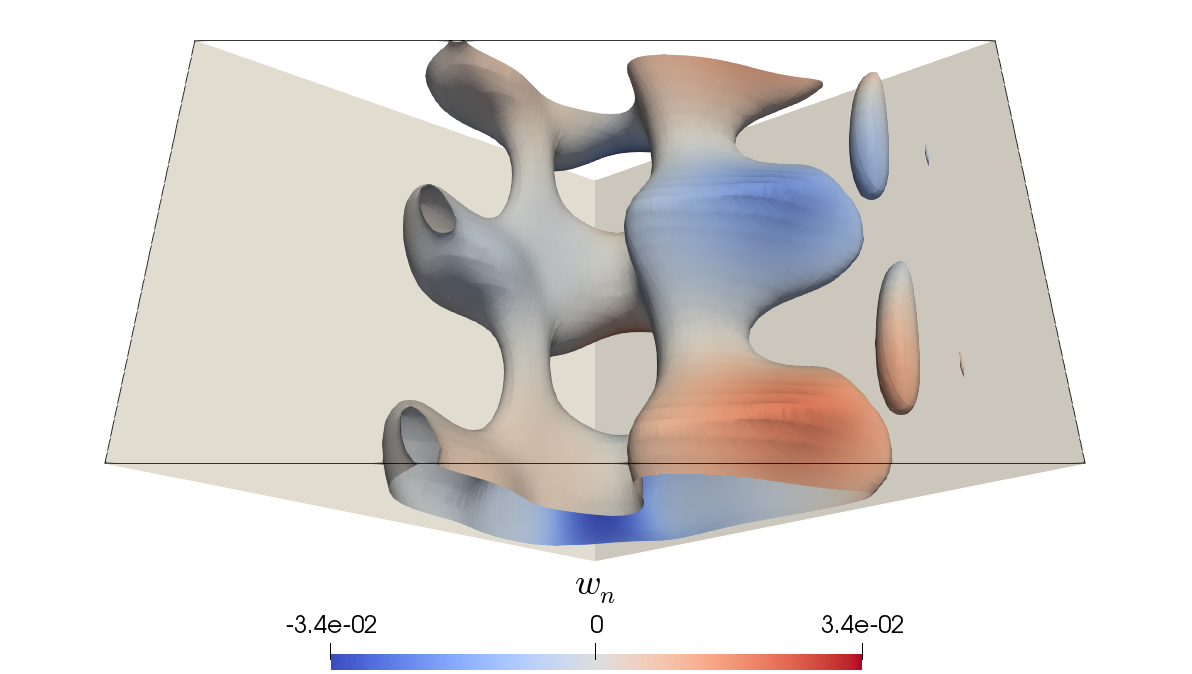}}
        \caption{}
        \label{fig:Hopf_a}
    \end{subfigure}%
    ~ 
    \begin{subfigure}[b]{0.5\textwidth}
        \centering
        \includegraphics[width=\textwidth]{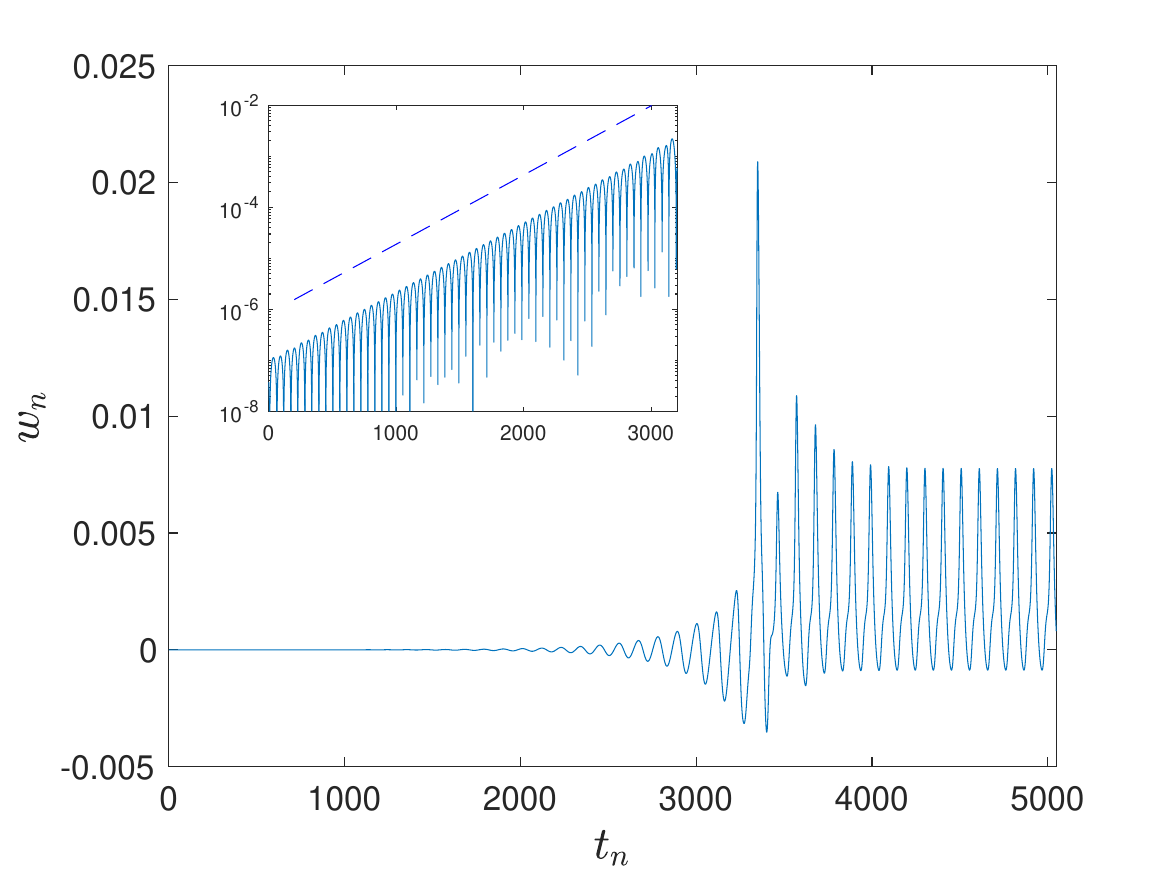}
        \caption{}
        \label{fig:Hopf_b}
    \end{subfigure}
    ~
    \begin{subfigure}[b]{0.5\textwidth}
        \centering
        \includegraphics[width=\textwidth]{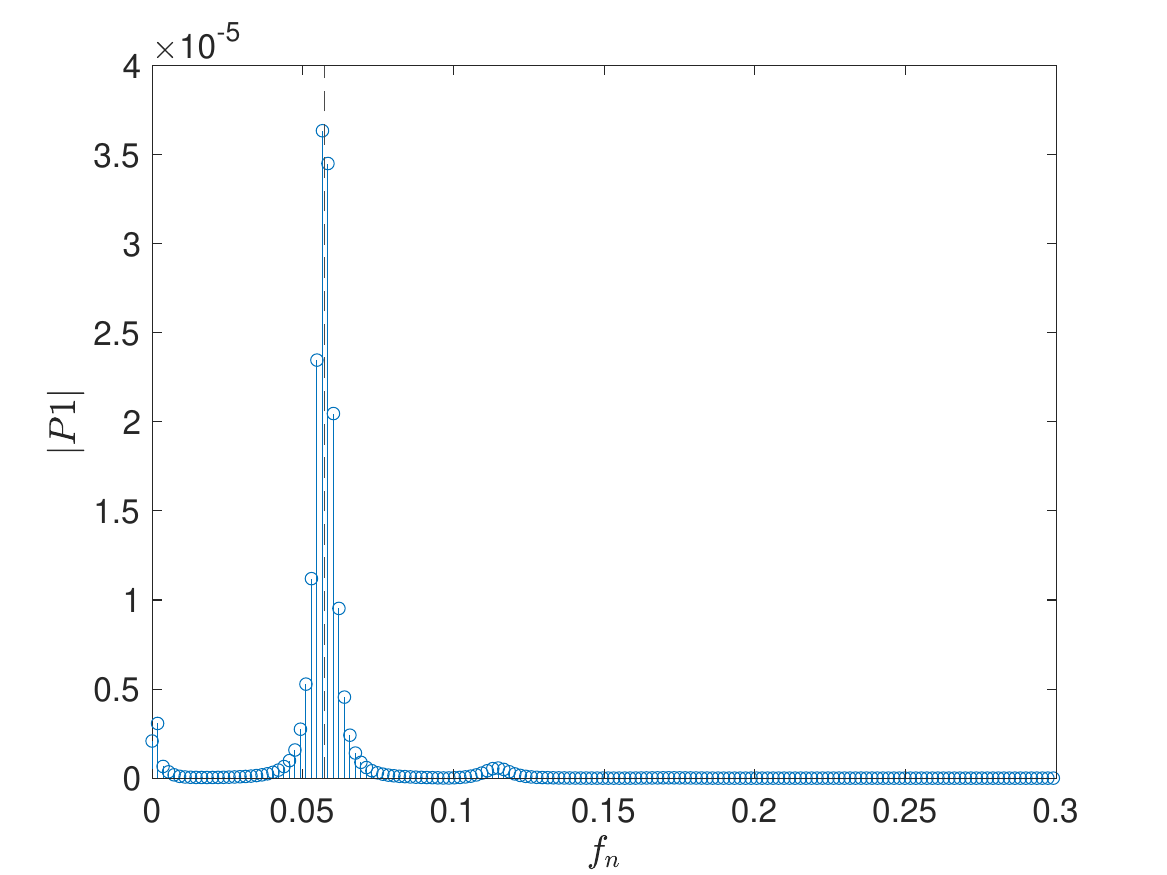}
        \caption{}
        \label{fig:Hopf_c}
    \end{subfigure}
    \caption{(a) Visualization of the eigenmode of the secondary instability to the 2D asymmetric state for $\Pi_c = 1178$, $\Pi_h = 1500$, and $L_z = 2$, (b) Time series of span-wise velocity $w$  at a point (0, 0.9, 0) for the 2D asymmetric state restarted with a small disturbance showing exponential growing oscillation. The inset shows the early times on a log-y axis, with the dashed line representing the growth rate for LSA. (c) Frequency spectrum of initial exponentially growing oscillation with the frequency predicted by LSA shown as a vertical dashed line.}
\end{figure*}

We now perform secondary linear stability analysis on the 2D asymmetric base flow. As $\Pi_c$ is increased, we obtain an unstable 3D mode with nonzero frequency, indicating the occurrance of a Hopf bifurcation. The visualization of the corresponding eigenvector is shown in figure \ref{fig:Hopf_a} for the case of $\Pi_c = 1178$, $\Pi_h = 1500$, and $L_z = 2$, corresponding to a growth rate of $3.13 \times 10^{-3}$ and frequency of $5.72 \times 10^{-2}$. To confirm the results of the LSA, we run a DNS with the 2D asymmetric base flow plus a small multiple of the 3D unstable eigenvector, and the span-wise velocity $w$  at a point over time is shown in figure \ref{fig:Hopf_b}. The flow evolution follows the expected trajectory, beginning with an exponentially growing oscillation that ultimately transitions to a steady oscillatory state. The inset confirms that the initial growth rate aligns with the prediction from LSA. Additionally, figure \ref{fig:Hopf_c} shows the frequency spectrum of the initially exponentially growing oscillation along with the frequency predicted from LSA with the vertical dashed line, and we see that there is good agreement between these two.

\begin{figure*}
    \centering
    \begin{subfigure}[b]{0.5\textwidth}
        \centering
        \includegraphics[width=\textwidth]{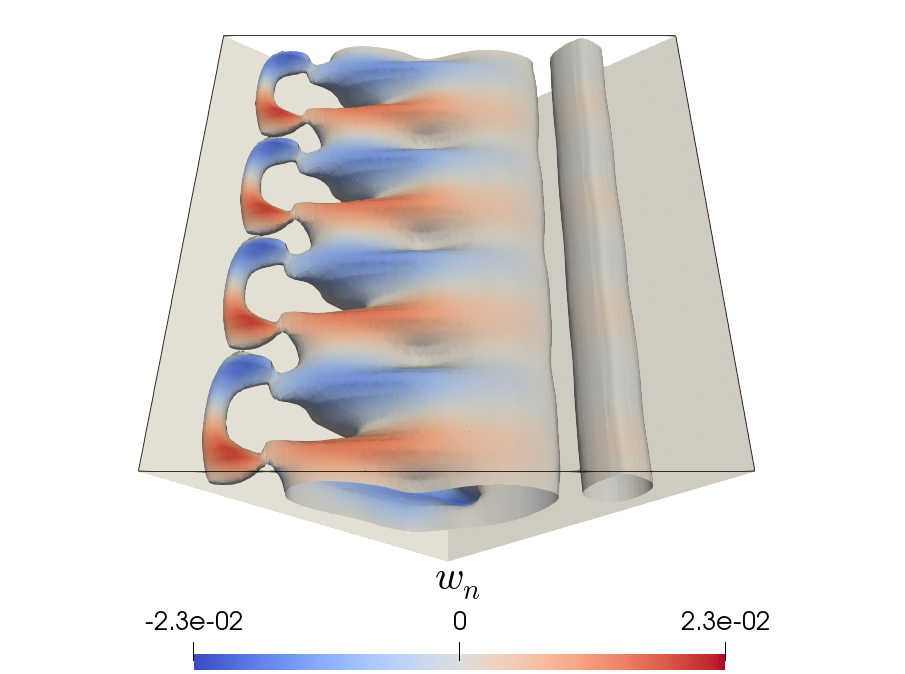}
        \caption{}
        \label{fig:Hopf_Qcrit}
    \end{subfigure}%
    ~ 
    \begin{subfigure}[b]{0.5\textwidth}
        \centering
        \includegraphics[width=\textwidth]{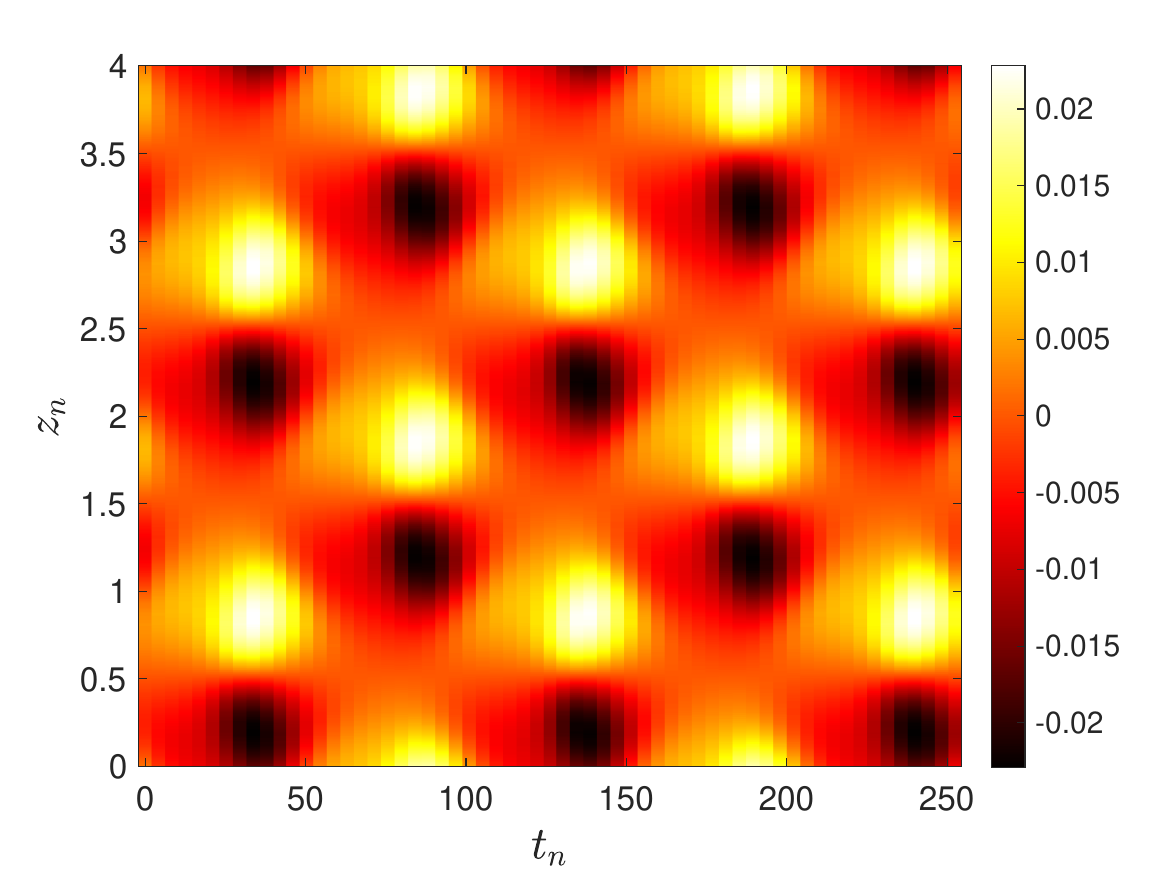}
        \caption{}
        \label{fig:Hopf_w_vs_t}
    \end{subfigure}
    ~ 
    \begin{subfigure}[b]{0.5\textwidth}
        \centering
        \includegraphics[width=\textwidth]{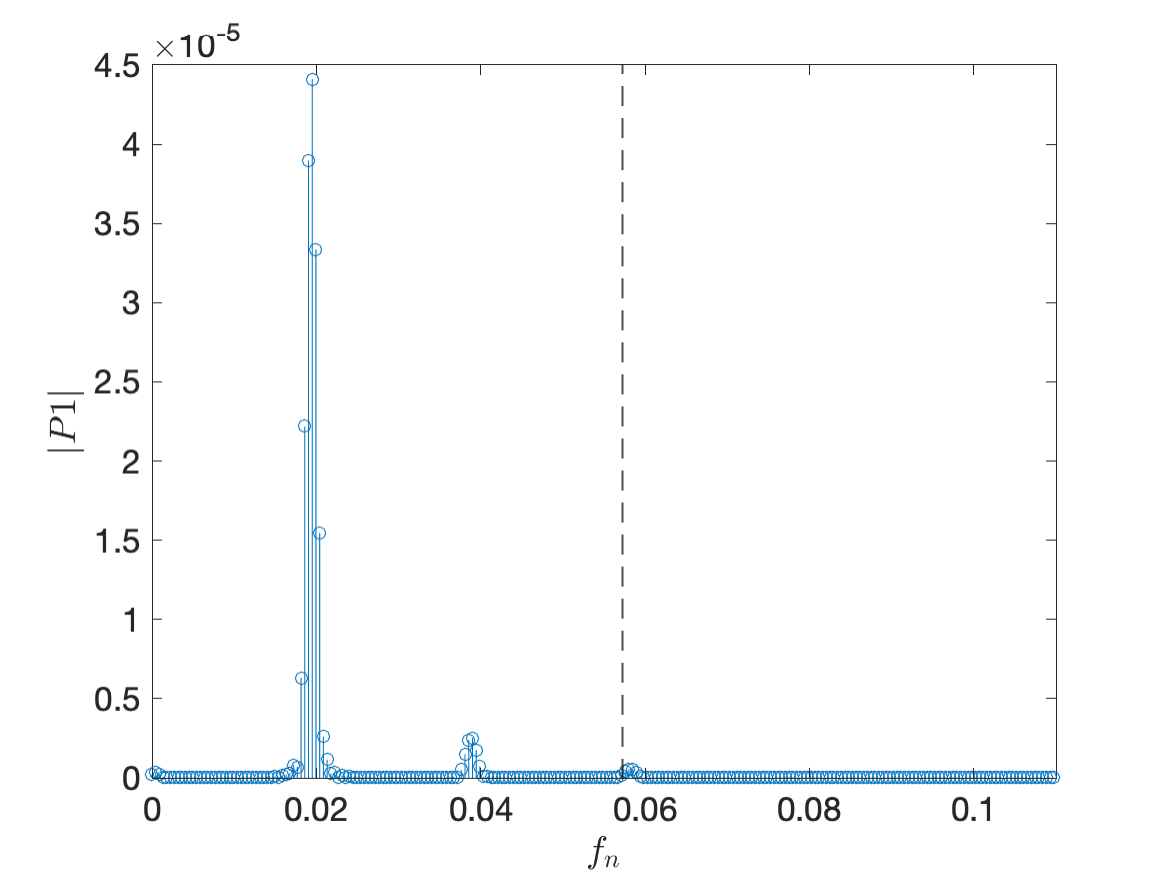}
        \caption{}
        \label{fig:Hopf_FFT}
    \end{subfigure}
    \caption{Oscillating asymmetric circulation state. (a) Visualization of time-averaged flow for $\Pi_c = 1178$ and $\Pi_h = 1500$, (b) Color plot of span-wise velocity $w$  along the $z$ direction versus time at point (0, 0.9, 0), (c) Frequency spectrum of kinetic energy data, with frequency of the initial instability marked by the vertical dashed line.}
\end{figure*}

We now consider the steady oscillating state resulting from this instability. The time-averaged state for multiple periods of oscillation is depicted with the $Q$ criterion in figure \ref{fig:Hopf_Qcrit}, and a color map of the span-wise velocity $w$ along the $z$ direction over time is shown in figure \ref{fig:Hopf_w_vs_t}. 
From visualizations of this oscillating state, it is clear that the Hopf bifurcation is caused by an interaction between two of the circulation rolls present in the 2D asymmetric state, namely the main central circulation and the adjacent, smaller corner circulation, causing an oscillation in the homogeneous direction. The third circulation in the opposite corner remains largely two-dimensional, with little influence from the strong 3D oscillating behavior in the other half of the valley. The frequency spectrum of the temporal signal of the steady oscillating state is shown in figure \ref{fig:Hopf_FFT}, indicating that there is one dominant frequency of approximately 0.02. We note that this frequency is significantly lower than the frequency of the initial instability of approximately 0.057, but this is not unexpected because LSA only predicts the frequency of the initial linear growth of the eigenmode. As shown in figure \ref{fig:Hopf_b}, once this disturbance is strong enough, nonlinear effects come into play and the flow evolves to a steady oscillation with a frequency different from that of the initial growth. Based on LSA, the critical value for this instability is approximately $\Pi_c = 1130$, while our DNS results show steady oscillating cases in the range of $1170 \lesssim \Pi_c \lesssim 1560$.

\subsubsection{Three-dimensional, steady, asymmetric circulation state}

\begin{figure*}
    \centering
    \begin{subfigure}[b]{0.32\textwidth}
        \centering
        \includegraphics[width=\textwidth]{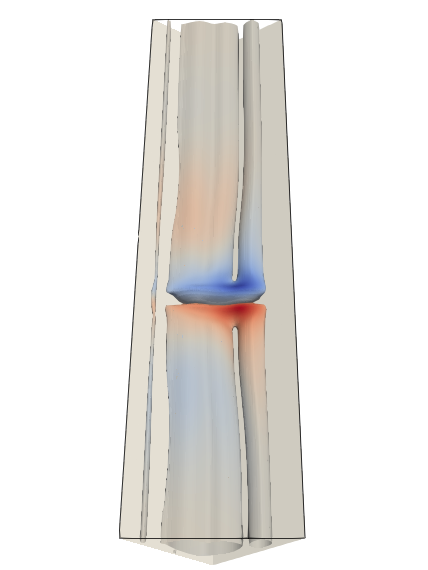}
        \caption{}
        \label{fig:DanPasStates_a}
    \end{subfigure}%
    ~ 
    \begin{subfigure}[b]{0.32\textwidth}
        \centering
        \includegraphics[width=\textwidth]{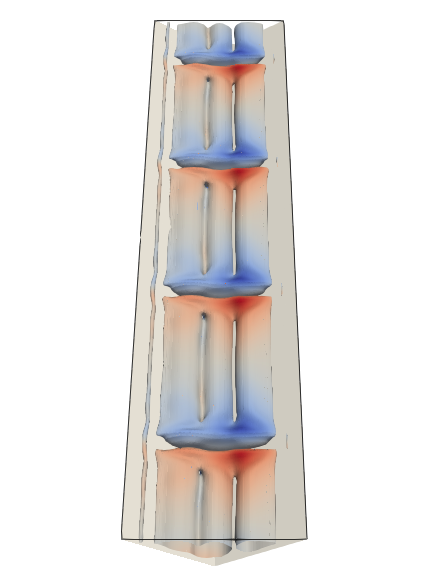}
        \caption{}
        \label{fig:DanPasStates_b}
    \end{subfigure}
    ~
    \begin{subfigure}[b]{0.32\textwidth}
        \centering
        \includegraphics[width=\textwidth]{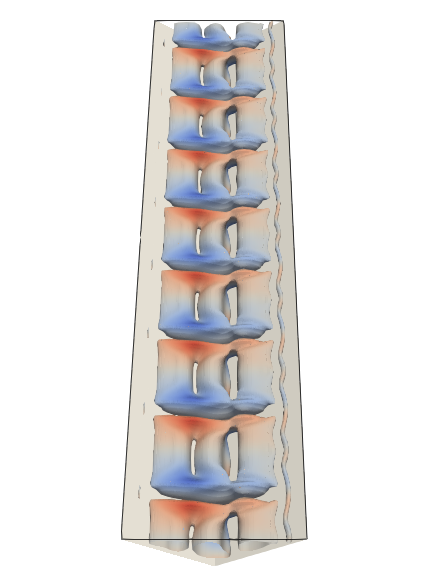}
        \caption{}
        \label{fig:DanPasStates_c}
    \end{subfigure}
    ~
    \begin{subfigure}[b]{0.99\textwidth}
        \centering
        \includegraphics[width=0.5\textwidth]{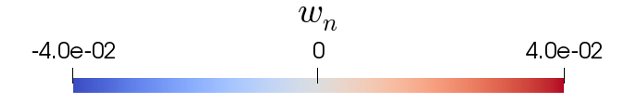}
    \end{subfigure}
    \caption{Three-dimensional, steady, asymmetric state, or Danish-pastry state. Visualization of $Q$ criterion for (a) $\Pi_c = 1040$, $\Pi_h = 2000$, (b) $\Pi_c = 1559$, $\Pi_h = 3000$, and (c) $\Pi_c = 1819$, $\Pi_h = 3500$. Flow structures are colored by span-wise velocity $w$ , and $L_z$ is 16 for each case.}
    \label{fig:DanPasStates}
\end{figure*}

\begin{figure*}
    \centering
    \begin{subfigure}[b]{0.5\textwidth}
        \centering
        \includegraphics[width=\textwidth]{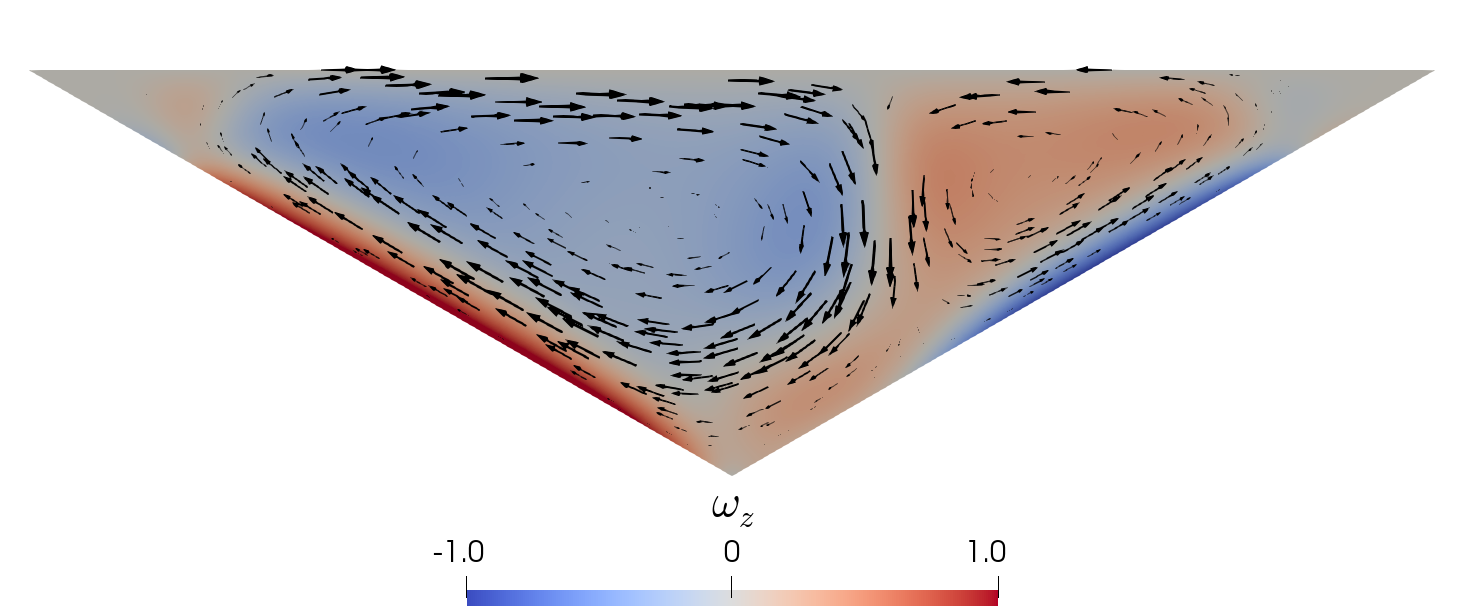}
        \caption{}
        \label{fig:DanishPastry_2dViz_a}
    \end{subfigure}%
    ~ 
    \begin{subfigure}[b]{0.5\textwidth}
        \centering
        \includegraphics[width=\textwidth]{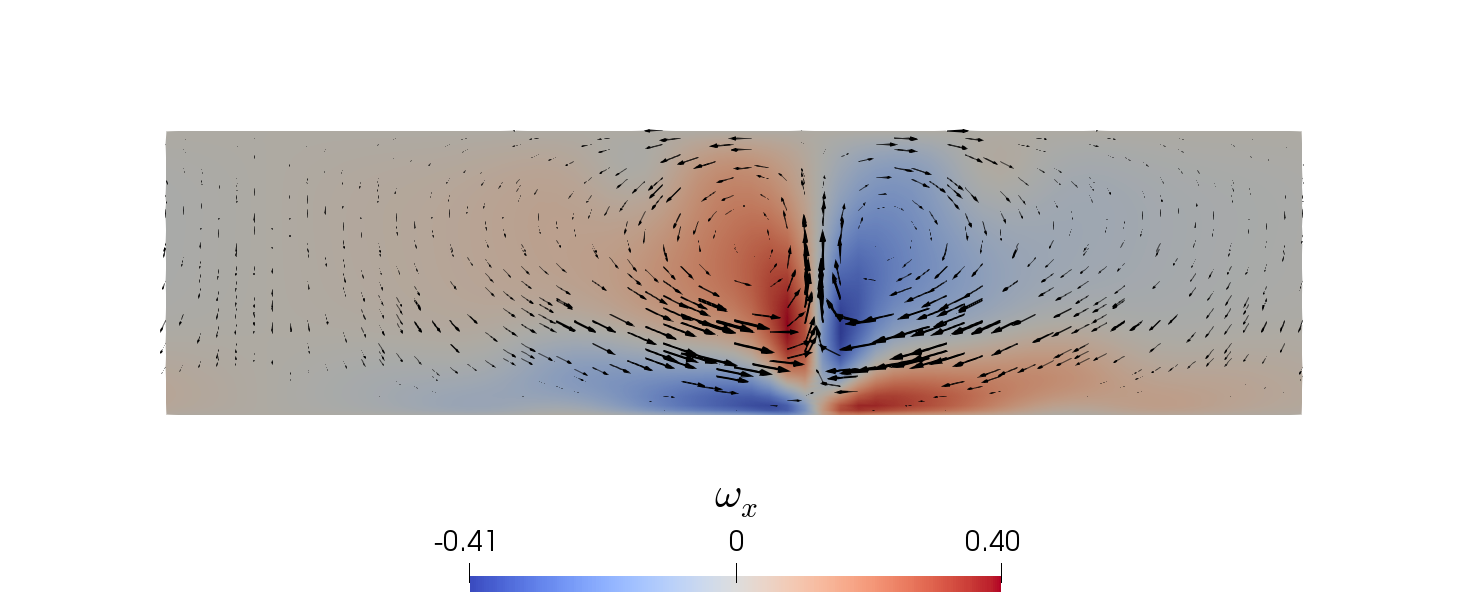}
        \caption{}
        \label{fig:DanishPastry_2dViz_b}
    \end{subfigure}
    \caption{Visualization of (a) 2D velocity profile in $x$-$y$ plane at $z = 0$ and (b) 2D velocity profile in $z$-$y$ plane at $x=0$ and $z$ from 0 to 4, both of the 3D, steady, asymmetric circulation state shown in figure \ref{fig:DanPasStates_b}. Visualization in panel (a) is colored by $z$-vorticity and panel (b) is colored by $x$-vorticity, and both show 2D velocity vectors.}
    \label{fig:DanishPastry_2dViz}
\end{figure*}

In our previous study \citep{stofanak2025self}, we demonstrated that within a certain range of $\Pi_c$, the flow asymptotically converges to the 2D asymmetric state despite an initial 3D instability, regardless of the domain length in the homogeneous direction, with cases examined up to $L_z = 16$. However, as $\Pi_c$ continues to increase, this behavior no longer holds. Beyond a certain threshold, a new 3D steady state emerges as an evolution of the 2D asymmetric state.
Figure \ref{fig:DanPasStates} presents three examples of these states for increasing values of $\Pi_c$. This new state results from the breakup of the 2D circulation rolls of the asymmetric state, leading to the formation of vortex ring structures with strong 3D circulation between them. The $Q$-criterion visualization in figure \ref{fig:DanPasStates}, colored by span-wise velocity $w$ , highlights this structure. In the center of each structure, the span-wise velocity $w$  is close to zero, and the flow remains nearly 2D, resembling the 2D asymmetric state with its characteristic three-roll structure, similar to figure \ref{fig:2dSS_3d_b}. However, between these vortex structures, the maximum and minimum values of $w$ appear, indicating significant circulation in the homogeneous direction.
This is further illustrated in figure \ref{fig:DanishPastry_2dViz_a}, which shows the 2D flow profile in the $x$-$y$ plane at $z=0$ for the case in figure \ref{fig:DanPasStates_b}. Here, the flow remains close to 2D, and the velocity field strongly resembles the 2D asymmetric state. Figure \ref{fig:DanishPastry_2dViz_b} displays the 2D flow in the $z$-$y$ plane at $x=0$ over one wavelength for the same case. This view reveals the strong 3D circulation occurring between each of the vortex ring structures.

The three visualizations of this state in figure \ref{fig:DanPasStates} show the change in the state for increasing $\Pi_c$. Most notably, the wavelength of the vortex-ring structure decreases with increasing wavelength, with values close to the critical value exhibiting a wavelength of 16, while values with larger $\Pi_c$ begin to exhibit structures with wavelengths of 2 for 4, likely due to the increased dynamical instability resulting from increased surface heating at larger $\Pi_c$ values. We note that if the wavelength is not long enough to capture the extent of the structure, such as using $L_z=8$ for the case $\Pi_c=1040$, the case will converge to the 2D asymmetric state. Therefore, within the reported range for the 2D asymmetric state, it may be the case that some of these cases will be 3D for sufficiently long $L_z$, but we limit our simulations to a maximum $L_z = 16$. 

In addition to the change in the wavelength, the structure of the flow changes with increasing $\Pi_c$ as well. For relatively low $\Pi_c$ values, the 3D flow structures are very close to the structures shown in the 2D case in figure \ref{fig:2dSS_3d_b}, only with an additional 3D component at the break in the structure. As $\Pi_c$ increases, the structure of the central circulation roll begins to gain more strength and extend over a larger area of the 2D valley. Consequently, visualization of this circulation with $Q$ criterion in figures \ref{fig:DanPasStates_b} and \ref{fig:DanPasStates_c} show a separation or gap between the main downwelling motion in the center and the upwelling motion in the corner. As a result, the secondary circulation in the same corner as the dominant circulation weakens. This phenomenon, observed in our visualizations, produces a row of distinct, well-separated vortex-ring structures. Due to their resemblance to the patterns of a Danish pastry, we refer to this flow state as the Danish-pastry state.

This state does not appear as a linear instability to the 2D asymmetric base flow in our LSA, and therefore we cannot give a precise critical value at which this state becomes an equilibrium state. However, from nonlinear simulations, we are able to obtain this state with $L_z = 16$ at a $\Pi_c$ value as low as approximately 940, and as large as 1870. We note that this lower bound for the Danish-pastry state is lower than that of the oscillating asymmetric state discussed previously in \S \ref{sec:Hopf}, and our simulations indeed show that both states coexist in the same range of $\Pi_c$ values. In fact, we have observed a number of cases in which both the steady Danish-pastry and the oscillating asymmetric state are obtained at the same parameter values with only differences in initial conditions. This indicates a sensitivity with respect to initial conditions and bistability in this region. However, in this paper we focus only on the final states, and the path leading to these final states deserves its own attention in a future work. 

\subsubsection{Quasi-steady Danish-pastry state} \label{sec:OscDanPast}

\begin{figure*}
    \centering
    \begin{subfigure}[b]{0.5\textwidth}
        \centering
        \includegraphics[width=.85\textwidth]{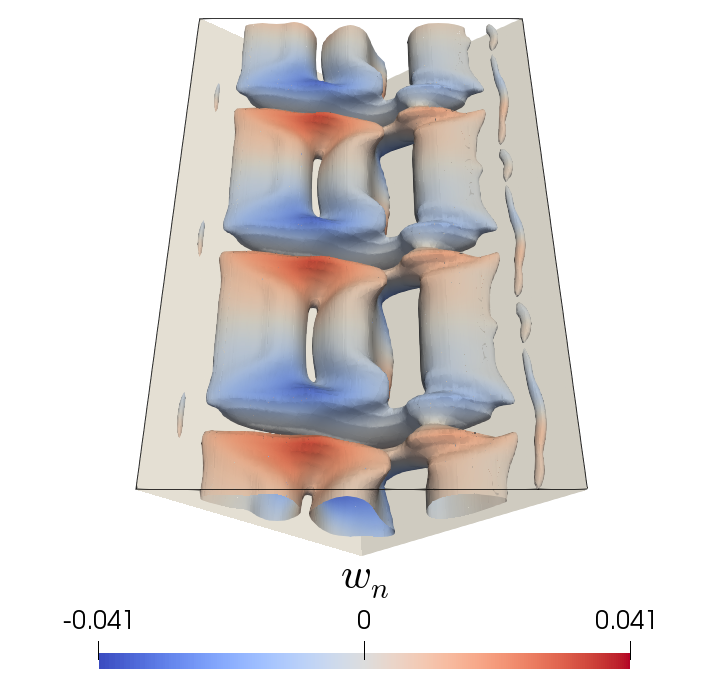}
        \caption{}
        \label{fig:OscDanPas_a}
    \end{subfigure}%
        ~ 
    \begin{subfigure}[b]{0.5\textwidth}
        \centering
        \includegraphics[width=\textwidth]{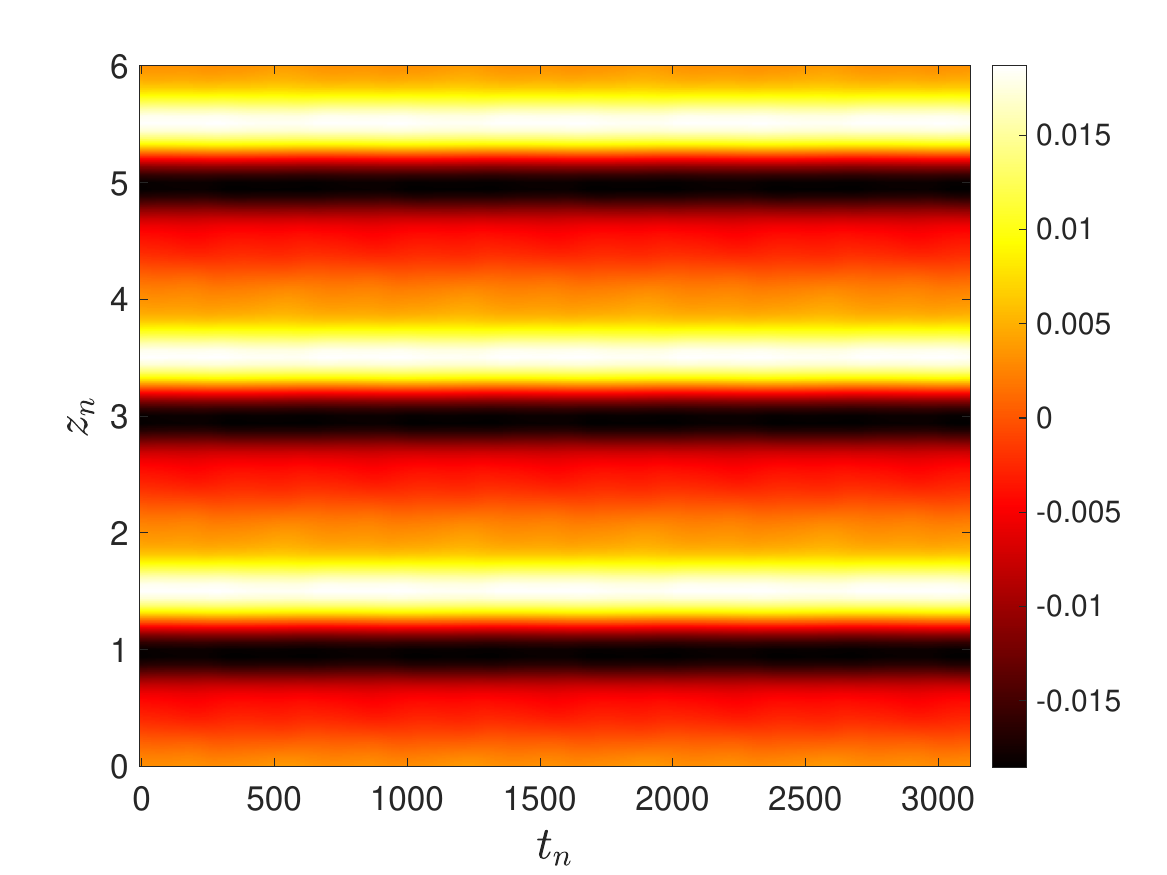}
        \caption{}
        \label{fig:OscDanPas_b}
    \end{subfigure}
    ~ 
    \begin{subfigure}[b]{0.5\textwidth}
        \centering
        \includegraphics[width=\textwidth]{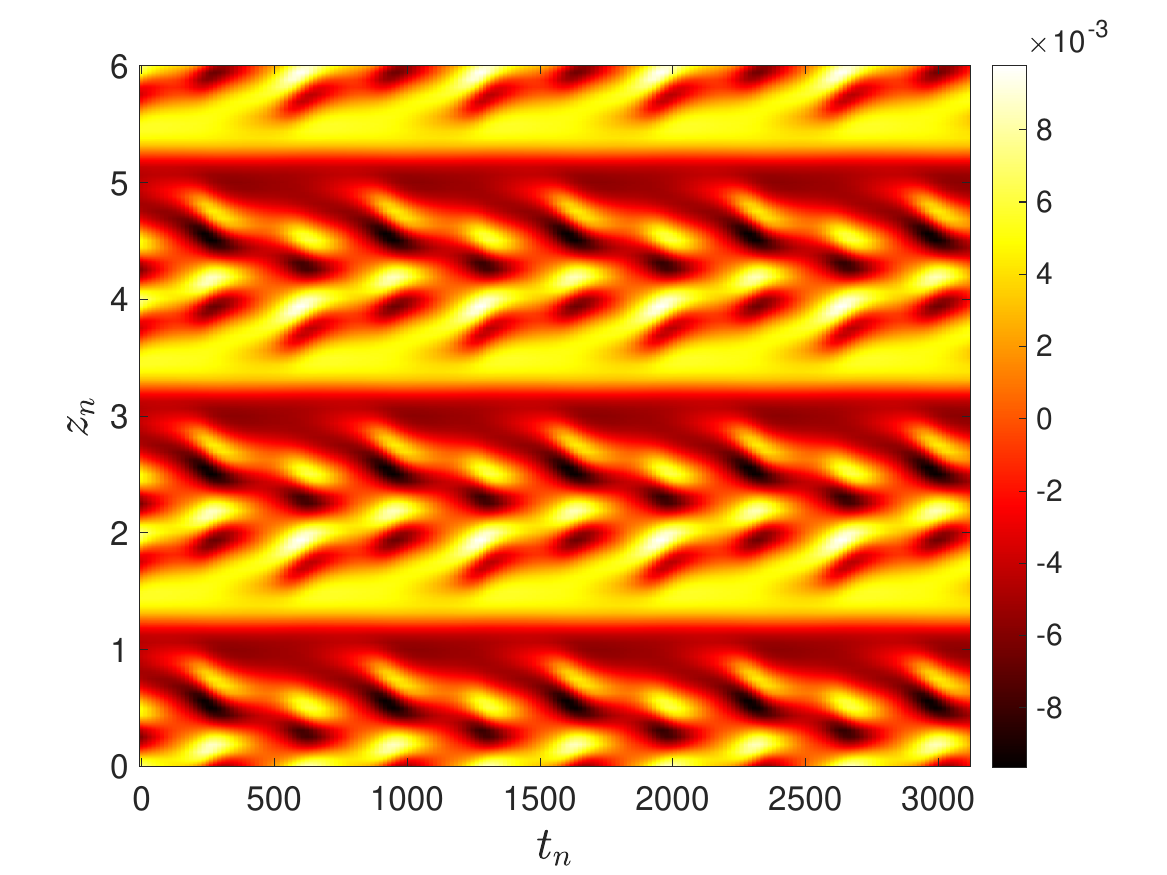}
        \caption{}
        \label{fig:OscDanPas_c}
    \end{subfigure}%
    ~ 
    \begin{subfigure}[b]{0.5\textwidth}
        \centering
        \includegraphics[width=\textwidth]{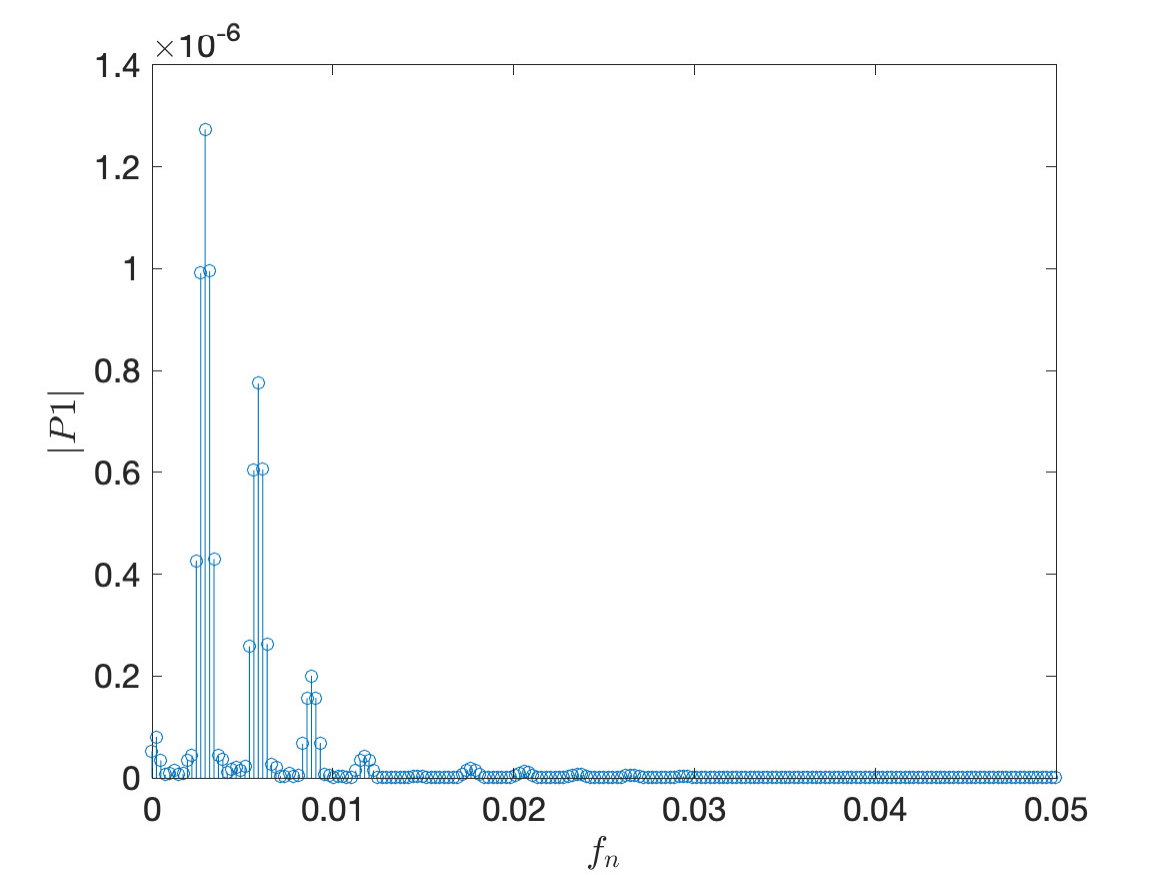}
        \caption{}
        \label{fig:OscDanPas_d}
    \end{subfigure}
    \caption{Quasi-steady Danish-pastry state. (a) Visualization of time-averaged flow for $\Pi_c = 2180$, $\Pi_h = 7000$, and $L_z = 6$, (b) Color plot of span-wise velocity $w$  along the $z$ direction versus time at point (0, 0.9, 0), (c) Color plot of span-wise velocity $w$  along the $z$ direction versus time at point (1.2, 0.85, 0), (d) Frequency spectrum of kinetic energy data.}
\end{figure*}

As $\Pi_c$ increases, the steady Danish-pastry state loses stability and an oscillating instability arises. Similar to the Hopf bifurcation in \S \ref{sec:Hopf}, this instability is due to the interaction of a corner vortex with a larger, more central vortex, but this time the instability arises in the corner opposite the main circulation and interacts with the secondary circulation. A visualization of the time-average state for multiple periods is shown with $Q$ criterion in figure \ref{fig:OscDanPas_a}.
The oscillation is very localized to the right corner of the valley. This can be seen in figures \ref{fig:OscDanPas_b} and \ref{fig:OscDanPas_c}; \ref{fig:OscDanPas_b} shows the color map of the span-wise velocity $w$  at a point in the center of the domain, $\left(0, 0.9, 0\right)$, whereas \ref{fig:OscDanPas_c} shows the color map of $w$ at a point in the right corner, $\left(1.2, 0.85, 0\right)$. This comparison shows that in the center of the domain there is little to no oscillation, and the state resembles the steady Danish-pastry, whereas in the corner there is a strong, steady oscillation. Figure \ref{fig:OscDanPas_d} shows the frequency spectrum of the kinetic energy over time data, showing that the dominant frequency is approximately 0.003, as well as a significant contribution from the second and third harmonic frequencies of the fundamental frequency. 
The fundamental frequency is significantly smaller than the dominant frequency of the prior oscillating state, which exhibited a dominant frequency of approximately 0.02. 
This may be due to the fact that the motion is confined to a corner of the domain where two relatively weaker vortices interact, whereas the prior instability resulted from an oscillation of the primary circulation roll which carries a greater amount of energy and heat transfer than the secondary vortices.
Nonlinear simulations suggest that this oscillation to the Danish-pastry state begins at a value of $\Pi_c \approx 1850$, and this steady oscillation seems to exist up to $\Pi_c \approx 2600$. 

\subsubsection{Unsteady flow} \label{sec:Unsteady}

As $\Pi_c$ increases further, the strong vortex rings that formed the Danish-pastry state break up completely, and the flow becomes fully unsteady. In this section, we analyze the unsteady flow at two different $\Pi_c$ values, one with significantly larger $\Pi_c$, to show how the unsteady flow changes with increase in $\Pi_c$.

\begin{figure*}
    \centering
    \begin{subfigure}[b]{0.45\textwidth}
        \centering
        \includegraphics[width=\textwidth]{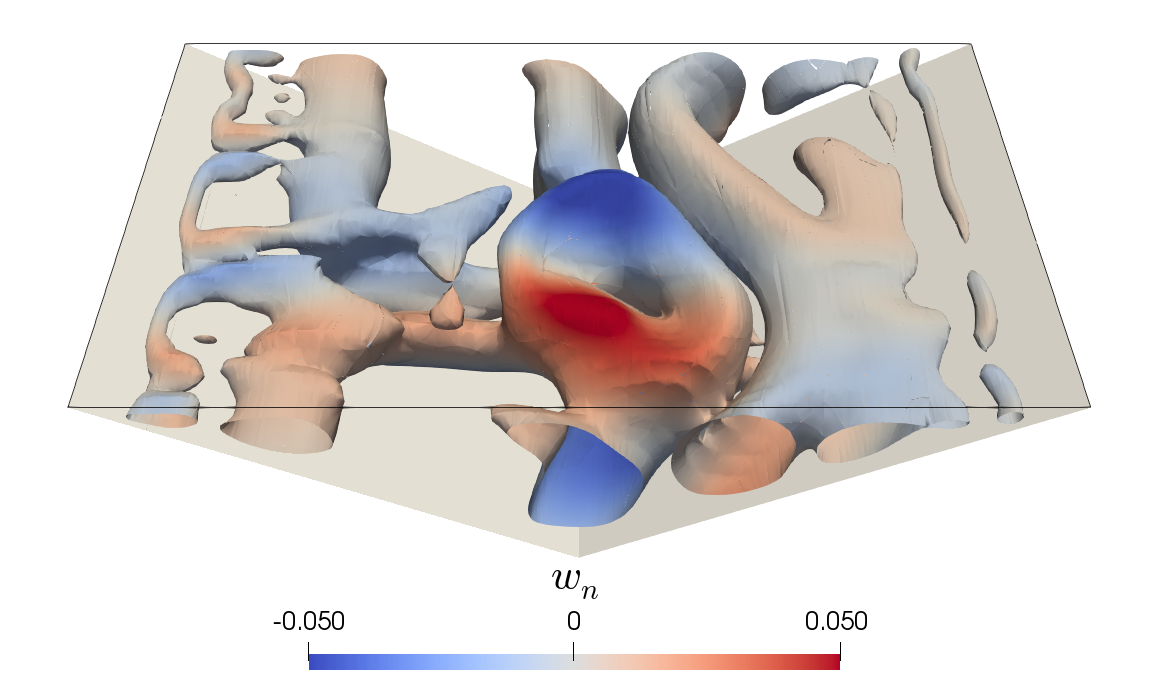}
        \caption{}
        \label{fig:Unsteady3700PIc_a}
    \end{subfigure}%
    ~ 
    \begin{subfigure}[b]{0.45\textwidth}
        \centering
        \includegraphics[width=\textwidth]{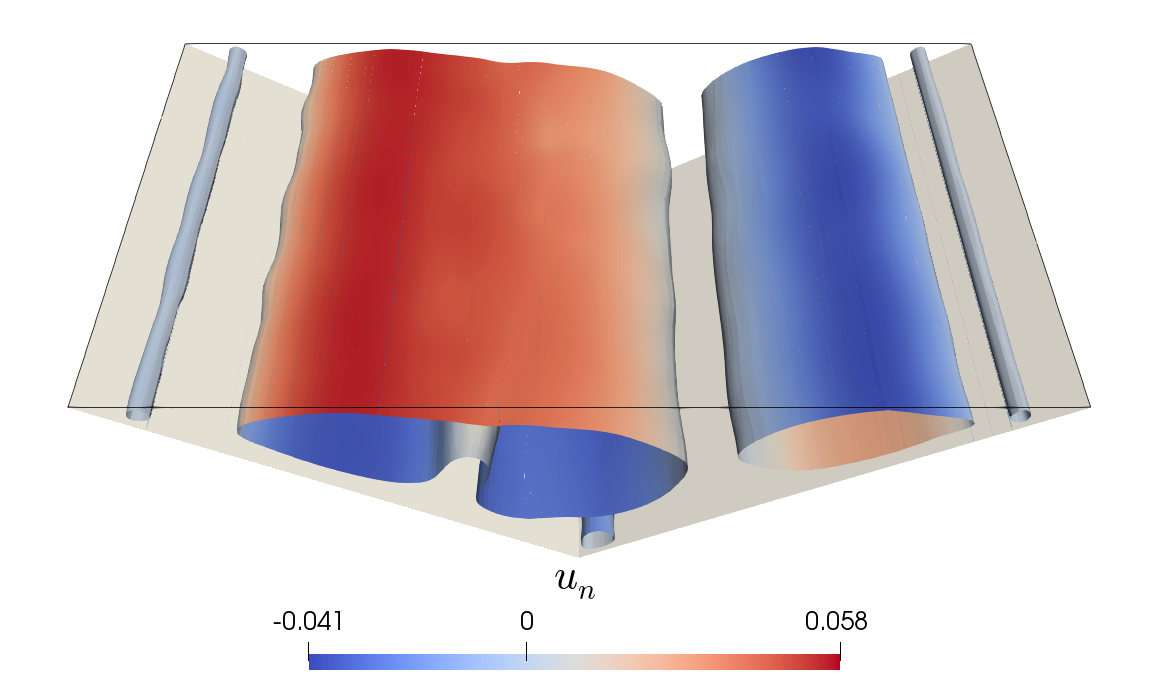}
        \caption{}
        \label{fig:Unsteady3700PIc_b}
    \end{subfigure}
    ~
    \begin{subfigure}[b]{0.45\textwidth}
        \centering
        \includegraphics[width=\textwidth]{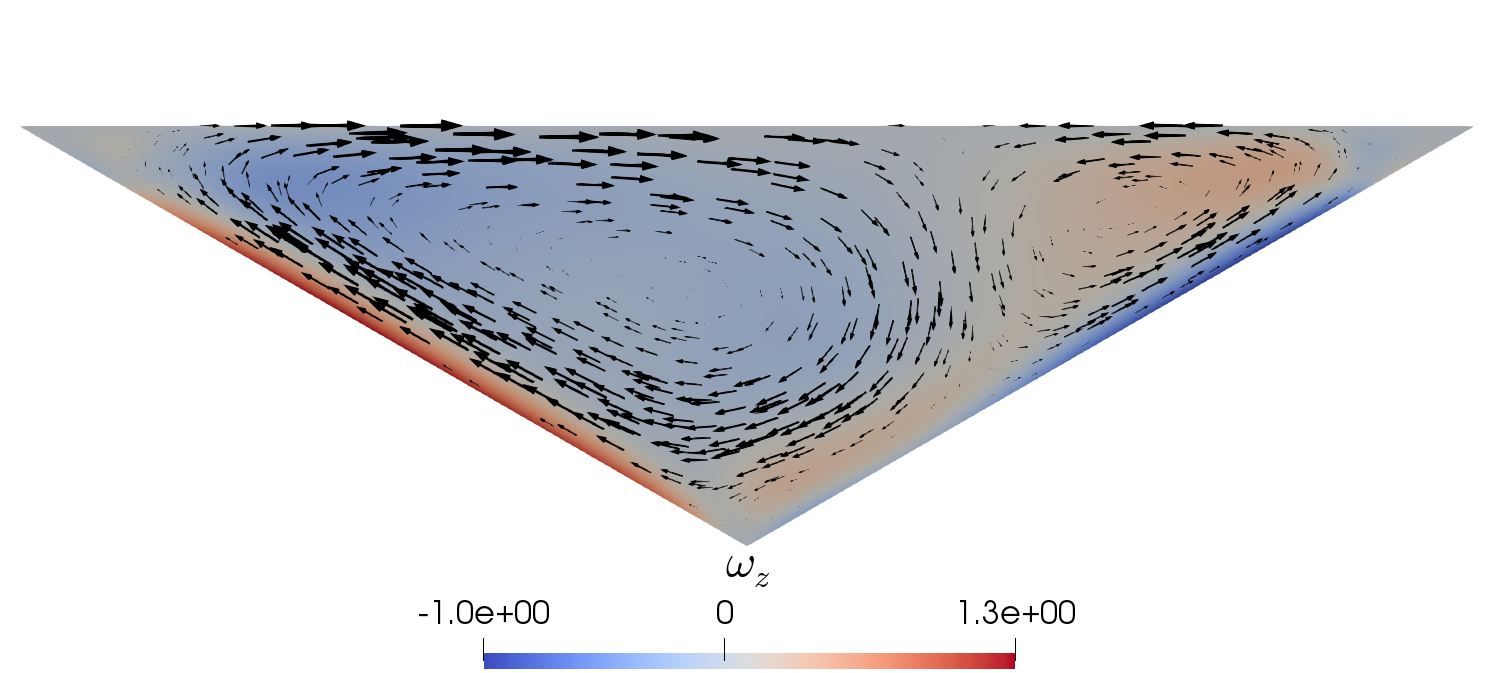}
        \caption{}
        \label{fig:Unsteady3700PIc_c}
    \end{subfigure}
    \caption{Visualization of 3D unsteady flow at parameter values $\Pi_c = 3725$, $\Pi_h = 1705$, (a) $Q$ criterion of instantaneous field,  (b) $Q$ criterion of time-averaged field for ten diffusion timescales, and (c) $z$-vorticity and velocity vectors of 2D-averaged field of time averaged field.}
    \label{fig:Unsteady3700PIc}
\end{figure*}

\begin{figure*}
    \centering
    \begin{subfigure}[b]{0.5\textwidth}
        \centering
        \includegraphics[width=\textwidth]{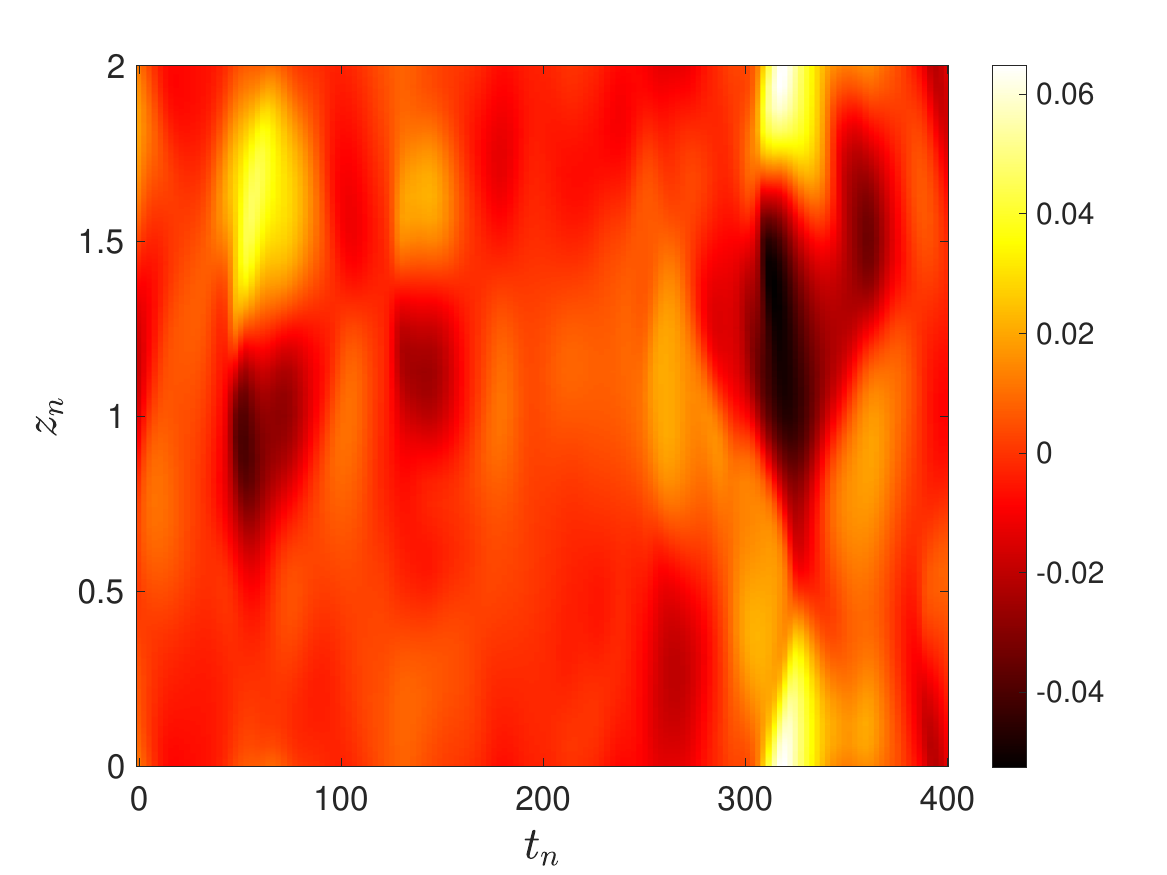}
        \caption{}
        \label{fig:Unsteady3700PIc_colormap}
    \end{subfigure}%
    ~ 
    \begin{subfigure}[b]{0.5\textwidth}
        \centering
        \includegraphics[width=\textwidth]{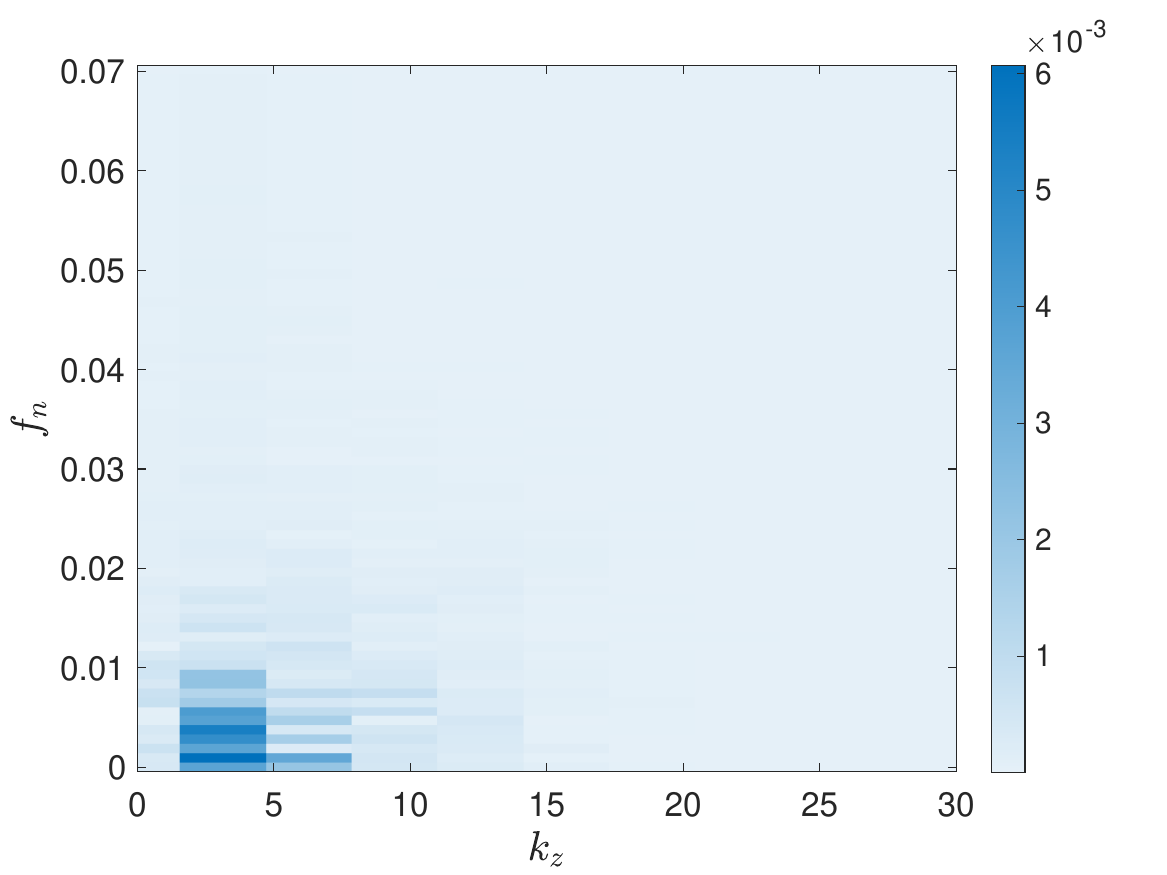}
        \caption{}
        \label{fig:Unsteady3700PIc_2dFFT}
    \end{subfigure}
    \caption{(a) Color plot of span-wise velocity $w$  along the $z$ direction versus time at point (0, 0.9, 0) for case with $\Pi_c = 3725$, $\Pi_h = 1705$, corresponding to flow structures in figure \ref{fig:Unsteady3700PIc_a}, (b) spatio-temporal spectrum of data in (a).}
\end{figure*}

First, figure \ref{fig:Unsteady3700PIc_a} depicts the instantaneous flow field with $Q$ criterion for $\Pi_c = 3725$, $\Pi_h = 1705$, and $L_z = 2$. This instantaneous visualization shows that the remnants of the convection rolls are still present when the flow becomes unsteady, only now with large 3D periodic bursts and more instability and interaction between the rolls, including bending and twisting of the rolls. This can also be seen in figure \ref{fig:Unsteady3700PIc_colormap}, showing a color map of $w$ versus time; this shows that there are periods when the w velocity is not very large, but there are periodic bursts in which there is a large transient 3D structure. The instant shown in figure \ref{fig:Unsteady3700PIc_a} represents one of these instants, at $t_n \approx 50$. 

Figure \ref{fig:Unsteady3700PIc_b} shows the $Q$ criterion of the time-averaged state for ten diffusion timescales. In order to determine the appropriate length to average each state over, we considered the three main time scales relevant to our problem: the convection timescale $t_{0,\mathrm{convection}} = H/u_0$, the diffusion timescale $t_{0,\mathrm{diffusion}} = H^2/\beta$, and the buoyancy or stratification timescale $t_{0,\mathrm{buoyancy}} = 1/N$. Of the three of these, the diffusion timescale is the largest, and thus to capture the physical processes of all aspects of the problem, we averaged each simulation over a number of diffusion timescales. 
The time-averaged state shows that for long times, all 3D fluctuations are averaged out, and the flow resembles the 2D asymmetric state. The 2D averaged state is shown in \ref{fig:Unsteady3700PIc_c} to further show that the 2D profile matches with that of the 2D asymmetric profile shown previously. 

Figure \ref{fig:Unsteady3700PIc_2dFFT} shows the spatio-temporal spectrum of the color map data; this shows both the frequency and the wavenumber content of the signal over time. As can be seen, given $L_z = 2$ for the simulation, the dominant wavenumber is 3.14. This makes sense given the strong 3D bursts which occur with one wavelength in the domain corresponding to this wavenumber. There is also a noticeable signal at wavenumber of 6.27 and 9.42, corresponding to wavelengths of 1 and 0.67 respectively. This sets the current case apart from the prior simulations described in which the signal was largely contained by one wavenumber. As we move into the unsteady regime, a broader spectrum of wavenumbers appear, although there is little signal for wavenumbers above 9.42. Regarding the temporal signal, the dominant frequencies are observed to be low frequencies at a wavenumber of 3.14; specifically the frequency peaks between 0.001 and 0.004. This represents a slow oscillation, similar to the fundamental frequency of oscillating Danish-pastry state described in \S \ref{sec:OscDanPast}. For frequencies of greater than approximately 0.1, the frequency content drops quickly signaling we are still far from the turbulent regime.

\begin{figure*}
    \centering
    \begin{subfigure}[b]{0.45\textwidth}
        \centering
        \includegraphics[width=\textwidth]{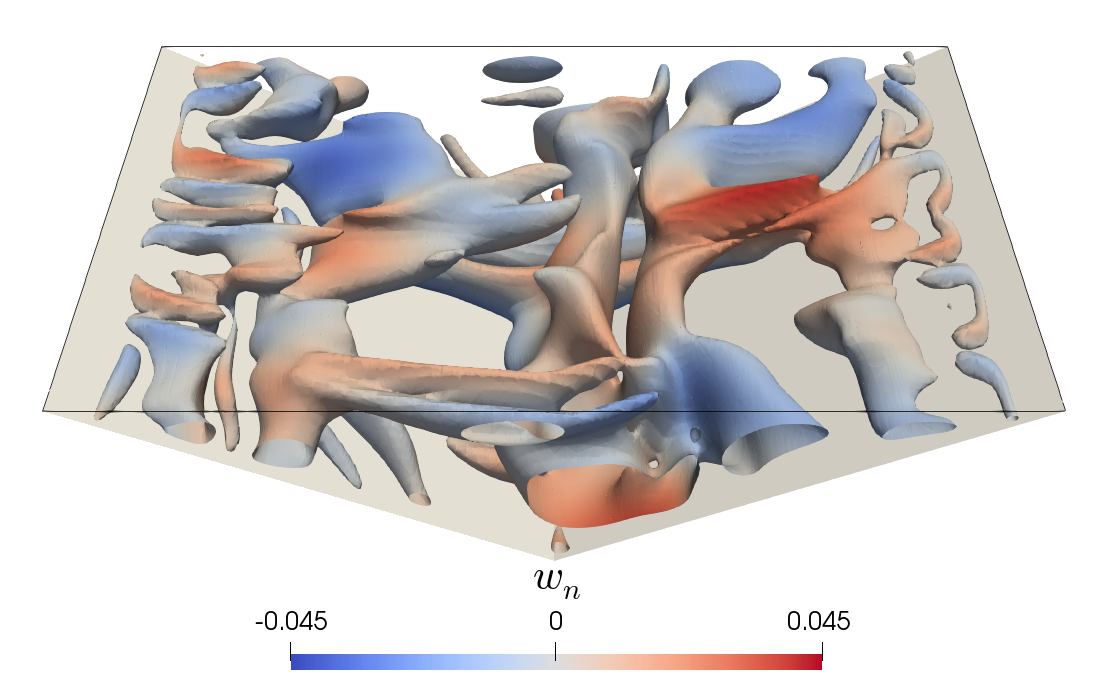}
        \caption{}
        \label{fig:Unsteady6554PIc_a}
    \end{subfigure}%
    ~ 
    \begin{subfigure}[b]{0.45\textwidth}
        \centering
        \includegraphics[width=\textwidth]{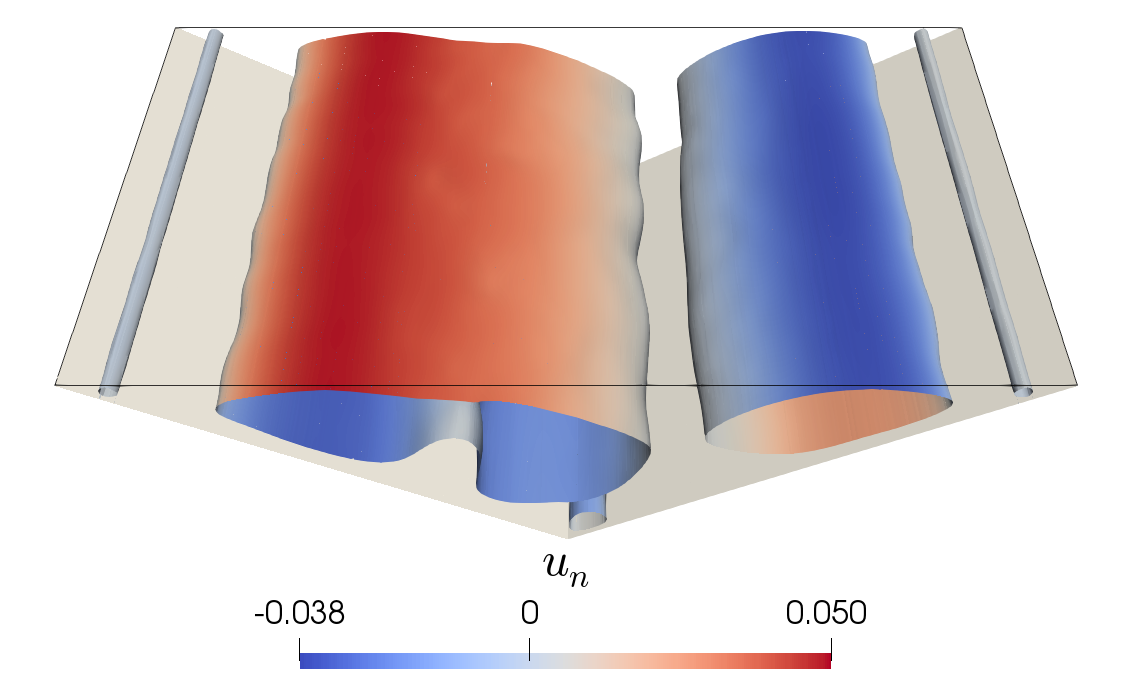}
        \caption{}
        \label{fig:Unsteady6554PIc_b}
    \end{subfigure}
    ~
    \begin{subfigure}[b]{0.45\textwidth}
        \centering
        \includegraphics[width=\textwidth]{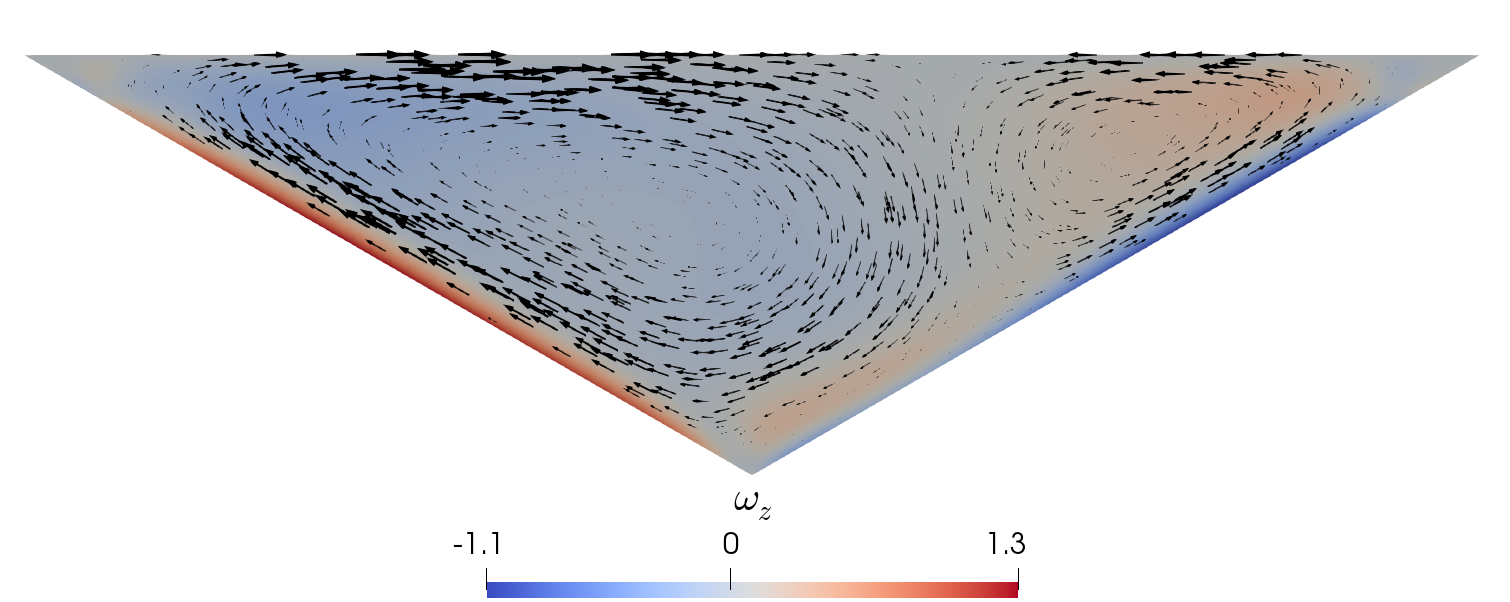}
        \caption{}
        \label{fig:Unsteady6554PIc_c}
    \end{subfigure}
    \caption{Visualization of 3D unsteady flow at parameter values $\Pi_c = 6554$, $\Pi_h = 7661$, (a) Q-criterion of instantaneous field,  (b) Q-criterion of time-averaged field for ten diffusion timescales, and (c) vorticity and velocity vectors of the spanwise-averaged and time-averaged field.}
    \label{fig:Unsteady6554PIc}
\end{figure*}

\begin{figure*}
    \centering
    \begin{subfigure}[b]{0.5\textwidth}
        \centering
        \includegraphics[width=\textwidth]{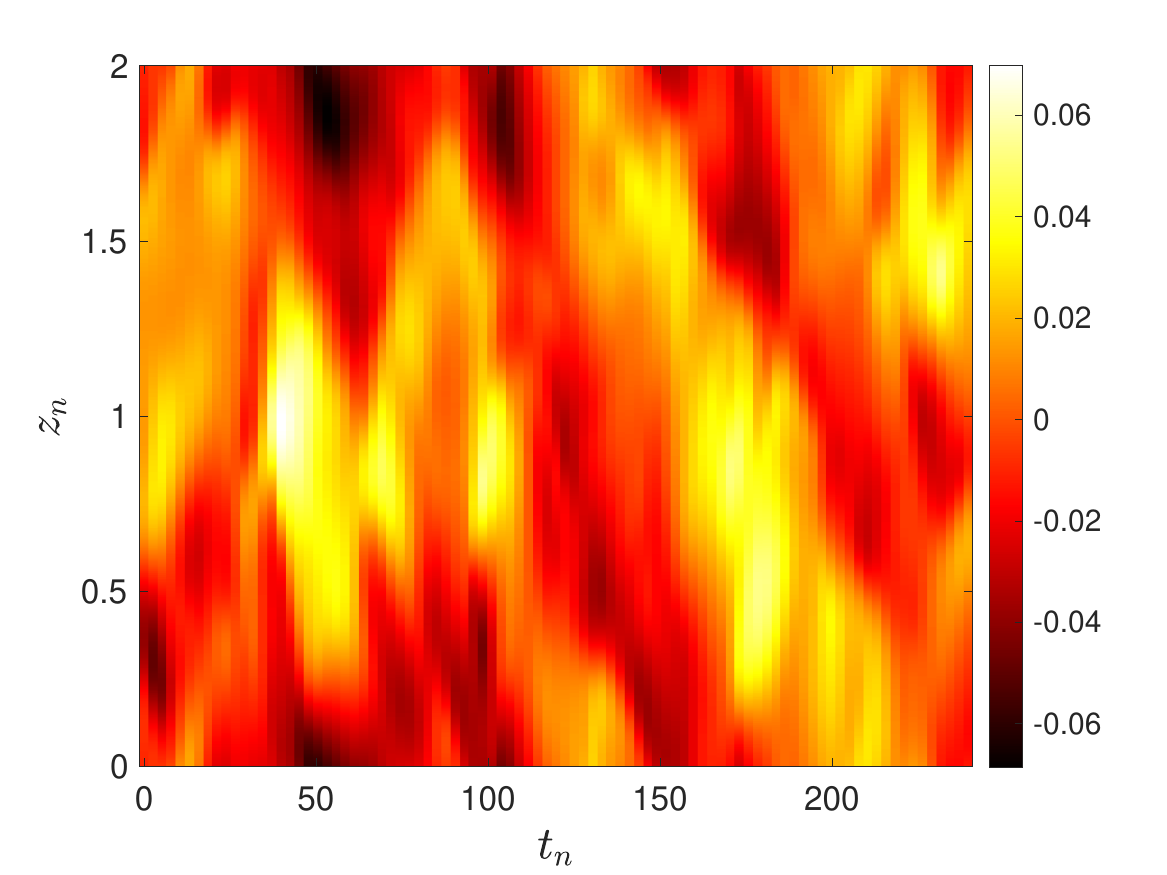}
        \caption{}
        \label{fig:Unsteady6500PIc_colormap}
    \end{subfigure}%
    ~ 
    \begin{subfigure}[b]{0.5\textwidth}
        \centering
        \includegraphics[width=\textwidth]{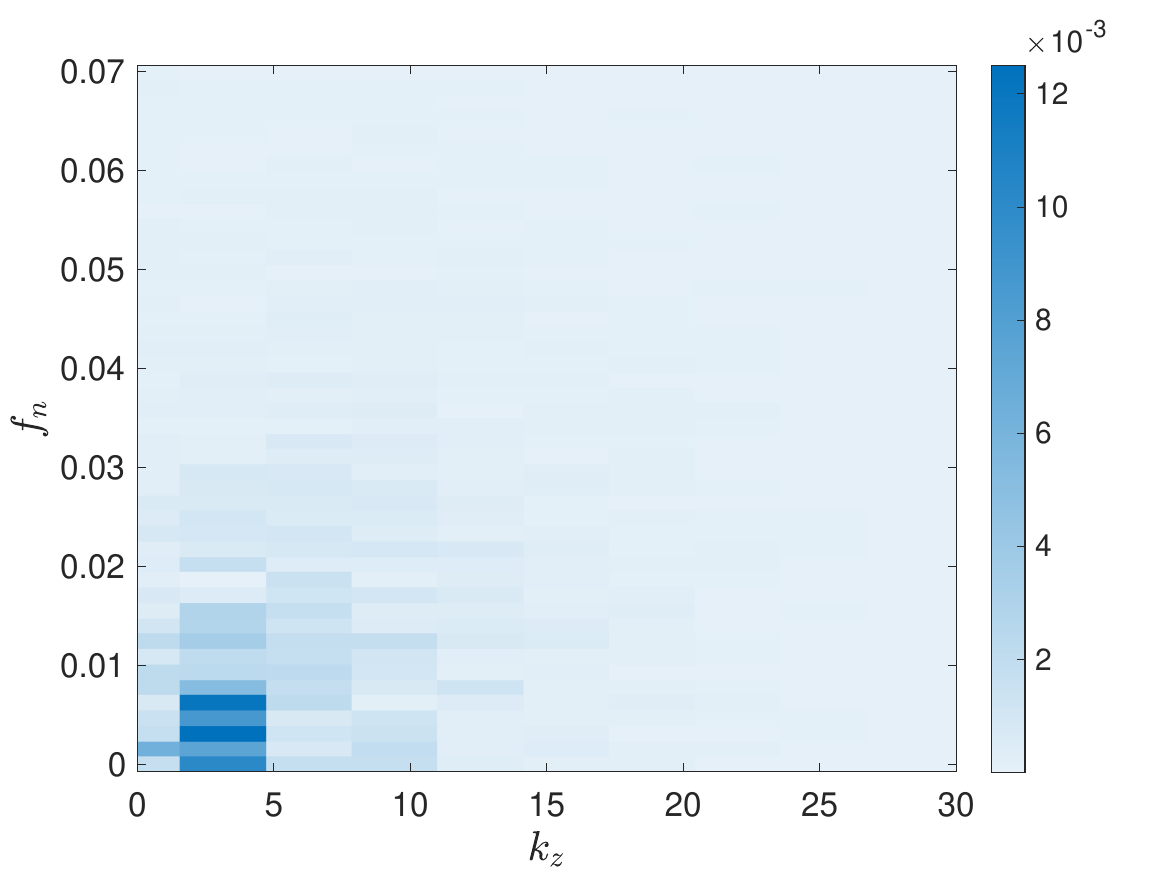}
        \caption{}
        \label{fig:Unsteady6500PIc_2dFT}
    \end{subfigure}
    \caption{(a) Color plot of span-wise velocity $w$  along the $z$ direction versus time at point (0, 0.9, 0) for case with $\Pi_c = 6554$, $\Pi_h = 7661$, corresponding to flow structures in figure \ref{fig:Unsteady6554PIc}, (b) spatio-temporal spectrum of data in (a).}
\end{figure*}

 We now increase the $\Pi_c$ to 6554 to observe further evolution of the unsteady flow state. Figure \ref{fig:Unsteady6554PIc_a} shows the instantaneous flow state at $\Pi_c = 6554$, $\Pi_h = 7661$, and $L_z = 2$, and \ref{fig:Unsteady6554PIc_b} shows the time-averaged state for ten diffusion timescales. Once again, traces of the 2D rolls are still present in the instantaneous flow field, but it is considerably more unsteady with smaller structures than that pictured at $\Pi_c = 3725$. This can also be seen in figure \ref{fig:Unsteady6500PIc_colormap}, showing the color map of the span-wise velocity $w$  over time. Unlike the prior case, the 3D bursts occur more frequently and are more unsteady than observed in the lower $\Pi_c$ case. However, in figure \ref{fig:Unsteady6554PIc_b}, the time-averaged state remains largely unchanged from the lower $\Pi_c$ case. The 2D asymmetric state continues to prevail for long time-averaging windows. This confirms the consistent presence of the 2D rolls in the instantaneous state, although they are not always clear. Figure \ref{fig:Unsteady6554PIc_c} shows the 2D average of the time-averaged state, again showing the same 2D asymmetric flow profile.

 Figure \ref{fig:Unsteady6500PIc_2dFT} shows the spatio-temporal spectrum of the data in figure \ref{fig:Unsteady6500PIc_colormap}. The dominant wavenumber is still 3.14, corresponding to the wavelength of 2, but when compared with the lower $\Pi_c$ value in figure \ref{fig:Unsteady3700PIc_2dFFT}, it is clear that there is now a greater signal at larger wavenumbers, indicating the presence of some smaller structures. Likewise, while the dominant temporal frequencies do not change drastically, the dominant frequency being approximately 0.003 to 0.006, there is a noticeable increase in the strength of higher frequency signals above 0.02. This demonstrates the greater presence of instability and faster motions in the fluid as a result of greater surface heating.

\subsubsection{Unsteady flow at $Pr = 0.7$} \label{sec:Unsteady_0_7Pr}

\begin{figure*}
    \centering
    \begin{subfigure}[b]{0.45\textwidth}
        \centering
        \includegraphics[width=\textwidth]{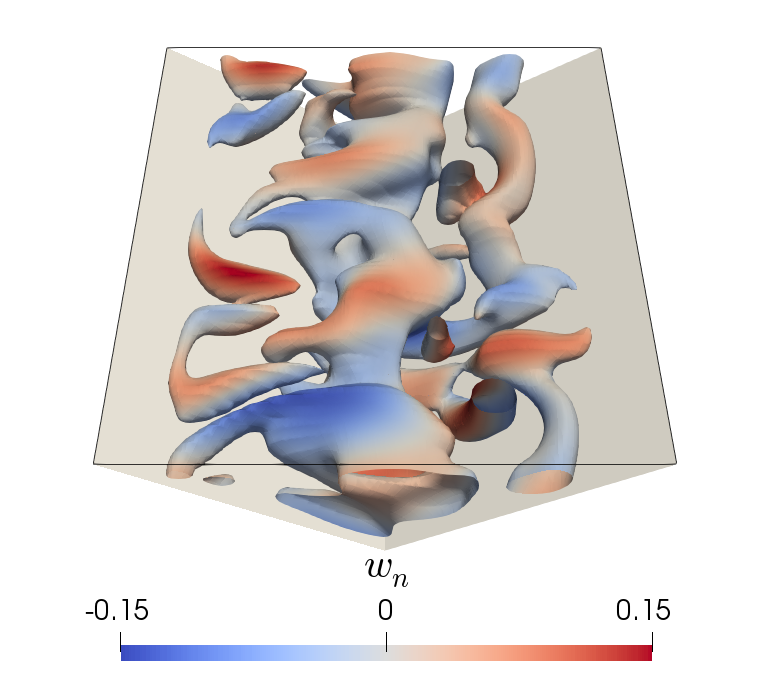}
        \caption{}
        \label{fig:UnsteadyPr0_7_a}
    \end{subfigure}%
    ~ 
    \begin{subfigure}[b]{0.45\textwidth}
        \centering
        \includegraphics[width=\textwidth]{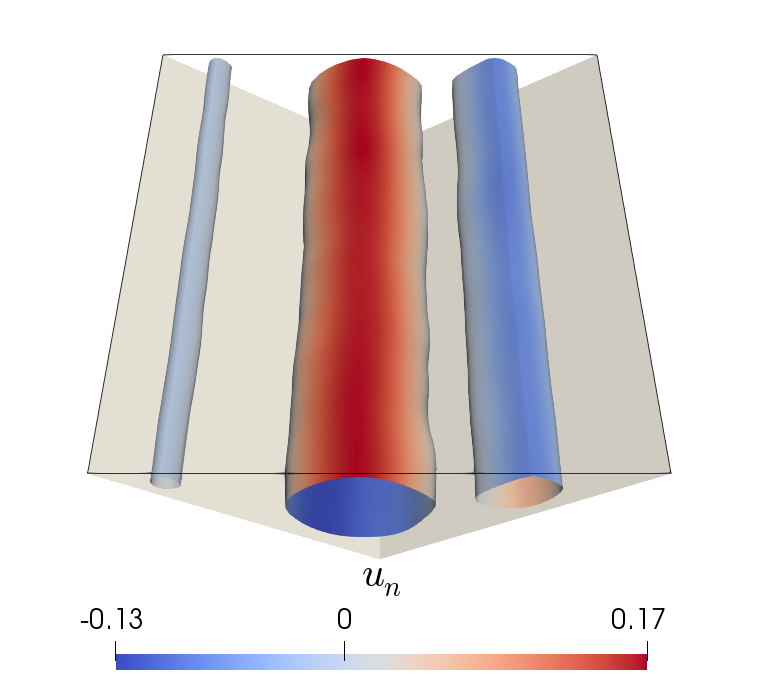}
        \caption{}
        \label{fig:UnsteadyPr0_7_b}
    \end{subfigure}
    ~
    \begin{subfigure}[b]{0.55\textwidth}
        \centering
        \includegraphics[width=\textwidth]{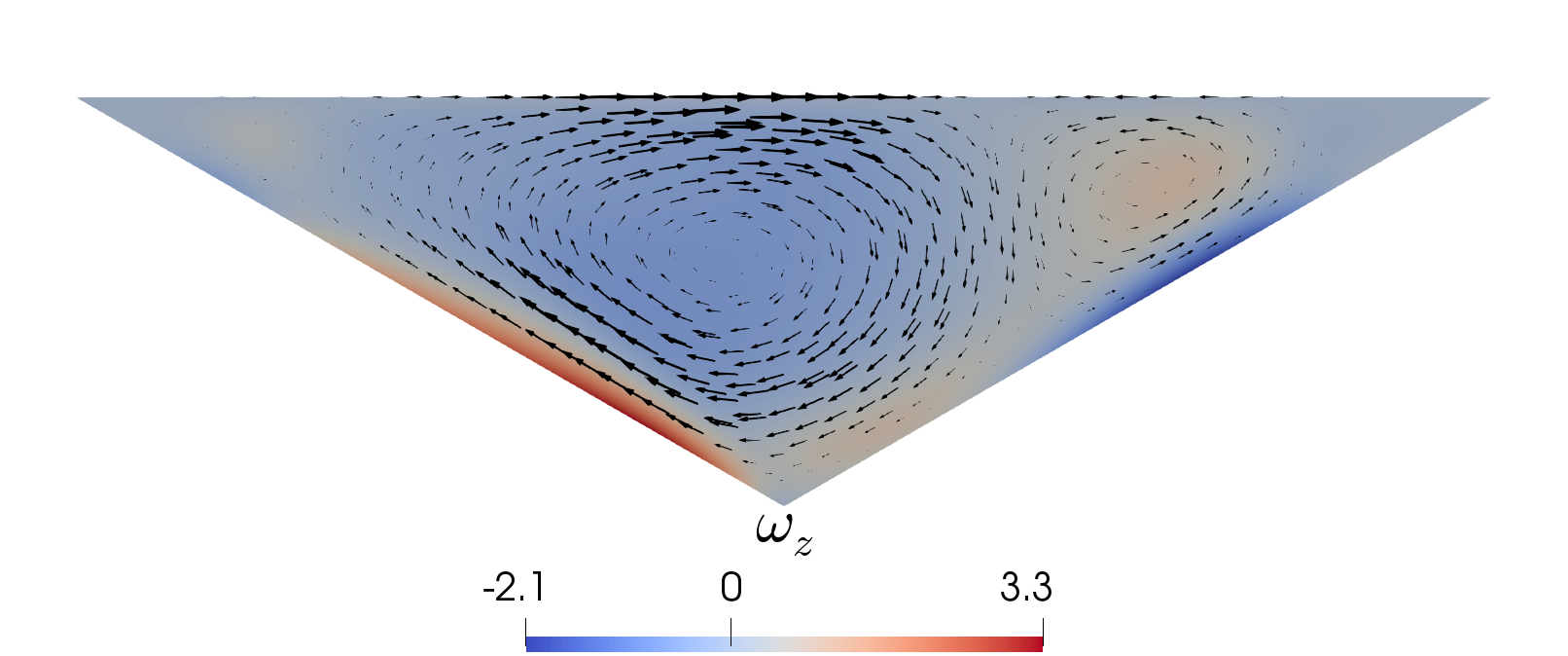}
        \caption{}
        \label{fig:UnsteadyPr0_7_c}
    \end{subfigure}
    \caption{Visualization of 3D unsteady flow at parameter values $\Pi_c = 297$, $\Pi_h = 1500$, and $Pr = 0.71$, (a) $Q$ criterion of instantaneous field,  (b) $Q$ criterion of time-averaged field for twenty diffusion timescales, and (c) vorticity and velocity vectors of the spanwise-averaged and time-averaged field.}
    \label{fig:UnsteadyPr0_7}
\end{figure*}

We now consider the case of unsteady flow at $Pr = 0.7$, more typical of air in a valley. Unlike at $Pr = 7$, we do not conduct an exhaustive characterization of all possible flow states at $Pr = 0.7$, but we have confirmed that at low $\Pi_c$ values, directly above the threshold for instability, the 2D asymmetric state continues to be the dominant state. Here however, we consider a case in the unsteady regime to compare to our prior cases at $Pr = 7$. Figure \ref{fig:UnsteadyPr0_7_a} shows the instantaneous flow field for the case of $\Pi_c = 297$, $\Pi_h = 1500$, and $Pr = 0.7$, while figure \ref{fig:UnsteadyPr0_7_b} shows the time-averaged state for 20 diffusion timescales. First, we point out that the lower Prandtl number makes the flow field considerably more unstable for the same $\Pi_c$ and $\Pi_h$. At the same $\Pi_c$ and $\Pi_h$ at $Pr = 7$, the flow exhibits the 2D steady asymmetric state, and is depicted in figure \ref{fig:2dSS_a}. This behavior is expected because decreasing the Prandtl number reduces momentum diffusion while increasing thermal diffusion. As a result, velocity perturbations are less damped, and thermal variations spread more quickly, making the flow more sensitive to disturbances. Considering the structure of the instantaneous state, similar to the prior cases, we see the prevalence of the central circulation in the middle of the valley, but with significant instability and interaction with the other rolls causing twisting and bending of the rolls along the $z$ direction. In contrast to the prior unsteady cases at $Pr = 7$, the motion is more concentrated in the center of the valley, with little effects from the small circulations in the corners. Looking at the time-averaged state in figure \ref{fig:UnsteadyPr0_7_b} and the 2D-averaged state in figure \ref{fig:UnsteadyPr0_7_c}, we see that the 2D asymmetric state is still prevalent for the smaller Prandtl number. The structure of the 2D state differs, with the central circulation being more confined to the center of the valley, but this may be due to the relatively smaller $\Pi_c$ value, as the same thing is seen at low $\Pi_c$ at $Pr = 7$, like the case shown in \ref{fig:2dSS_a} and \ref{fig:2dSS_3d_a}.

\begin{figure*}
    \centering
    \begin{subfigure}[b]{0.5\textwidth}
        \centering
        \includegraphics[width=\textwidth]{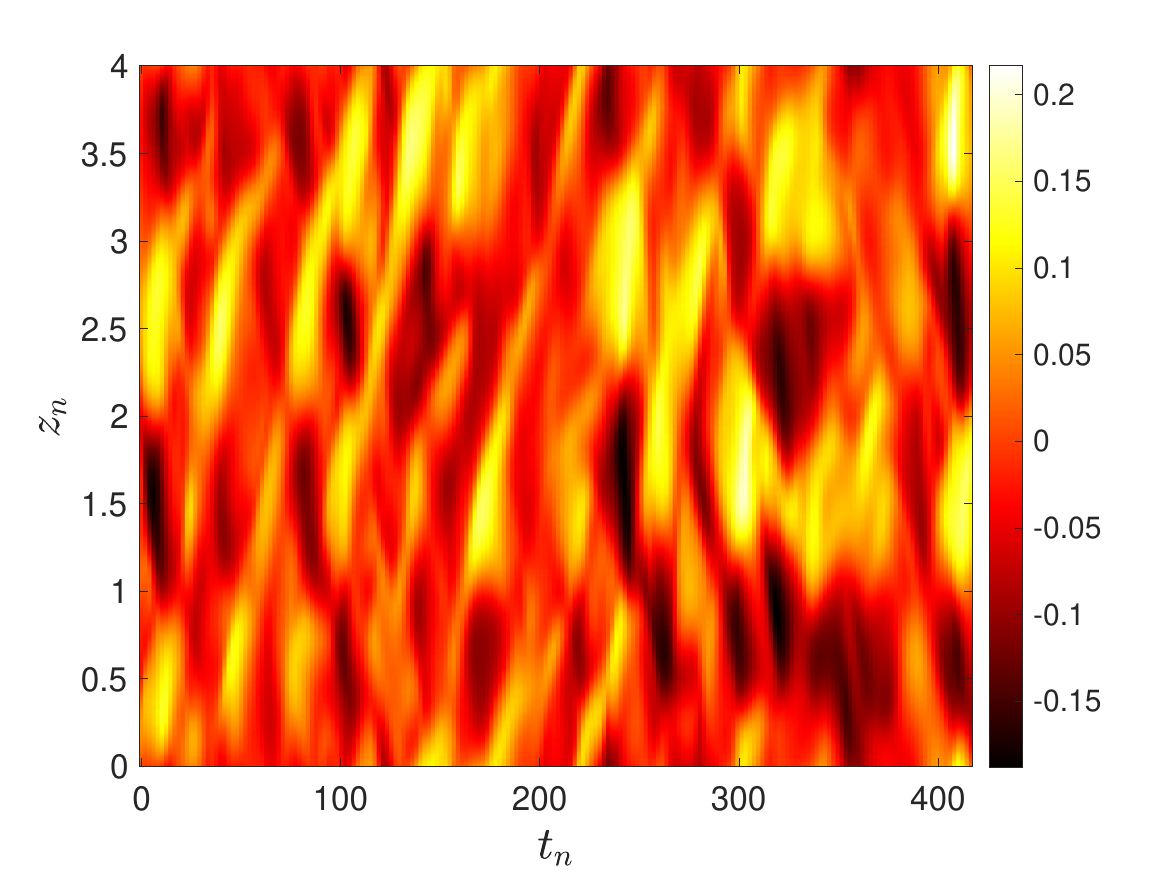}
        \caption{}
        \label{fig:UnsteadyPr0_7_colormap}
    \end{subfigure}%
    ~ 
    \begin{subfigure}[b]{0.5\textwidth}
        \centering
        \includegraphics[width=\textwidth]{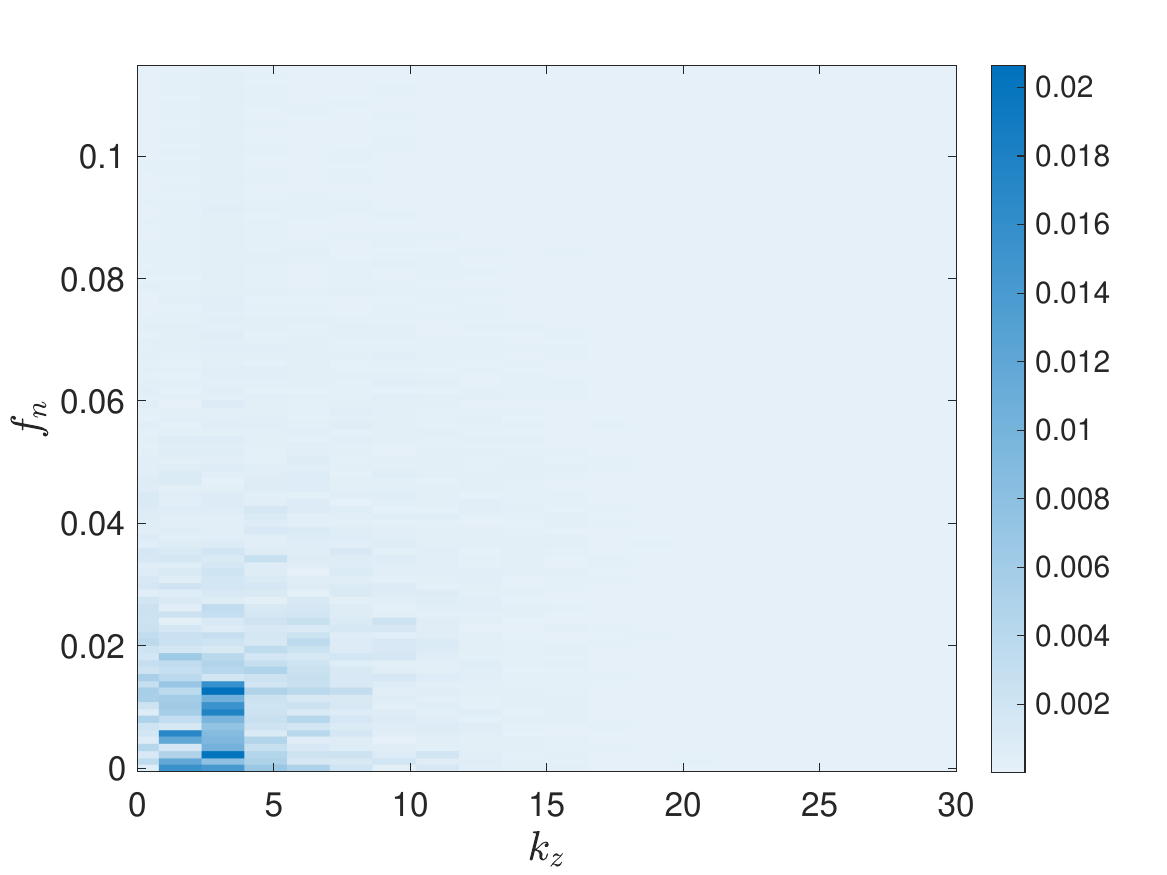}
        \caption{}
        \label{fig:UnsteadyPr0_7_2dFT}
    \end{subfigure}
    \caption{(a) Color plot of span-wise velocity $w$  along the $z$ direction versus time at point (0, 0.9, 0) for case with $\Pi_c = 297$, $\Pi_h = 1500$, and $Pr = 0.71$, corresponding to flow structures in figure \ref{fig:UnsteadyPr0_7}, (b) spatio-temporal spectrum of data in (a).}
\end{figure*}

The color map of span-wise velocity $w$  over time is shown in figure \ref{fig:UnsteadyPr0_7_colormap}, and the spatio-temporal spectrum is shown in figure \ref{fig:UnsteadyPr0_7_2dFT}. 
Unlike the two unsteady case analyzed at $Pr = 7$, the color map appears to show some quasi-periodic behavior, such as from $0 \leq t_n \leq 100$. However after this the periodicity breaks up and the flow becomes more unsteady and unpredictable.
In the spatio-temporal spectrum, the dominant wavenumber is once again 3.14, although because we use $L_z = 4$ for this simulation, the next most dominant wavenumber is 1.57, corresponding to a wavelength of 4. The signal at larger wavenumbers appears similar to the case of $\Pi_c = 3725$ at $Pr=7$, quickly dropping off at wavenumbers greater than 7. 
However, there is a notable difference in the temporal frequency content. While a significant signal remains at very low frequencies—such as a peak at 0.002 for a wavenumber of 3.14—there is a much stronger contribution from higher frequencies. For instance, another peak appears at approximately 0.012 for the same wavenumber, and frequencies in the range of 0.02 to 0.04 show noticeable contributions across the wavenumber spectrum. This behavior is consistent with the lower Prandtl number, as reduced momentum diffusion allows velocity fluctuations to persist, while increased thermal diffusivity enhances buoyancy-driven instability. Together, these effects lead to more dynamically unstable flow with a broader distribution of higher-frequency motions.

\subsection{$\Pi_h$ Dependence}

\begin{figure*}
    \centering
    \includegraphics[width=0.6\textwidth]{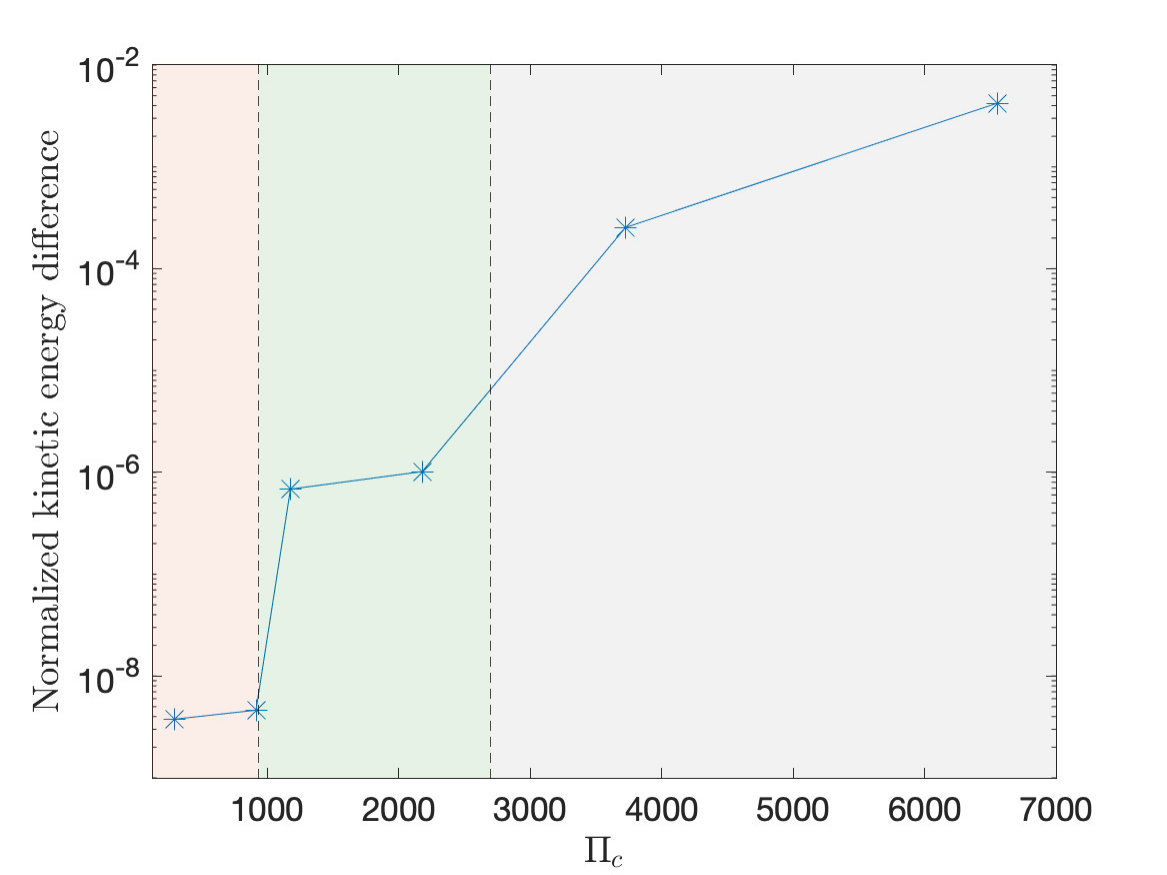}
    \caption{Total normalized kinetic energy difference versus $\Pi_c$. Kinetic energy difference is calculated by subtracting the steady states or time-averaged and 2D-averaged states of two simulations with equal $\Pi_c$ but different $\Pi_h$, and integrating over the valley area. The kinetic energy difference is normalized by the kinetic energy of one of the steady or time-averaged states at each $\Pi_c$ value.}
    \label{fig:KE_vs_PIc}
\end{figure*}

Thus far, we have focused only on the dependence on the $\Pi_c$ parameter, and have not addressed the dependence of each state on the $\Pi_h$ parameter. As shown in \S \ref{sec:prob_des}, the linearized equations under the pure conduction base state depend only on $\Pi_c$, and therefore the solution to the linearized equations is identical for all $\Pi_h$ given the same $\Pi_c$, Prandtl number, and slope angle. The nonlinear equations, on the other hand, are found to depend on both $\Pi_c$ and $\Pi_h$. However, our simulations show only a weak dependence on $\Pi_h$ throughout the parameter space we explore here. To depict this, we select a number of values of $\Pi_c$, and at each value we run two simulations with different $\Pi_h$ values to steady state or time averaged over a sufficiently long time. We then calculate the difference kinetic energy in the 2D-averaged flow field as $E_{K,\mathrm{diff}} = 0.5 \left( (u_2 - u_1)^2 + (v_2 - v_1)^2 \right)$, where subscripts denote the different states at the same $\Pi_c$ but different $\Pi_h$. The difference kinetic energy is then integrated over the 2D area and normalized by the total kinetic energy of one of the 2D flow profiles at the given $\Pi_c$. This normalized kinetic energy difference is plotted for increasing $\Pi_c$ in figure \ref{fig:KE_vs_PIc}, and regimes are colored based on the character of the flow state. In the left-most region, when $\Pi_c < 1000$, the flow is the 2D asymmetric steady state. For these case, the difference between the states with differing $\Pi_h$ is minimal, and there is essentially no dependence on the $\Pi_h$ parameter. In this region, we can assume the nonlinear effects are small, and thus the flow state depends only on the $\Pi_c$ parameter, similar to the case of the linearized equations. In the middle region, from $1000 < \Pi_c < 2700$, the flow becomes 3D and unsteady, with one point representing a steady Danish-pastry state, and the other representing an oscillating Danish-pastry state. In this region, we see an immediate and drastic increase in the kinetic energy difference, but the absolute value is still small, below $10^{-6}$, and thus $\Pi_h$ does not play a significant role. With the onset of three-dimensional instabilities, nonlinearity in the flow becomes more significant, leading to an increase in dependence on $\Pi_h$, but the flow is still laminar and steady, so large nonlinear effects are not observed. Finally, in the last region, for $\Pi_c > 2700$, the flow becomes unsteady, and we observe another large increase in the difference kinetic energy. This indicates that in more dynamically unstable regimes, the nonlinearity of the equations begins to have an increasingly important role on the flow dynamics which leads to greater dependence on the $\Pi_h$ parameter.

In the calculation of difference kinetic energy for all oscillating and unsteady states, we confirm that averaging for a longer period of time does not impact the kinetic energy difference substantially. For the steady oscillating state at $\Pi_c = 2180$, the state was time-averaged for 10 periods of the dominant frequency, and we confirmed that larger time-averaging periods do not lead to substantial decreases in the kinetic energy difference. For the unsteady cases at $\Pi_c = 3726$ and $\Pi_c = 6554$, the simulations are time-averaged for 10 diffusion timescales, defined by $t_0 = H^2/\beta$, which was chosen because it is the longest timescale in our problem. This period corresponds to approximately 37,000 and 65,000 convective timescales ($t_0 = H/u_0$) for the $\Pi_c = 3726$ and $\Pi_c = 6554$ cases respectively. Because of the unsteadiness of these cases, the values of kinetic energy difference are not independent of averaging window, but our tests show that increasing averaging window does not lead to significant differences in total kinetic energy difference.

\begin{figure*}
    \centering
    \begin{subfigure}[b]{0.6\textwidth}
        \centering
        \includegraphics[width=\textwidth]{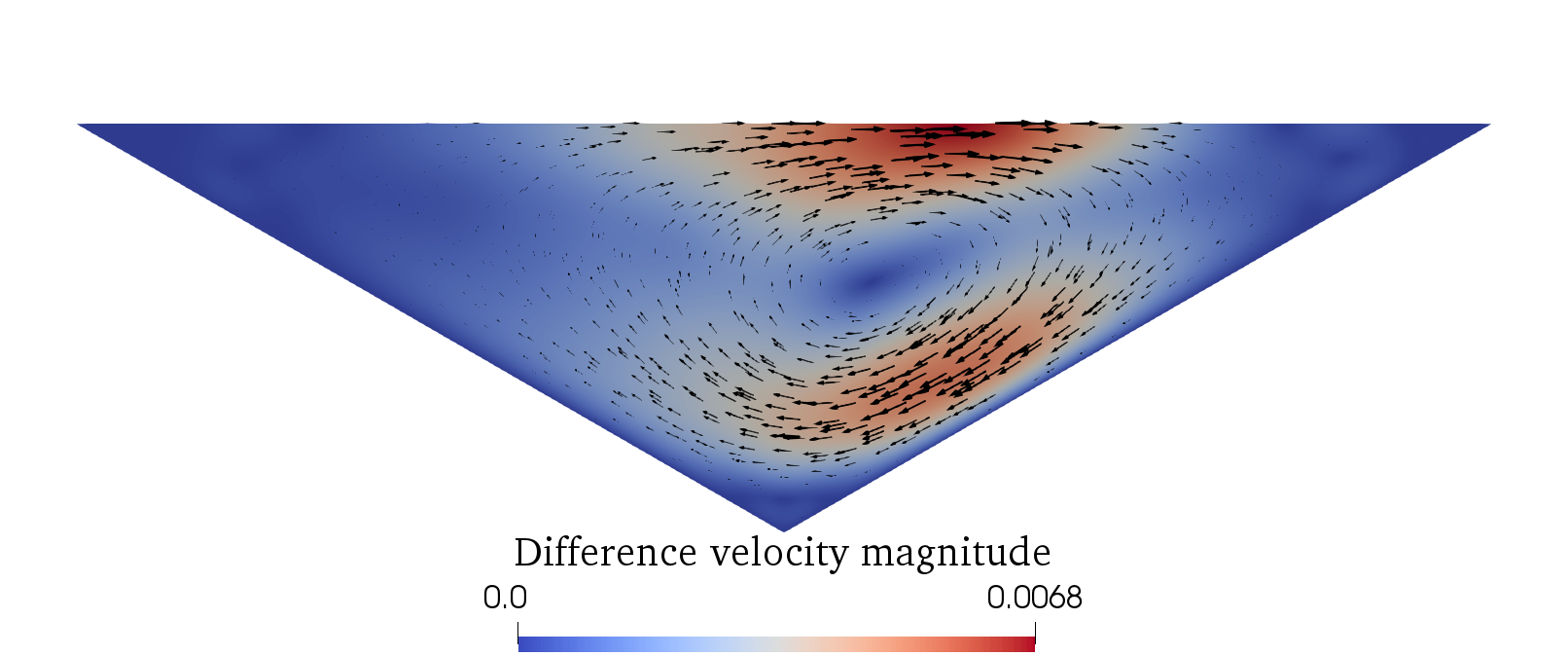}
        \caption{}
        \label{fig:VelDiff_6500PIc_a}
    \end{subfigure}    
    ~
    \begin{subfigure}[b]{0.95\textwidth}
        \centering
        \includegraphics[width=\textwidth]{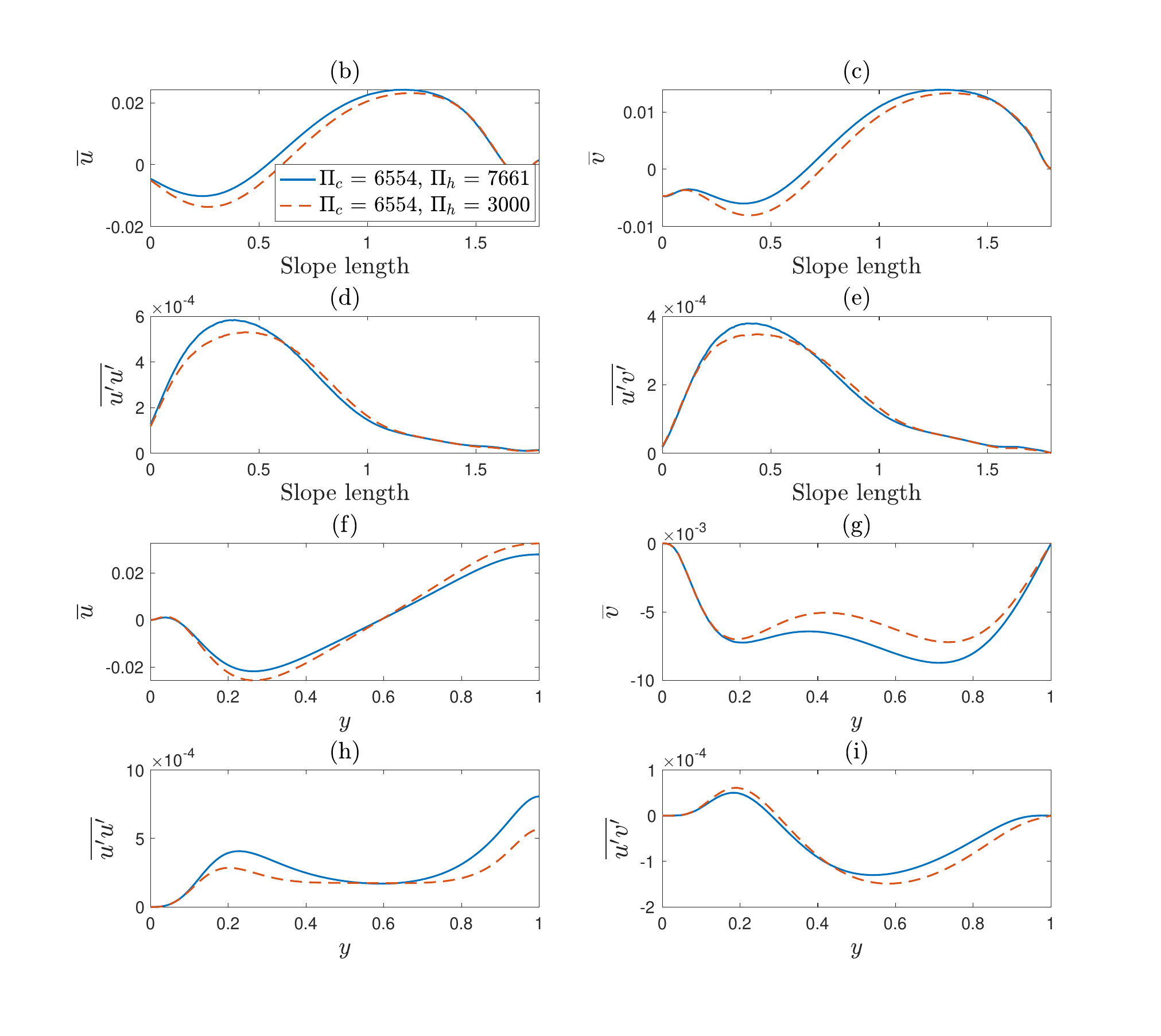}
        \label{fig:VelDiff_6500PIc_subplot}
    \end{subfigure}
    \caption{Comparison between two states with same $\Pi_c = 6554$ and $Pr = 7$, but varying $\Pi_h$. (a) Difference in velocity magnitude between $\Pi_h = 7661$ and $\Pi_h = 3000$, both cases were averaged for 10 diffusion timescales and the 2D average fields were subtracted to obtain the difference in velocity magnitude. Panels (b) and (c) show comparison of time-averaged velocity terms $\overline{u}$ and $\overline{v}$, and (d) and (e) show Reynolds stress terms $\overline{u^{\prime}u^{\prime}}$ and $\overline{u^{\prime}v^{\prime}}$, plotted along the right slope with a vertical distance of 0.1 above the boundary. Panels (f)-(i) show the vertical profiles of $\overline{u}$, $\overline{v}$, $\overline{u^{\prime}u^{\prime}}$, and $\overline{u^{\prime}v^{\prime}}$ at $x=0$.}
    \label{fig:VelDiff_6500PIc}
\end{figure*}

Figure \ref{fig:VelDiff_6500PIc_a} depicts the difference in velocity magnitude for two cases with $\Pi_c = 6554$ and $Pr = 7$, one with $\Pi_h = 7661$ and the other with $\Pi_h = 3000$. This case corresponds to the rightmost point in figure \ref{fig:KE_vs_PIc}. Since both simulations result in a time-averaged state resembling the 2D asymmetric circulation, as depicted in figure \ref{fig:Unsteady6554PIc_c}, the velocity difference highlights variations in the 2D asymmetric flow structure. Figure \ref{fig:VelDiff_6500PIc_a} indicates significant differences in the 2D flow fields throughout the valley, particularly along the right slope and near the top of the domain. Comparing this region to the time-averaged flow field in figure \ref{fig:Unsteady6554PIc_c}, the largest differences are observed in the region between the dominant and secondary circulations.
Figures \ref{fig:VelDiff_6500PIc}b and \ref{fig:VelDiff_6500PIc}c show the time-averaged $u$ and $v$ velocity components along the slope. The largest differences occur near the negative peak of both velocity components, with the $\Pi_h = 7661$ case exhibiting consistently higher velocities than the $\Pi_h = 3000$ case over much of the slope length. Figures \ref{fig:VelDiff_6500PIc}d and \ref{fig:VelDiff_6500PIc}e display the Reynolds stress terms $\overline{u^{\prime}u^{\prime}}$ and $\overline{u^{\prime}v^{\prime}}$ along the right slope, revealing additional notable differences. In particular, the $\Pi_h = 7661$ case exhibits larger peaks with a slightly faster decay, whereas the $\Pi_h = 3000$ case has flatter peaks and a more gradual decline along the slope. 
The vertical profile of the time averaged $u$ and $v$ velocity as well as the Reynolds stress terms $\overline{u^{\prime}u^{\prime}}$ and $\overline{u^{\prime}v^{\prime}}$ are shown in figure \ref{fig:VelDiff_6500PIc}f-i at the location $x=0$. This further shows some divergences in the flow field, especially in the peaks of the $u$ and $v$ velocity as well as in the trend of the variance $\overline{u^{\prime}u^{\prime}}$ along the height of the valley.
These results indicate that variations in $\Pi_h$ at sufficiently high $\Pi_c$ lead to observable differences and further support that $\Pi_h$ has a greater influence in dynamically more unstable regimes.

\begin{figure*}
    \centering
    \includegraphics[width=0.8\textwidth]{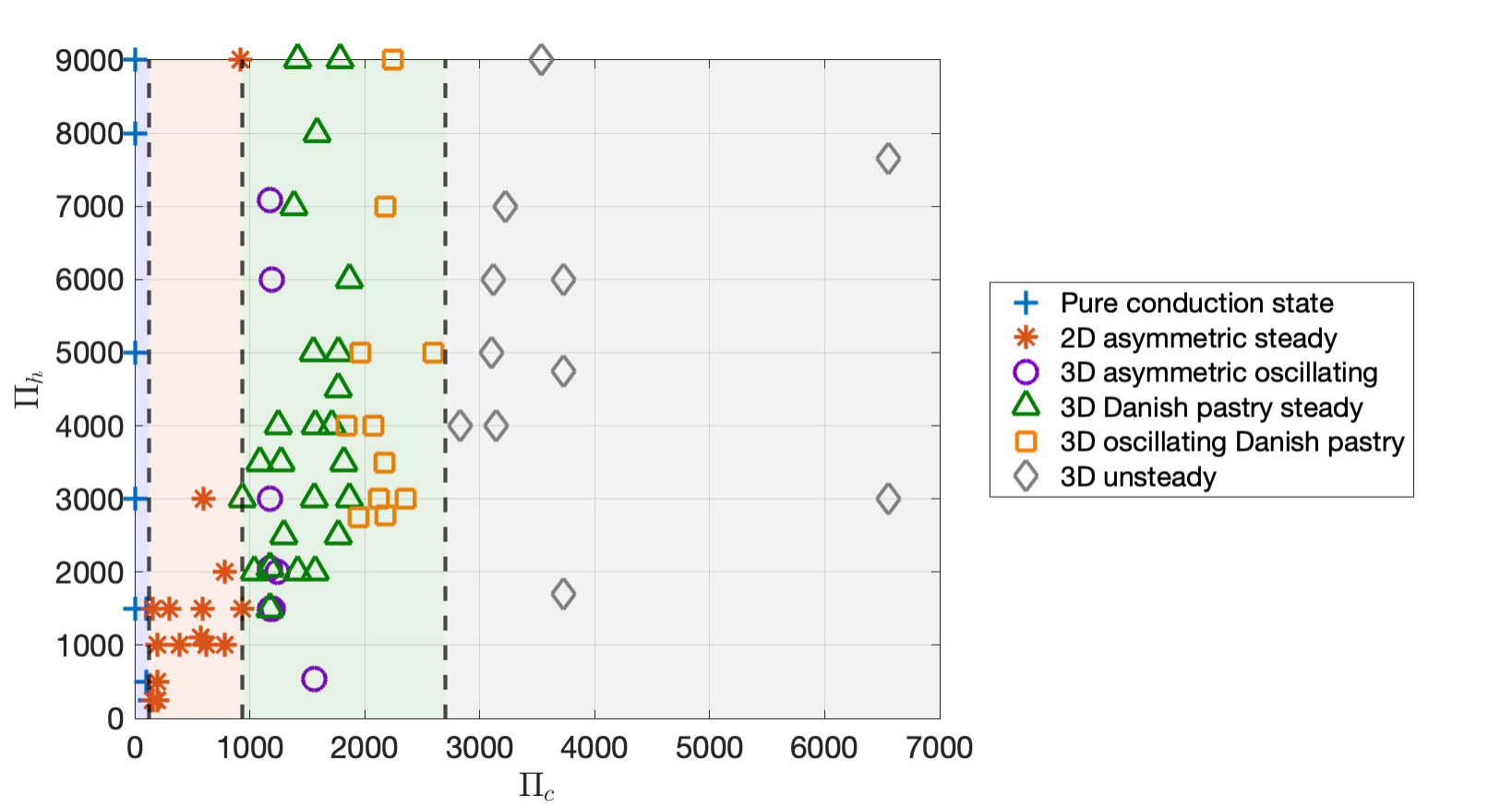}
    \caption{Regime map as a function of $\Pi_h$ and $\Pi_c$ parameters for $Pr = 7$. Regime boundaries and background colors are used to mark significant changes in the flow regime, with the blue region representing the pure conduction state, orange representing 2D flow, green representing the different 3D steady and oscillating states, and finally the gray region representing 3D unsteady flow states. }
    \label{fig:RegimeMap}
\end{figure*}

A regime map depicting all flow states as a function of $\Pi_h$ and $\Pi_c$ is shown in figure \ref{fig:RegimeMap}. Each marker on the regime map represents a simulation and the dashed lines represent approximate boundaries to each regime. The left most regime represents the regime in which the pure conduction state is stable, which is followed by the 2D asymmetric flow state regime. For $\Pi_c > 1000$, the 3D Hopf bifurcation occurs and the 3D Danish-pastry state emerges. This regime, representing 3D steady and oscillating flow state, extends up to $\Pi_c \approx 2700$. Above this value, the flow becomes 3D and fully unsteady, representing the final regime explored in this study. We note that based on the parameter values we have explored here, we see no dependence on the $\Pi_h$ value in our regime map. We do however observe slight dependence on $\Pi_h$ in the 3D unsteady cases with large $\Pi_c$, manifesting as small differences in time-averaged states and Reynolds stress terms. Therefore, it is expected that as the flow becomes more unsteady and eventually turbulent, the dependence on the $\Pi_h$ parameter will increase further. However in this study we do not investigate large enough $\Pi_c$ values to observe such large differences.

\subsection{Transition to chaotic flow}

\begin{figure*}
    \centering
    \begin{subfigure}[b]{0.49\textwidth}
        \centering
        \includegraphics[width=\textwidth]{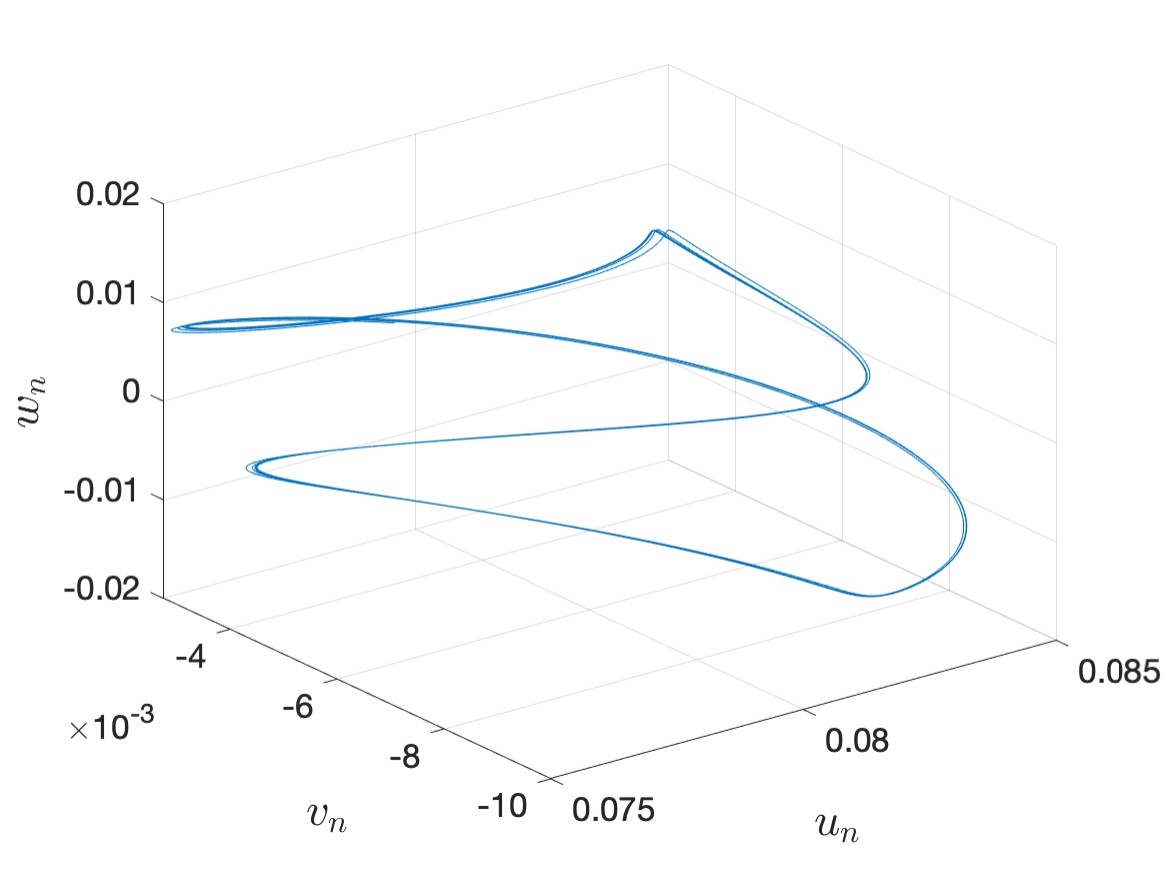}
        \caption{}
        \label{fig:phaseDiag_1178}
    \end{subfigure}%
    ~ 
    \begin{subfigure}[b]{0.49\textwidth}
        \centering
        \includegraphics[width=\textwidth]{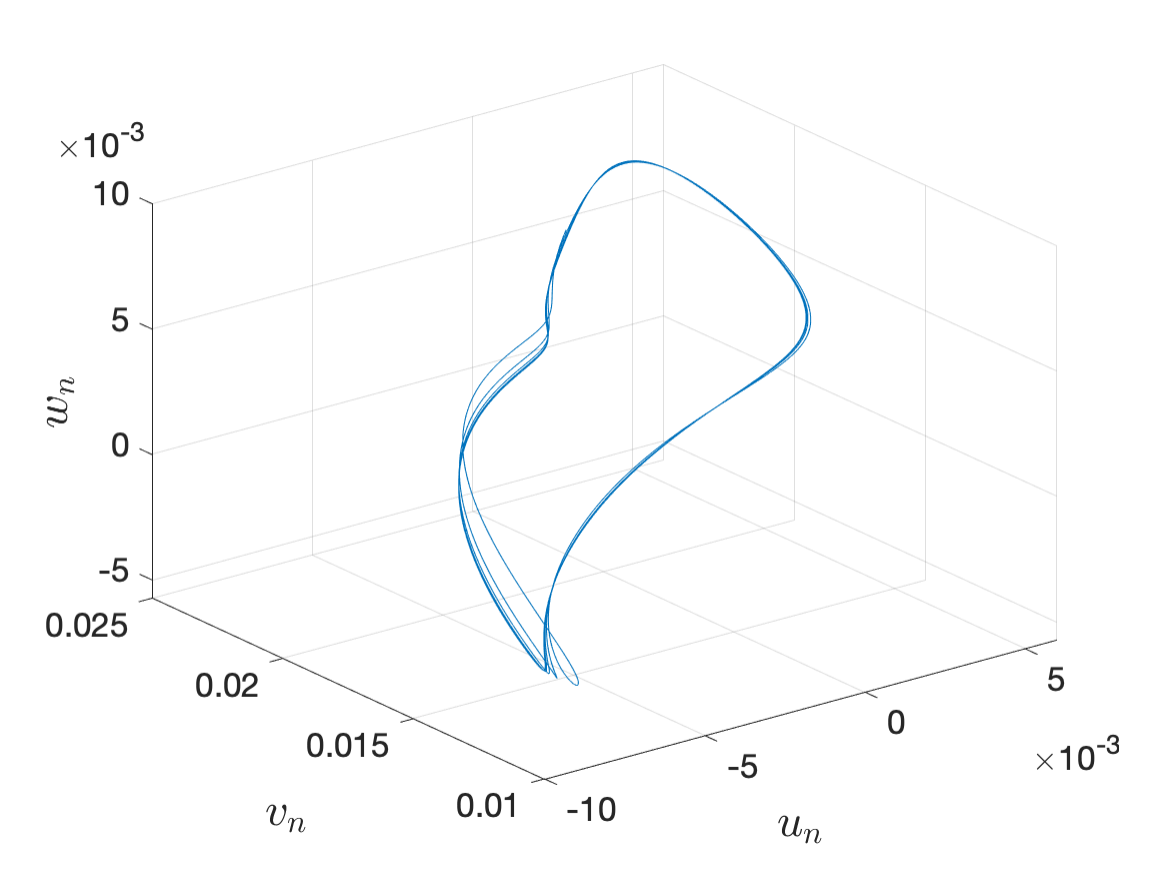}
        \caption{}
        \label{fig:phaseDiag_2180}
    \end{subfigure}
    ~ 
    \begin{subfigure}[b]{0.49\textwidth}
        \centering
        \includegraphics[width=\textwidth]{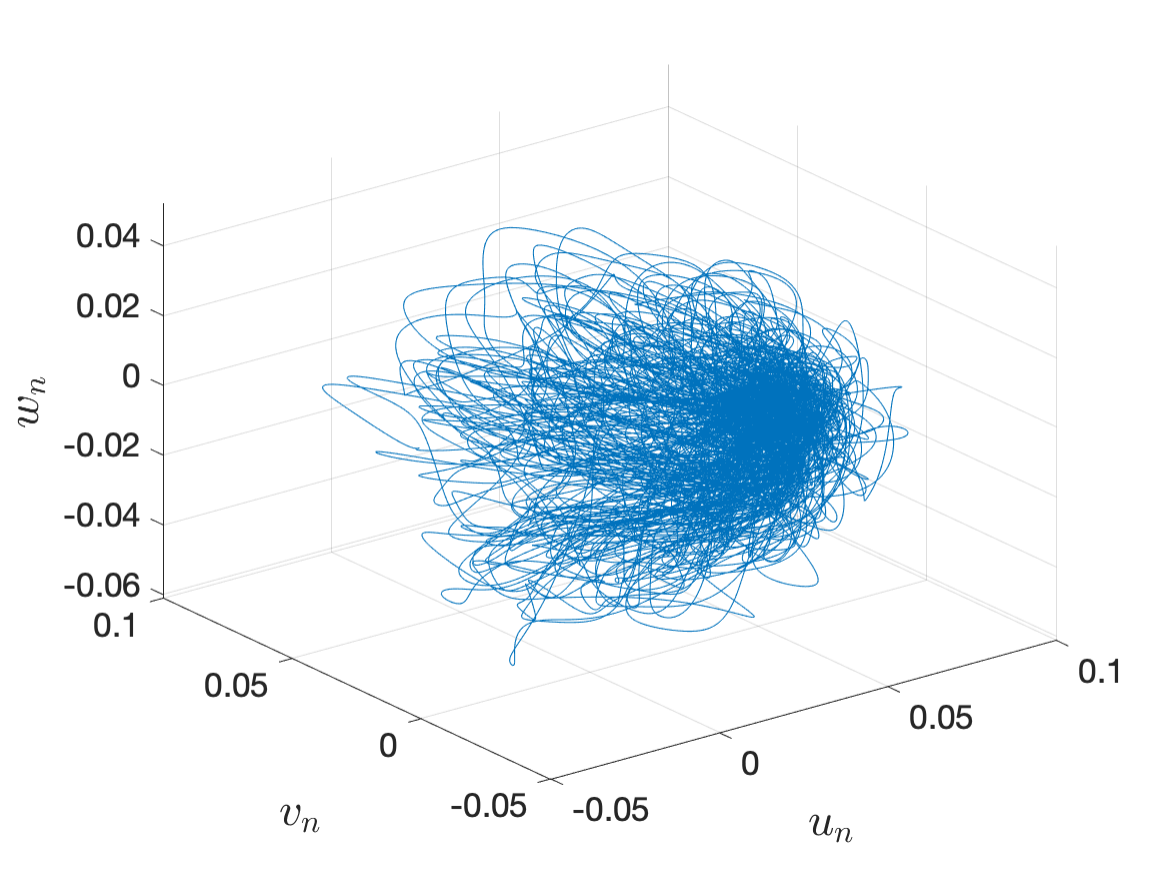}
        \caption{}
        \label{fig:phaseDiag_3700}
    \end{subfigure}%
    ~ 
    \begin{subfigure}[b]{0.49\textwidth}
        \centering
        \includegraphics[width=\textwidth]{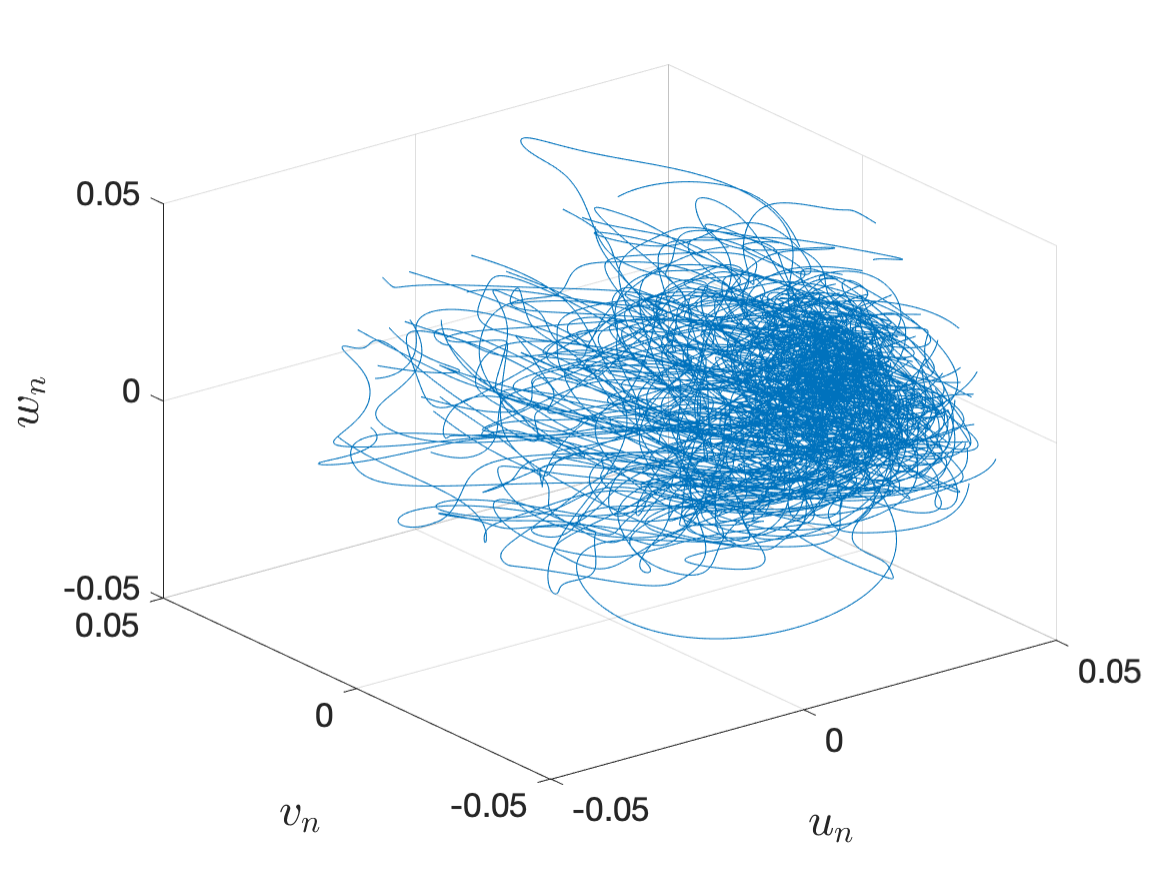}
        \caption{}
        \label{fig:phaseDiag_6500}
    \end{subfigure}
    ~ 
    \begin{subfigure}[b]{0.49\textwidth}
        \centering
        \includegraphics[width=\textwidth]{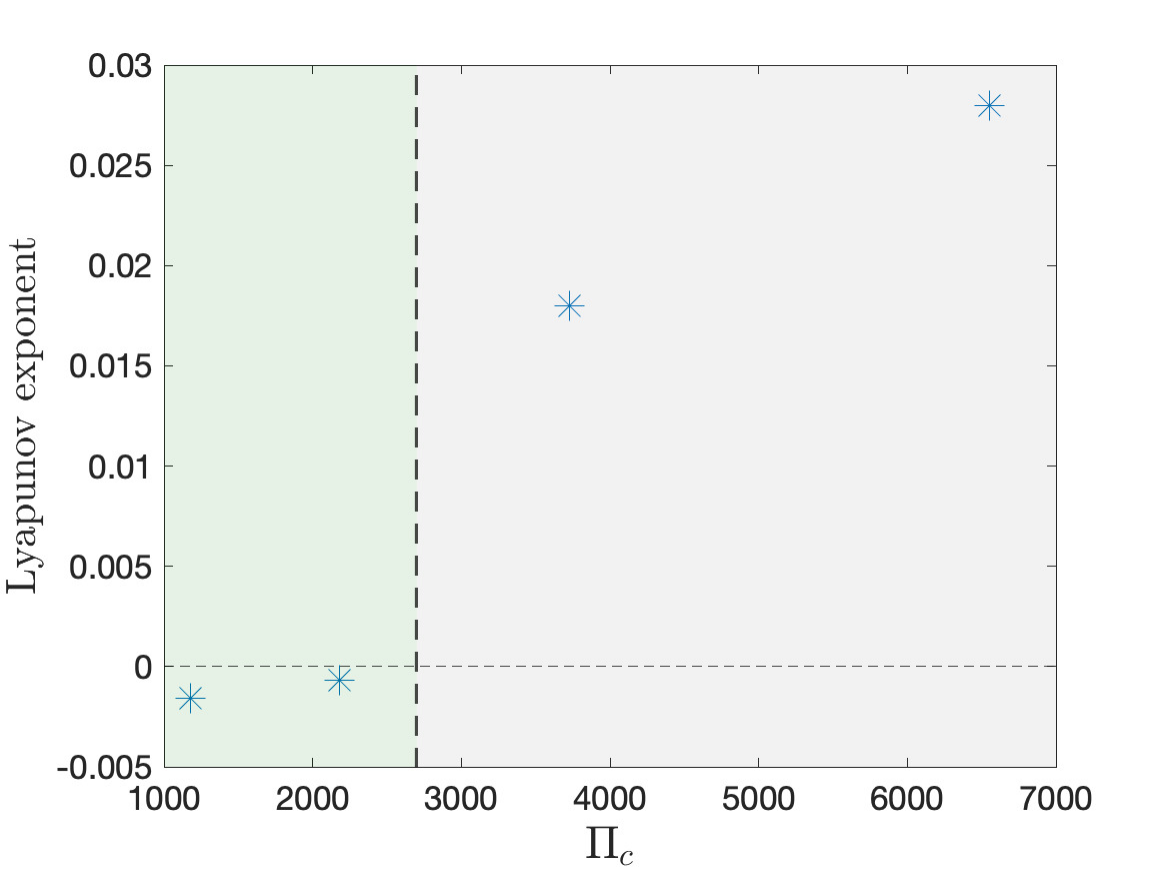}
        \caption{}
        \label{fig:lyaExp_vs_PIc}
    \end{subfigure}
    \caption{Phase diagrams for (a) oscillating asymmetric state at $\Pi_c = 1178$, (b) oscillating Danish-pastry state at $\Pi_c = 2180$, (c) unsteady state at $\Pi_c = 3726$, and (d) unsteady state at $\Pi_c = 6554$ and $\Pi_h = 1705$, using point (0, 0.9, 0). Lyapunov exponent for each state is plotted versus $\Pi_c$ in (e), with the vertical dashed line representing the approximate boundary between periodic and chaotic flow states.}
    \label{fig:phaseDiagrams}
\end{figure*}

To illustrate the transition of flow states as $\Pi_c$ increases, we plot the trajectories of all oscillating and unsteady states analyzed in this study in the $u$-$v$-$w$ phase space, shown in figure \ref{fig:phaseDiagrams}. Each phase diagram is generated from data at the point $(0, 0.9, 0)$ in the domain, collected over many periods of the dominant frequency. The trajectories corresponding to the oscillating asymmetric state and the oscillating Danish-pastry state are shown in figures \ref{fig:phaseDiag_1178} and \ref{fig:phaseDiag_2180}, respectively. As expected, both exhibit closed-loop trajectories that form limit cycles in phase space. However, for $\Pi_c$ values between 2180 and 3726, the flow becomes fully unsteady and chaotic, as evidenced by the phase diagrams in figures \ref{fig:phaseDiag_3700} and \ref{fig:phaseDiag_6500}.

\begin{figure*}
    \centering
    \begin{subfigure}[b]{0.49\textwidth}
        \centering
        \includegraphics[width=\textwidth]{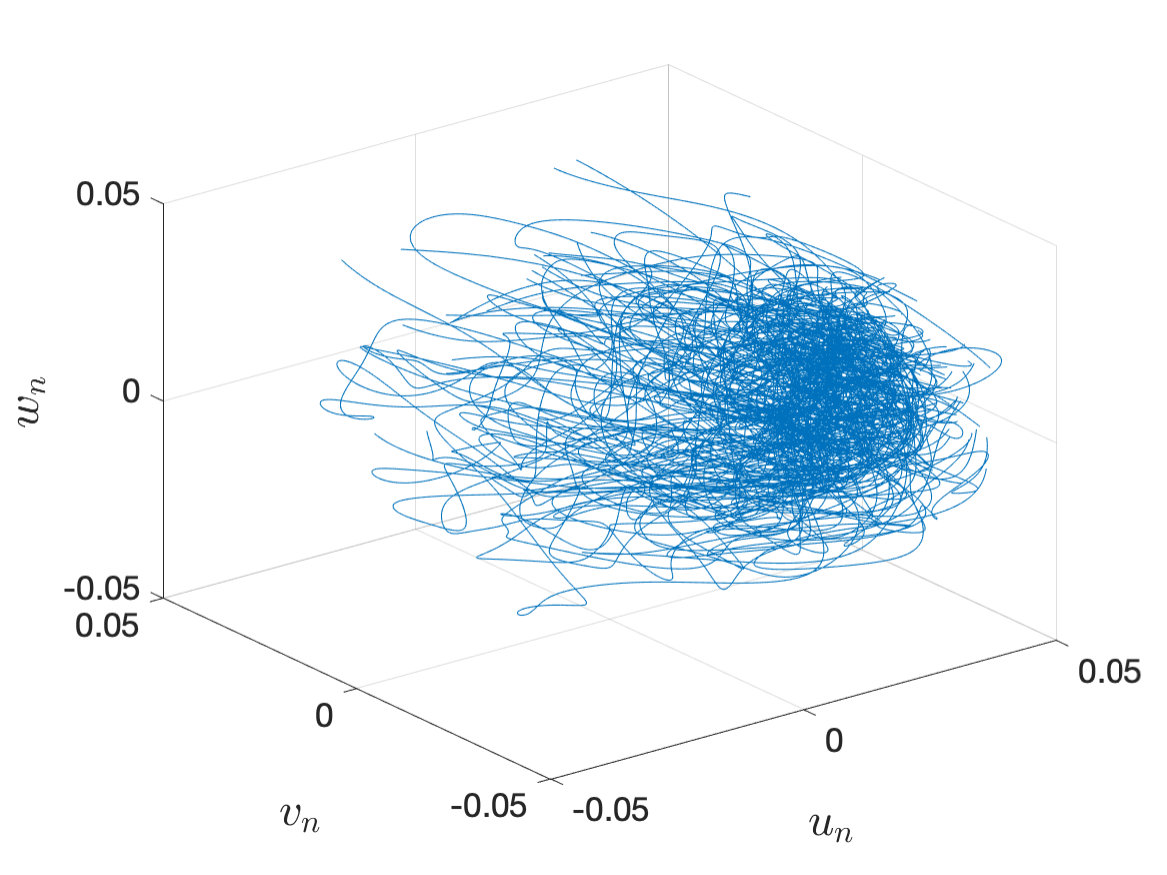}
        \caption{}
        \label{fig:phaseDiag_6500PIc1_5PIs}
    \end{subfigure}%
    ~ 
    \begin{subfigure}[b]{0.49\textwidth}
        \centering
        \includegraphics[width=\textwidth]{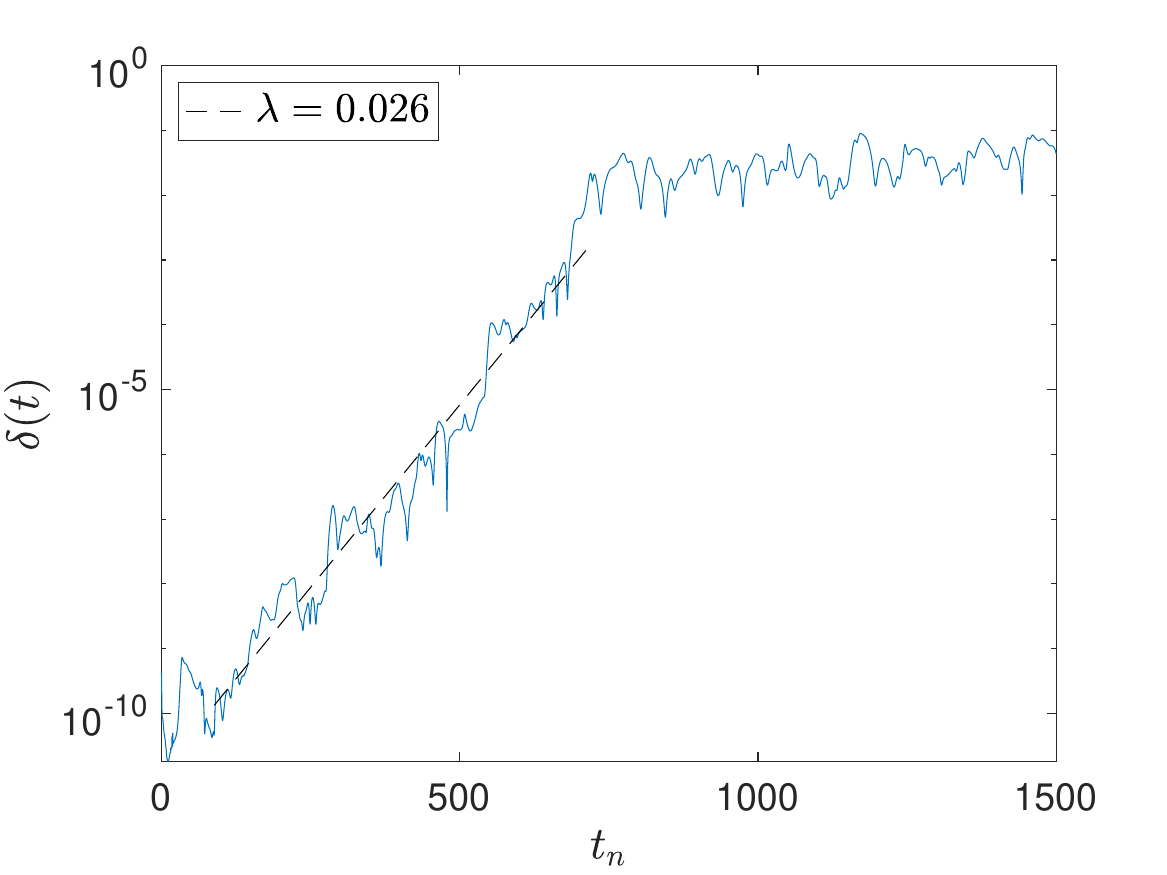}
        \caption{}
        \label{fig:delta_vs_t_1_5PIs}
    \end{subfigure}
    \caption{(a) Phase diagram and (b) Lyapunov exponent for $\Pi_c = 6554$ and $\Pi_h =7661$. Panel (b) shows the distance between trajectories $\delta$ versus time with the slope of the linear fit indicating the value of Lyapunov exponent.}
    \label{fig:LyaExpComp}
\end{figure*}

To determine the onset of chaotic dynamics, we compute the Lyapunov exponent for each periodic and unsteady flow state. The Lyapunov exponent characterizes the rate at which two initially close trajectories separate over time \citep{strogatz2024nonlinear}. For an initial separation $\delta_0$, the evolution of the separation is given by
\begin{equation} 
|\delta(t)| = |\delta_0| \exp^{\lambda t}, 
\end{equation}
where $\lambda$ is the Lyapunov exponent. A positive $\lambda$ indicates chaotic behavior, in which trajectories diverge exponentially.
To calculate $\lambda$ for our cases, we restart each simulation with a small perturbation—of order $10^{-9}$—added to the $u$ velocity field. We then monitor the evolution of the separation between the perturbed and unperturbed trajectories at the point $(0, 0.9, 0)$. Once any initial transient behavior subsides, we fit the resulting time series of $\delta(t)$ to an exponential curve and extract the slope as the maximum Lyapunov exponent. An example of this calculation is shown in figure \ref{fig:delta_vs_t_1_5PIs}.

Figure \ref{fig:lyaExp_vs_PIc} plots the Lyapunov exponent as a function of $\Pi_c$. These results confirm that the flow states at $\Pi_c = 3726$ and $\Pi_c = 6554$ are chaotic, and that the transition from periodic to chaotic flow occurs between $\Pi_c = 2180$ and $\Pi_c = 3726$. Based on our simulations, we estimate the critical value for this transition to be approximately 2700, although a greater number of simulations would be necessary to determine the precise value.

In figure \ref{fig:LyaExpComp}, we show the phase diagrams and Lyapunov exponent for the case $\Pi_c = 6554$ and $\Pi_h =7661$. This represents the same $\Pi_c$ but different $\Pi_h$ than the phase diagram depicted in figure \ref{fig:phaseDiag_6500}, which corresponds to $\Pi_h = 1705$.  
Comparing these two phase diagrams, both display similarly chaotic behavior. Although there are minor differences in the fluctuation patterns, the densest region in phase space is comparable between the two, corresponding to the average velocity at the domain midpoint. 
Figure \ref{fig:delta_vs_t_1_5PIs} plots the separation between perturbed and unperturbed trajectories over time for the state at $\Pi_c = 6554$ and $\Pi_h =7661$. The distance grows exponentially during an initial period, which is characterized by the Lyapunov exponent—indicated by a dashed line in the figure. 
Comparing the value of Lyapunov exponent for different values of $\Pi_h$, we observe only minor differences between the two $\Pi_h$ values: for $\Pi_h = 1705$, the exponent is 0.028, while for $\Pi_h = 7661$, it is 0.026, pointing to similar chaotic dynamics. 

\subsection{Nusselt number scaling}

We now examine the strength of heat transfer within the enclosure as a function of $\Pi_c$. To quantify this, we use the Nusselt number, defined by
\begin{equation}
    Nu = \frac{\frac{\partial b_{\mathrm{total}}}{\partial y}|_w}{\overline{\Delta b_{\mathrm{total}}}} H = \frac{\left(G_y - N^2 \right) H}{\overline{\Delta b_{\mathrm{total}}}},
\end{equation}
where $b_{\mathrm{total}}$ is the buoyancy field plus the background stratification profile given by $b_{\mathrm{total}} = b + (N^2y - N^2)$, and $\overline{\Delta b_{\mathrm{total}}}$ represents the averaged buoyancy difference between the bottom and top walls. We choose to use the average buoyancy difference due to the fact that we use constant heat flux conditions on the bottom walls, meaning the numerator of Nusselt number is imposed, while the temperature is inhomogeneous along the bottom walls. Assumption of averaged temperature difference between the plates has been previously used to calculate the Nusselt number in studies of Rayleigh-B\'enard (R-B) convection with constant heat flux conditions \citep{verzicco2008comparison}. To compare the scaling of our case against other convective flow problems, we define the Rayleigh number. Following the presentation of \citet{verzicco2008comparison}, we define a Rayleigh number based on the imposed heat flux as
\begin{equation}
    Ra_q = \frac{\frac{\partial b_{\mathrm{total}}}{\partial y}|_w H^4}{\nu \beta} = \frac{\Pi_{c}^{2}}{Pr}.
\end{equation}
Given this, Rayleigh number based on temperature difference can be determined by $Ra = Ra_q / Nu$. We use the $Ra$ based on temperature difference in the following scaling analysis.

\begin{figure*}
    \centering
    \begin{subfigure}[b]{0.5\textwidth}
        \centering
        \includegraphics[width=\textwidth]{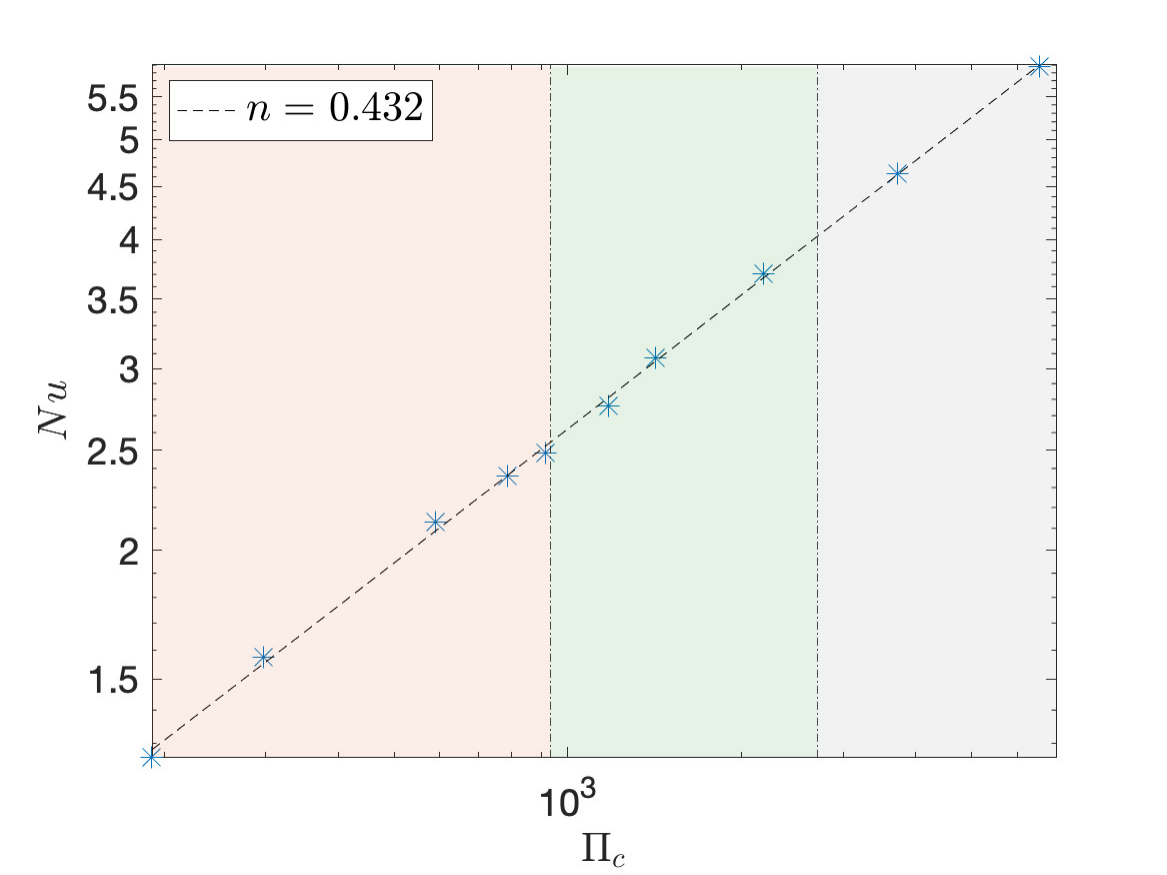}
        \caption{}
        \label{fig:Nu_vs_PIc}
    \end{subfigure}%
    ~ 
    \begin{subfigure}[b]{0.5\textwidth}
        \centering
        \includegraphics[width=\textwidth]{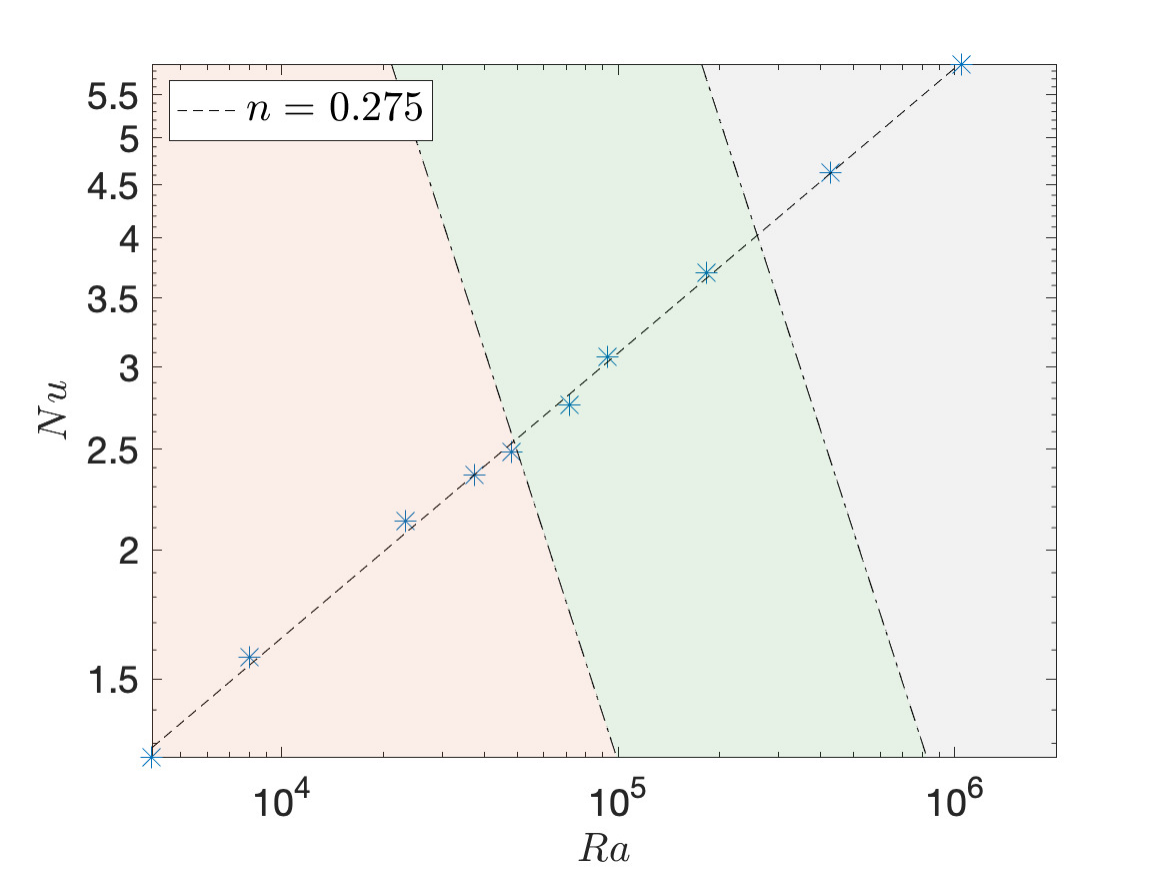}
        \caption{}
        \label{fig:Nu_vs_Ra}
    \end{subfigure}
    \caption{Nusselt number scaling versus (a) $\Pi_c$ and (b) Rayleigh number. The dashed lines indicate scaling as $Nu \sim \Pi_{c}^{n}$, where the value of $n$ is given in the legend. Figures are colored by flow regime in the same way as figure \ref{fig:RegimeMap}. }
    \label{fig:NuScaling}
\end{figure*}

Figure \ref{fig:Nu_vs_PIc} shows a log-log plot of $Nu$ versus $\Pi_c$ over the range of range of parameter values investigated, and \ref{fig:Nu_vs_Ra} shows the same data for $Nu$ versus $Ra$. 
We observe that the Nusselt number scales with $\Pi_c$ as $Nu \sim \Pi_{c}^{0.43}$ throughout the entire range of the parameter space investigated here. Converting this to the Rayleigh number, the scaling is given by $Nu \sim Ra^{0.275}$. This scaling is similar but somewhat smaller than that found in R-B convection for a variety of Prandtl numbers, which is usually bounded between $2/7$ and $1/3$ \citep{silano2010numerical}. On the other hand, our results indicate slightly larger scaling compared to prior studies of convection in triangular cavities. For convection in a V-shaped valley heated from below, \citet{bhowmick2019transition,bhowmick2022chaotic} find a scaling exponent of $1/4$ using Prandtl number corresponding to both air and water. For convection in an attic-shaped cavity (an inverted V-shaped cavity) heated from below, \citet{lei2008unsteady} find a scaling exponent of approximately 0.21 for water, while experimental results again find an exponent of approximately $1/4$ \citep{ridouane2005experimental}. While these studies used fixed temperature boundary conditions, it has been found in R-B convection that fixed heat flux conditions do not lead to significantly different scaling of the Nusselt number than fixed temperature conditions \citep{johnston2009comparison,verzicco2008comparison}. 
Additionally, prior numerical studies of convection in triangular cavities have only reported this scaling relationship in 2D simulations. 
Therefore, our inclusion of constant heat flux boundary conditions along with consideration of three-dimensional flow states may partly explain the slightly larger scaling seen here, although our results do not diverge greatly from the results of prior studies.
Additionally, we note that this power law relationship between Nusselt number and $\Pi_c$ holds across all of the flow regimes investigated here, as can be seen by the background color of the plots in figure \ref{fig:NuScaling}, representing the various flow regimes outlined in the regime map in figure \ref{fig:RegimeMap}. 
The regime boundary lines in figure \ref{fig:Nu_vs_Ra} are slanted because the definition of Rayleigh number depends on both $\Pi_c$ and Nusselt number. Therefore, the regime boundaries represent contours of constant $\Pi_c$ based on Rayleigh and Nusselt numbers.
Further work needs to be done to determine whether this scaling relationship will hold for increasingly large $\Pi_c$ values into the fully turbulent regime.

We note that at the range of $\Pi_c$ and $Pr$ values considered in figure \ref{fig:NuScaling}, the Nusselt number does not show a strong dependence on $\Pi_h$. For example, for the comparison at $\Pi_c = 6554$ shown in figure \ref{fig:VelDiff_6500PIc}, which is the largest $\Pi_c$ value considered here, the difference in Nusselt number is determined to be about 2\%. However, it is expected that at dynamically more unstable regimes and turbulent flow states, as well as for lower $Pr$, the difference in Nusselt number is expected to have a greater dependence on differences in $\Pi_h$. Because of this, future work must extend this analysis into the higher $\Pi_c$ range to determine if this scaling relationship persists into the turbulent regime, or if there is a transition to a new regime.

\section{Conclusion} \label{sec:conclusion}

In this study, we explored the parameter space of a stably stratified fluid confined within a V-shaped valley, subjected to symmetric heating along its sloping walls with an inclination angle of $30^{\circ}.$ Our analysis examines the progression from the initial instability of the quiescent base state to fully three-dimensional, chaotic flow regimes. We identified several distinct flow states that have not been previously reported in studies of heated triangular cavities, namely the oscillating asymmetric state and a newly characterized structure that we refer to as the Danish-pastry state. Furthermore, we delineated the boundaries of each flow regime in terms of the dimensionless parameters $\Pi_c$ and $\Pi_h$ and identify a scaling relation for the heat transfer across these regimes. The key findings of our study are summarized below.

Our investigation has revealed the predominant role of asymmetric circulation within the heated V-shaped valley. In our prior studies \cite{stofanak2024unusual, stofanak2025self}, we demonstrated that the two-dimensional asymmetric circulation was the primary instability that emerged from the quiescent base state and remained the sole steady solution within a certain range of parameter values. The present results extend this finding by showing that the asymmetric circulation is not confined to the two-dimensional regime; rather, it persists as a structural feature throughout the transition to three-dimensional, dynamically unstable, unsteady flow states. It is embedded in all observed three-dimensional steady and oscillatory states, each of which can be interpreted as a three-dimensional instability superimposed on the underlying two-dimensional asymmetric circulation. 

Interestingly, as we transition to increasingly unsteady and chaotic flow regimes--such as the unsteady cases discussed in \S \ref{sec:Unsteady} and \S \ref{sec:Unsteady_0_7Pr}--the time-averaged profiles consistently converge back to the asymmetric flow circulation. Notably, the direction of circulation in these regimes appears to be randomly selected, highlighting the spontaneous emergence of asymmetry despite the symmetric geometry and boundary conditions--a phenomenon also observed in other fluid systems \citep{williams2024asymmetries}. This finding is particularly significant for understanding flows in real valleys, where geometric irregularities and non-uniform surface heating are inherent. Such factors likely amplify asymmetric behavior in atmospheric or geophysical environments.

Another key result has been the reduction in the dimensionless parameters space of the governing nonlinear equations within a specific range of the parameters. Application of the Buckingham-$\pi$ theorem to the governing equations indicates that, in addition to the Prandtl number and slope angle, the flow dynamics should depend on two key dimensionless parameters: the composite stratification parameter $\Pi_c$, and the buoyancy parameter $\Pi_h$. 

Our results unequivocally demonstrate that, across most of the parameter space, the buoyancy parameter $\Pi_h$ has little to no influence. Only at large $\Pi_c$ values do variations in $\Pi_h$ introduce minor differences in time-averaged states and Reynolds stress terms. This finding aligns with the results of \citet{xiao2022impact}, who investigated the influence of stratification mechanisms on turbulent channel flows. Their analysis showed that the linearized equations depend on a single composite Froude number, whereas fully turbulent simulations exhibit sensitivity to two dimensionless parameters. Similarly, we propose that the range of $\Pi_c$ values investigated in the present study may be insufficient to fully capture the influence of $\Pi_h$, and that the sensitivity to $\Pi_h$ is likely to increase in more unstable and turbulent flow regimes. This hypothesis is supported by our unsteady, 3D simulations, which exhibit chaotic dynamics characterized by a positive Lyapunov exponent.  

Our investigation has established a necessary condition for the instability of the pure-conduction base state in a stably stratified V-shaped symmetric valley. For any slope angle of its walls, the quiescent conduction state is guaranteed when $0< \Pi_s < \cos \alpha$, where $\Pi_s$ is the stratification perturbation parameter \citep{xiao2019stability}. Additionally, we quantify the heat flux in the valley using the Nusselt number and establish that it follows a power law relationship as $Nu \sim \Pi_{c}^{0.43}$ across the flow regimes observed in the present study. By converting $\Pi_c$ to the classical Rayleigh number definition based on temperature difference, we derive the scaling relation: $Nu \sim Ra^{0.275}$. This trend is consistent with the scaling behavior observed in other convective flow problems, such as Rayleigh-B\'enard convection and convection in V-shaped and attic-shaped triangular cavities. The persistence of this scaling relation in turbulent regimes remains an open question for future investigations.

Finally, our results suggest that the Prandtl number has a moderate influence on the unsteady chaotic regimes than on stationary flow regimes. For example, at much lower $\Pi_c$ values, we observe a noticeable increase in unsteadiness for $Pr = 0.7$ compared to $Pr = 7$, along with differences in time-averaged flow structures and frequency content.
For unsteady flow regimes with a small Prandtl number, the dependence on $\Pi_h$ is more pronounced due to heightened dynamical instability and intensified nonlinear interactions.
In the context of real valley flows, the implication is that $\Pi_h$ dependence could be significant, as atmospheric flows exhibit very large $\Pi_c$ values and small Prandtl numbers. Conversely, in small-scale experimental studies of stratified flow instabilities using a water tank with $Pr\approx 7$, $\Pi_h$ and $Pr$ dependence can likely be neglected, because flow dynamics is primarily controlled by $\Pi_c$. Further computational studies will be necessary to determine how heat transfer characteristics depend on the Prandtl number.

\bibliographystyle{unsrtnat}
\bibliography{references}  






\end{document}